\documentclass[11pt]{article}
\usepackage{amsfonts,slashed}
\usepackage{arydshln}
\usepackage[greek,english]{babel}
\usepackage{cancel}
\usepackage{upgreek}
\usepackage{float}
\usepackage{url}
\usepackage{latexsym}
\usepackage{amsfonts}
\usepackage{amsmath}
\usepackage{epsfig}
\usepackage{latexsym,amssymb}
\usepackage{amsmath,amssymb,amsthm}
\usepackage{mathrsfs}
\usepackage{hyperref}
\usepackage{fontenc}
\usepackage{graphicx,tikz}
\usetikzlibrary{decorations.markings,arrows.meta,positioning,calc,patterns}
\usepackage{multirow}
\usepackage{xfrac}
\usepackage[T1]{fontenc}
\usepackage{textcomp}
\usepackage{lmodern}
\usepackage{stmaryrd}
\usepackage{cite}
\usepackage{lipsum}

\usepackage{subfigure}

\usepackage{stmaryrd}

\usepackage[margin=20pt,small]{caption}
\usepackage{subcaption}
\usepackage[framemethod=tikz]{mdframed}
\definecolor{mycolor}{RGB}{102,0,51}
\newmdenv[innerlinewidth=0.5pt, roundcorner=4pt,linecolor=mycolor,innerleftmargin=6pt,innerrightmargin=6pt,innertopmargin=6pt,innerbottommargin=6pt]{mybox}
\usepackage{ifpdf}
\ifx\pdfoutput\undefined
   \pdffalse
   \usepackage{cite}
 \else
   \pdfoutput=1
   \pdftrue
\fi
\DeclareFontFamily{OMS}{rsfs}{\skewchar\font'60}
\DeclareFontShape{OMS}{rsfs}{m}{n}{<-5>rsfs5 <5-7>rsfs7 <7->rsfs10 }{}
\DeclareSymbolFont{rsfs}{OMS}{rsfs}{m}{n}
\DeclareSymbolFontAlphabet{\Scr}{rsfs}

\numberwithin{equation}{section}
\def\be{\begin{equation}}
\def\ee{\end{equation}}
\def\ba{\begin{array}}
\def\ea{\end{array}}

\newcommand{\bea}{\begin{eqnarray}}
\newcommand{\eea}{\end{eqnarray}}

\def\={~=~}
\def\*{{}^*}

\def\M{\mathcal{M}}

\newcommand{\uK}{{\underline{K}}}
\newcommand{\uL}{{\underline{L}}}
\newcommand{\uM}{{\underline{M}}}
\newcommand{\uN}{{\underline{N}}}
\newcommand{\uP}{{\underline{P}}}

\newcommand{\uR}{{\underline{R}}}

\def\={~=~}
\def\*{{}^*}

\def\M{\mathcal{M}}

\usepackage{mathrsfs}
\newcommand{\dd}{\ensuremath{\mathfrak{d}}}
\newcommand{\bg}[1]{{#1}}

\newcommand{\down}[1]{\mathring{#1}}
\usepackage[bold=0.1]{xfakebold}
\newcommand{\fbseries}{\unskip\setBold\aftergroup\unsetBold\aftergroup\ignorespaces}
\newcommand{\g}{\text{\fbseries g}}

\begin{document}

\begin{titlepage}
\rightline{}
%\rightline\today
\rightline{July 2026}
\rightline{HU-EP-26/22-RTG}  
\begin{center}
\vskip 1.5cm
{\Large \bf{Kaluza-Klein Perturbation Theory from Exceptional Field Theory}}
\vskip 1.2cm

{\large\bf {Camille Eloy$^1$, Olaf Hohm$^2$, Camilla Lavino$^2$, \\[1ex]  Henning Samtleben$^{1,3}$ and  Yehudi Simon$^1$ }} \vskip 1.1cm

$^1$ {\it  ENS de Lyon, CNRS, LPENSL, UMR5672, 69342, Lyon cedex 07, France}\\[2.5ex]

$^2$ {\it  Institute for Physics, Humboldt University Berlin,\\
 Zum Gro\ss en Windkanal 6, D-12489 Berlin, Germany}\\[2.5ex]

$^3$ {\it  Institut Universitaire de France (IUF)}\\ [4ex]

 camille.eloy@ens-lyon.fr, ohohm@physik.hu-berlin.de, lavinoca@physik.hu-berlin.de, henning.samtleben@ens-lyon.fr, yehudi.simon@ens-lyon.fr

\vskip .1cm

\vskip .2cm

\end{center}

\bigskip\bigskip
\begin{center} 
\textbf{Abstract}

\end{center} 
\begin{quote}

We develop the perturbation theory of ten and eleven-dimensional supergravities on a large class of Kaluza-Klein backgrounds  including familiar AdS examples such as \mbox{AdS$_5\times S^5$}, but also more general manifolds such as black hole geometries.  
Employing ${\rm E}_{6(6)}$ exceptional field theory 
with the backgrounds characterized by a generalized Scherk-Schwarz ansatz, we determine the first 
order field equations for the fluctuations that are gauge invariant 
under linearized generalized diffeomorphisms. 
We then present the details of the Higgs mechanism for all fields including spin-2 for the subset of backgrounds in which higher-form gauge fields vanish. We use a recently established 
machinery based on homotopy transfer that allows one to separate the fields into gauge invariant physical modes  and pure gauge unphysical modes  to all orders in perturbation theory. 
Finally, as a first application for backgrounds with higher-form gauge fields switched on, 
we analyze part of the spectrum of ten-dimensional Kaluza-Klein modes of type IIB supergravity around a Kerr-Newman AdS$_5$ black hole that in the near horizon limit becomes a fibered product of AdS$_2$ and a squashed three-sphere.

\end{quote} 
\vfill
\setcounter{footnote}{0}
\end{titlepage}

\tableofcontents
%\newpage

\section{Introduction}

Our goal in this paper is to construct the perturbation theory for ten and eleven-dimensional supergravities on a large class of Kaluza-Klein backgrounds. 
 In particular, we obtain, to linear order in fields, the full equations for all massive Kaluza-Klein modes, in addition to the massless fields. 
More precisely, we obtain those equations for the class of generalized Scherk-Schwarz backgrounds in exceptional field theory  
with five external dimensions.

Exceptional field theory (ExFT) is  a reformulation of higher-dimensional maximal supergravity in which U-duality groups are manifest before compactification,  thanks to extended coordinates. This is in contrast to the standard formulation of supergravity, 
where the U-duality groups only arise upon dimensional reduction (torus compactification plus truncation to the massless modes) 
\cite{Hohm:2013pua,Hohm:2013vpa}.

We focus on compactifications to five-dimensional (5D)  supergravity, 
which, in the ungauged case, exhibits the global exceptional symmetry group   ${\rm E}_{6(6)}$. The corresponding ExFT rewrites ten or eleven-dimensional supergravity 
in an ${\rm E}_{6(6)}$ covariant  way that looks structurally like 5D  gauged supergravity, but with the dependence on all internal coordinates retained inside
covariant derivatives, non-abelian curvatures and other structures defining the theory. Concretely, this is achieved by introducing coordinates 
$Y^M$, $M=1,\ldots, 27$, in the fundamental representation of~${\rm E}_{6(6)}$, together with an ${\rm E}_{6(6)}$ covariant `section constraint' 
on the extended derivatives $\partial_M= \frac{\partial}{\partial Y^M}$. The solutions of this section constraint 
reduce the $Y^M$ to the five or six internal coordinates of ten or eleven-dimensional supergravity. 
The fields include the five-dimensional metric ${\g}_{\mu\nu}(x,Y)$, an internal generalized scalar metric ${\cal M}_{MN}(x,Y)$, together with  one- and two-form 
gauge fields. 
This is the field content of 5D  gauged supergravity, but 
here the fields still depend on all $5+27$ coordinates, 
modulo the section constraint.

One of the most powerful applications  of ExFT has been to prove the consistency of a large class of Kaluza-Klein truncations, 
using the technique of generalized Scherk-Schwarz compactifications~\cite{Lee:2014mla,Hohm:2014qga,Baguet:2015xha}. One expresses the full ExFT fields in terms of the 5D fields 
by means of a $Y$-dependent~${\rm E}_{6(6)}$ valued $27\times 27$ matrix $U$ and a scale factor $\rho$:
 \begin{equation} \label{IntroSSS}
    	\begin{aligned}
    		{\g}_{\mu\nu}(x,Y) & = \rho^{-2}(Y)\,{g}_{\mu\nu}(x) \;,\\
    		{\cal M}_{MN}(x,Y) & = U_{M}{}^{\underline{M}}(Y)\,U_{N}{}^{\underline{N}}(Y)\,{M}_{\underline{MN}}(x)\;,
        	\end{aligned}
    \end{equation}
with analogous formulas for the higher form gauge fields. In order to distinguish between the  ${\rm E}_{6(6)}$ indices of ExFT and those 
of 5D gauged supergravity, we underline the latter. The generalized Scherk-Schwarz data $U$ and  $\rho$ are subject to 
generalized parallelizability conditions. The general formalism of ExFT guarantees that any solution  $U$ and  $\rho$ of these  conditions 
yields, via the 
above formulas, a consistent uplift to ten- or eleven-dimensional supergravity. More precisely, this is a 
consistent Kaluza-Klein truncation: 
for any solution ${g}_{\mu\nu}(x)$, ${M}_{\underline{MN}}(x)$, etc.,   
of 5D gauged supergravity the above formulas yield a solution ${\g}_{\mu\nu}(x,Y) $, ${\cal M}_{MN}(x,Y)$, etc., of ExFT and hence 
of  ten- or eleven-dimensional supergravity.

Given this general relation between the solutions of 5D gauged supergravity and those of higher-dimensional supergravity it is  
natural to develop the perturbation theory in ten or eleven dimensions around those backgrounds that are obtained via the Scherk-Schwarz uplift 
from 5D gauged supergravity. To this end we add perturbations to the background solutions (\ref{IntroSSS}) as follows: 
\begin{equation} \label{INTRO:reductionansatz}
    \begin{aligned}
        {\g}_{\mu\nu}(x,Y) & =  \rho^{-2}(Y)\,\Big({g}_{\mu\nu}(x) + h_{\mu\nu}(x,Y)\Big)\;,\\
        \M_{MN}(x,Y) & = U_{M}{}^{\underline{M}}(Y)U_{N}{}^{\underline{N}}(Y)\Big({M}_{\underline{MN}}(x) + m_{\underline{MN}}(x,Y)\Big)\;,
    \end{aligned}
\end{equation}
again with similar ans\"atze for the higher form gauge fields. The fluctuations are functions of all~$5+27$ coordinates $(x^{\mu}, Y^M)$ subject to the section constraint. 
Expanding the fluctuations  in suitable harmonics of the internal space yields the higher Kaluza-Klein modes. For AdS backgrounds this ansatz leads to compact universal mass matrices for the entire Kaluza-Klein spectrum~\cite{Malek:2019eaz,Malek:2020yue}.
This has been exploited in many examples, see e.g.\
\cite{Guarino:2020flh,Eloy:2020uix,Cesaro:2020soq,Bobev:2020lsk,Cesaro:2021tna,Eloy:2021fhc,Eloy:2023acy,Eloy:2024lwn}.
In this paper 
we will work out the field equations for the fluctuations in ${\rm E}_{6(6)}$  ExFT to first order around generalized  
Scherk-Schwarz backgrounds~(\ref{IntroSSS}).

As a first step towards a general analysis of the Kaluza-Klein spectrum we will then work out, 
for the first time, the details of the Higgs mechanism, 
not yet for the most general class of backgrounds, but for those where the higher-form gauge fields vanish in the background and 
the background scalar metric ${M}_{\underline{MN}}$ is constant. These assumptions still leave a large class of backgrounds including AdS$_5\times S^5$ but 
going far  beyond it by allowing for  general Einstein manifolds in the external space with metric $g_{\mu\nu}(x)$. 
The gauge symmetries of ExFT, governed by a higher gauge algebra of generalized diffeomorphisms and their higher-form descendants, 
give rise to subtle gauge redundancies among the fluctuation fields that, as usual, make it hard to determine the physical spectrum.
 In past studies on AdS backgrounds \cite{Malek:2020yue}, the physical spectrum was determined by reading off the naive mass matrices and counting the Goldstone mode directions in order to eliminate by hand certain eigenvalues. 
In this paper, following the recent treatment of the torus toy model \cite{Eloy:2025ebd}, we give a completely  systematic analysis of 
the Higgs mechanism, i.e., the rearrangement of fields into gauge invariant and hence physical massive modes plus unphysical pure gauge modes that 
can be eliminated from the field content. To this end we use the formulation of perturbative field theories in terms of $L_{\infty}$ algebras 
and the technique of homotopy transfer \cite{Zwiebach:1992ie,Zeitlin:2007vv,Hohm:2017pnh,Arvanitakis:2020rrk}, employing and generalizing the results of \cite{Chiaffrino:2020akd}. 
(See also \cite{Hohm:2022pfi} for the closely related case of double field theory on cosmological backgrounds.) 
While in the present paper we only work out the linear theory, the power of the homotopy transfer theorem 
is that it provides an algorithm to determine the rearrangement into physical fields to any order in perturbation theory.

As a second application to the fluctuation equations around generalized Scherk-Schwarz backgrounds we illustrate our formulas for a class of extremal Kerr-Newman-${\rm AdS}_{5}$ black holes in type IIB supergravity. These solutions break the 5D covariance and feature a ${\rm U}(1)$ vector in the background, giving rise to non-trivial couplings within and between the fluctuations equations. We explore these couplings in the near-horizon limit in which the black holes isometries are given by a squashed three-sphere $S^{3}$ fibered over an ${\rm AdS}_{2}$ spacetime. We focus here our study on fields that do not couple to modes of different five-dimensional spin, thus reducing the couplings to those arising from the black hole backgrounds. We study the stability of this large class of modes thanks to an analysis of the fluctuations equations from the perspective of the ${\rm AdS}_{2}$ factor. Such a study has already been conducted in~\cite{Ezroura:2024xba} but restricted to the fluctuations of the five-dimensional theory, whereas our analysis include the fluctuations of the full ten-dimensional Kaluza-Klein towers.

\medskip

The paper is organized as follows. We display in section~\ref{sec:ExFT5dsugra} the main aspects of ${\rm E}_{6(6)}$ Exceptional Field Theory and five-dimensional gauged supergravity. In particular, we review how the latter is described in ExFT using a generalized Scherk-Schwarz ansatz. 
Then, in section~\ref{sec:pertubtheory}, we develop the perturbation theory around five-dimensional generalized Scherk-Schwarz backgrounds, with a particular emphasis on the linearized fluctuation equations and gauge transformations.
 This general framework is then applied for  two different examples. In section~\ref{sec:homoyopytransfer}, we  focus on backgrounds with vanishing higher-form gauge fields and constant scalars for which the Higgs mechanism is worked out in details using homotopy transfer. 
 To help the reader navigate through this section, we summarize the main formulas in table \ref{tab:homotopytransfer}.
 In section~\ref{sec:BHbackground} we perform the analysis of the Kaluza-Klein spectrum of Kerr-Newman-${\rm AdS}_{5}$ black holes in type IIB supergravity. We conclude in section~\ref{sec:ccl} with prospects on future extensions of this work, and some technical details in appendix~\ref{app:Formulas}.

\section{\texorpdfstring{\boldmath ${\rm E}_{6(6)}$}{E6(6)} Exceptional Field Theory and 5D Gauged Supergravity} \label{sec:ExFT5dsugra}

${\rm E}_{6(6)}$ Exceptional Field Theory is a reformulation of type II and 11d supergravities that is explicitly covariant under the ${\rm E}_{6(6)}$ duality symmetry that arises when these theories are compactified down to five dimensions. We review in this section its main features and how to connect it to maximal five-dimensional supergravity.

    \subsection[{Basics of \texorpdfstring{${\rm E}_{6(6)}$}{E6(6)} Exceptional Field Theory}]{Basics of \texorpdfstring{\boldmath ${\rm E}_{6(6)}$}{E6(6)} Exceptional Field Theory} \label{sec:ExFTrecap}

    ${\rm E}_{6(6)}$ ExFT was first constructed in ref.~\cite{Hohm:2013vpa} (see also ref.~\cite{Baguet:2015xha} for a review). It reformulates the bosonic dynamics of type II and 11d supergravities in terms of the following fields: 

    \begin{equation} \label{eq:ExFTfields}
        \Big\{\g_{\mu\nu}, {\cal M}_{MN}, {\cal A}_{\mu}{}^{M}, {\cal B}_{\mu\nu\,M}\Big\},
    \end{equation}
    where $\mu\in\llbracket1,5\rrbracket$ are external five-dimensional indices and $M\in\llbracket1,27\rrbracket$ labels the (anti-)fundamental representations $\boldsymbol{27}$ and $\boldsymbol{\bar 27}$ of ${\rm E}_{6(6)}$. The field $\g_{\mu\nu}$ is the five-dimensional metric, the scalars ${\cal M}_{MN}$, sometimes called the generalized metric, parametrize the coset ${\rm E}_{6(6)}/{\rm USp(8)}$, and ${\cal A}_{\mu}{}^{M}$ and ${\cal B}_{\mu\nu\, M}$ are one-form and two-form gauge fields in the representations $\boldsymbol{27}$ and $\boldsymbol{\bar 27}$ of ${\rm E}_{6(6)}$. They all depend on the $5$ external coordinates $x^{\mu}$ and on 27 internal coordinates $Y^{M}$. The dependence on $Y^{M}$ is constrained by the section condition
    \begin{equation} \label{eq:sectioncond}
        d^{MNP}\,\partial_{M} \otimes \partial_{N} = 0,
    \end{equation}
    where the $\otimes$ product means that the derivatives act on any product of fields, and $d^{MNP}$ and $d_{MNP}$ are the totally symmetric invariant tensors of ${\rm E}_{6(6)}$ with normalisation
    \begin{equation}\label{eq:norm}
    d_{MPQ}\,d^{NPQ} = \delta_M{}^N.
    \end{equation}
    The section condition~\eqref{eq:sectioncond} ensures that only 5 (respectively 6) of the 27 internal coordinates $Y^{M}$ are physical, corresponding to the internal coordinates of the type II (respectively 11d) supergravities compactified down to five-dimensions.

     The ${\rm E}_{6(6)}$ ExFT is gauge invariant under so-called internal ${\rm E}_{6(6)}$ generalized diffeomorphims, which makes it compatible with the ${\rm E}_{6(6)}$ duality symmetry of five-dimensional supergravity.
     These diffeomorphisms are defined on vectors ${\cal V}^{M}$ and co-vectors ${\cal W}_{M}$ of weight $\lambda$ as
    \begin{equation}  \label{eq:genLie}
        \begin{aligned}
            \mathbb{L}_{\Lambda}{\cal V}^M &= \Lambda^K \partial_K {\cal V}^M - 6\, \mathbb{P}_N{}^M{}_L{}^K\,\partial_K \Lambda^L\,{\cal V}^N + \lambda\,\partial_P \Lambda^P\,{\cal V}^M,\\
            \mathbb{L}_{\Lambda}{\cal W}_M &= \Lambda^K \partial_K {\cal W}_M + 6\, \mathbb{P}_M{}^N{}{}_L{}^K\,\partial_K \Lambda^L\,{\cal W}_N + \lambda\,\partial_P \Lambda^P\,{\cal W}_M,
        \end{aligned}
    \end{equation}
    where 
    \begin{equation} \label{eq:projadj}
    	\mathbb{P}_{M}{}^{K}{}_{N}{}^{L} = \kappa^{\alpha\beta} t_{\alpha\,M}{}^{K}t_{\beta\,N}{}^{L} = \frac{1}{18}\,\delta_{M}{}^{K}\delta_{N}{}^{L} + \frac{1}{6}\,\delta_{M}{}^{L}\delta_{N}{}^{K} - \frac{5}{3}\,d_{MNR}d^{KLR}
    \end{equation}
    is the projector onto the adjoint representation of ${\rm E}_{6(6)}$. In~\eqref{eq:projadj}, $t_{\alpha\,M}{}^{N}$ are the generators of the ${\rm E}_{6(6)}$ algebra and $\kappa^{\alpha\beta}$ the corresponding inverse Cartan-Killing metric.  With the definition~\eqref{eq:genLie}, the tensor $d_{MNP}$ is an invariant tensor of weight $0$, \textit{i.e.}
    \begin{equation}\label{eq:tinv}
        t_{\alpha\,(M}{}^{N}d_{PQ)N} = 0 \quad \implies \quad \mathbb{L}_{\Lambda} d_{MNP}=0,
    \end{equation}
    and the generalized diffeomorphisms action on ${\cal M}_{MN}$ preserves its group property. The ${\rm E}_{6(6)}$~generalized diffeomorphims are thus compatible with the ${\rm E}_{6(6)}$ Lie algebra.

    ${\rm E}_{6(6)}$ ExFT is a gauge invariant theory under the internal generalized diffeomorphisms~\eqref{eq:genLie}. The vector fields ${\cal A}_{\mu}{}^{M}$ play the role of a gauge connection and enter the covariant derivative
    \begin{equation} \label{eq:ExFTcovder}
        {\cal D}_{\mu} = \nabla_{\mu} - \mathbb{L}_{{\cal A}_{\mu}},
    \end{equation}
    whereas the two-form tensors $\mathcal{B}_{\mu\nu\,M}$ are identified as gauge potentials. In~\eqref{eq:ExFTcovder}, $\nabla_{\mu}$ is the covariant derivative with respect to the metric $\g_{\mu\nu}$. The internal gauge symmetries are parametrized by gauge parameters $\Lambda^{M}(x,Y)$ and $\Xi_{\mu\,M}(x,Y)$. They take the form
    \begin{equation} \label{eq:internalgaugetransfo}
        \begin{aligned}
            \delta_{\Lambda} \g_{\mu\nu} &= \Lambda^M \partial_M \g_{\mu\nu} + \frac{2}{3}\,\partial_M \Lambda^M\,\g_{\mu\nu},\\[5pt]
            \delta_{\Lambda}{\cal M}_{MN} &= \mathbb{L}_{\Lambda}{\cal M}_{MN},\\[5pt]
            \delta_{\Lambda} {\cal A}_\mu{}^M &= {\cal D}_\mu \Lambda^M  -10\, d^{MNK} \partial_K\Xi_{\mu N},\\[5pt]
            \delta_{\Lambda} {\cal B}_{\mu\nu M} &= 2\,{\cal D}_{[\mu}\Xi_{\nu]\,M} + d_{MKL}\,\Big(\Lambda^K{\cal F}_{\mu\nu}{}^{L}-{\cal A}_{[\mu}{}^K\, \delta_{\Lambda} {\cal A}_{\nu]}{}^L\Big) + \dots
        \end{aligned}
    \end{equation}
    The variation of ${\cal B}_{\mu\nu\,M}$ is defined up to terms that vanish under the projection with $d^{MNK}\partial_{K}$. We will not need these terms in the following. The weights of the fields and gauge parameters with respect to the generalized Lie derivative~\eqref{eq:genLie} are summarized in the following table:
    \begin{equation}	\label{eq:exftweights}
    	\begin{tabular}{c|cccccc}
    			& $\g_{\mu\nu}$		& $\mathcal{M}_{MN}$	& $\mathcal{A}_{\mu}{}^{M}$	& $\mathcal{B}_{\mu\nu\,M}$	& $\Lambda^{M}$	& $\Xi_{\mu\,M}$	\\[5pt]
                \hline	
    		$\lambda$	&	$2/3$	&	$0$	&  $1/3$ &	$2/3$	&	$1/3$	&	$2/3$	
    	\end{tabular}
    	\,.
    \end{equation}

    The full Lagrangian is
    \begin{equation} \label{eq:E6lag}
        \mathscr{L} = \sqrt{\vert \g \vert}\,\bigg(\widehat{R} + \frac{1}{24}\,{\cal D}_{\mu}{\cal M}^{MN}{\cal D}^{\mu}{\cal M}^{MN} - \frac{1}{4}\,{\cal M}_{MN}{\cal F}_{\mu\nu}{}^{M}{\cal F}^{\mu\nu\,N}-V_{\rm ExFT}\bigg) + \mathscr{L}_{\rm top}.
    \end{equation}
    The first term is the scalar curvature covariantized with respect to ${\rm E}_{6(6)}$ generalized diffeomorphisms. It is defined as $\widehat{R}=\widehat{R}_{\mu\nu}{}^{ab}e_{a}{}^{\mu}e_{b}{}^{\nu}$ from the fünfbein $e_{\mu}{}^{a}$ that defines the metric as $\g_{\mu\nu} = e_{\mu}{}^{a}\,\eta_{ab}\,e_{\nu}{}^{b}$, with the ${\rm SO}(1,4)$ invariant tensor $\eta_{ab}$, and the covariantized Riemann tensor
    \begin{equation}
        \widehat{R}_{\mu\nu}{}^{ab} = R_{\mu\nu}{}^{ab}  + {\cal F}_{\mu\nu}{}^{M}\,e^{a\rho}\partial_{M}e_{\rho}{}^{b}.
    \end{equation}
    The second and third terms in~\eqref{eq:E6lag} are kinetic and Yang-Mills terms for the scalar and the vector fields, with the field strength
    \begin{equation} \label{eq:ExFTF}
        {\cal F}_{\mu\nu}{}^{M} = 2\,\partial_{[\mu}{\cal A}_{\nu]}{}^{M} - 2\,{\cal A}_{[\mu}{}^{K}\partial_{K}{\cal A}_{\nu]}{}^{M} + 10\,d^{MKR}d_{NLR}\,{\cal A}_{[\mu}{}^{N}\partial_{K}{\cal A}_{\nu]}{}^{L} + 10\,d^{MKL}\partial_{K}{\cal B}_{\mu\nu\,L}.
    \end{equation}
    It transforms under internal gauge transformations as a vector of weight $1/3$. Finally, the potential~$V_{\rm ExFT}$ is defined as
    \begin{equation}
        \begin{aligned}
            V_{\rm ExFT} &= -\frac{1}{24}\,{\cal M}^{MN}\partial_{M}{\cal M}^{KL}\partial_{N}{\cal M}_{KL} + \frac{1}{2}\,{\cal M}^{MN}\partial_{M}{\cal M}^{KL}\partial_{L}{\cal M}_{NK} \\
            & \quad -\frac{1}{2}\,\g^{-1}\partial_{M}\g\partial_{N}{\cal M}^{MN} -\frac{1}{4}\,{\cal M}^{MN}\,\g^{-1}\partial_{M}\g\,\g^{-1}\partial_{N}\g -\frac{1}{4}\,{\cal M}^{MN}\,\partial_{M}\g_{\mu\nu}\partial_{N}\g^{\mu\nu},
        \end{aligned}
    \end{equation}
    and the topological term $\mathscr{L}_{\rm top}$ is defined by its exterior derivative
    \begin{equation}
        {\rm d}\mathscr{L}_{\rm top} = \sqrt{\frac{5}{32}}\,\bigg(d_{MNP}\,{\cal F}^{M}\wedge{\cal F}^{N}\wedge{\cal F}^{P} - 40\,d^{MNP}\,{\cal H}_{M}\wedge\partial_{N}{\cal H}_{P}\bigg),
    \end{equation}
    where ${\cal H}_{\mu\nu\rho\,M}$ is the three-form field strength associated to ${\cal B}_{\mu\nu\,M}$:
    \begin{equation}
        \begin{aligned}
            {\cal H}_{\mu\nu\rho\,M} & = 3\,{\cal D}_{[\mu}{\cal B}_{\nu\rho]\,M} -3\,d_{MKL}\,{\cal\,A}_{[\mu}{}^{K}\partial_{\nu}{\cal A}_{\rho]}{}^{L} + 2\,d_{MKL}\,{\cal A}_{[\mu}{}^{K}{\cal A}_{\nu}{}^{P}\partial_{P}{\cal A}_{\rho]}{}^{L} \\
            & \quad -10\,d_{MKL}d^{LPR}d_{RNQ}\,{\cal A}_{[\mu}{}^{K}{\cal A}_{\nu}{}^{N}\partial_{P}{\cal A}_{\rho]}{}^{Q}+\ldots
        \end{aligned}
    \end{equation}
    The ellipsis denote terms that are projected out by the operator $d^{MNK}\partial_{K}$. ${\cal H}_{\mu\nu\rho\,M}$ can equivalently be defined via the Bianchi identity
    \begin{equation} \label{eq:ExFTbianchi}
        3\,{\cal D}_{[\mu} {\cal F}_{\nu\rho]}{}^{M} = 10\,d^{MKL}\partial_{K}{\cal H}_{\mu\nu\rho\,L}.
    \end{equation}
    This identity defines a duality between the vector fields ${\cal A}_{\mu}{}^{M}$ and the two-form tensors ${\cal B}_{\mu\nu\,M}$. 

    All the terms of the ${\rm E}_{6(6)}$ action constructed from the Lagrangian~\eqref{eq:E6lag} are separately gauge invariant under the internal gauge transformations~\eqref{eq:internalgaugetransfo}. The theory is additionally invariant under five-dimensional external diffeomorphisms of parameter $\xi^{\mu}(x,Y)$:\footnote{Here and in the following $\varepsilon_{\mu\nu\rho\sigma\tau}$ denotes the Levi-Civita symbol, and $\epsilon_{\mu\nu\rho\sigma\tau}$ is the associated tensor.}
    \begin{equation} \label{eq:externalgaugetransfo}
        \begin{aligned}
             \delta_{\xi} \g_{\mu\nu} &= \xi^\rho {\cal D}_\rho \g_{\mu\nu} + 2\,{\cal D}_{(\mu} \xi^\rho \g_{\nu)\rho},\\[5pt]
            \delta_{\xi} {\cal M}_{MN} &= \xi^{\mu}\,{\cal D}_{\mu}{\cal M}_{MN},\\[5pt]
            \delta_{\xi} {\cal A}_{\mu}{}^M &= \xi^\nu\,{\cal F}_{\nu\mu}{}^M + {\cal M}^{MN}\,\g_{\mu\nu} \,\partial_N \xi^\nu, \\[5pt]
            \delta_{\xi} {\cal B}_{\mu\nu\,M} &= \frac{1}{2\,\sqrt{10}}\,\xi^\rho\, \sqrt{|\g|}\,\varepsilon_{\mu\nu\rho\sigma\tau}\, {\cal F}^{\sigma\tau\,N} {\cal M}_{MN} - d_{MKL}\,{\cal A}_{[\mu}{}^K\, \delta_{\xi} {\cal A}_{\nu]}{}^L\,.
        \end{aligned}
    \end{equation}
    It is this symmetry that fixes the relative factors in the Lagrangian~\eqref{eq:E6lag}. ${\rm E}_{6(6)}$ ExFT is the unique invariant theory under the internal and external gauge transformations of~\eqref{eq:internalgaugetransfo} and~\eqref{eq:externalgaugetransfo}. Upon solving the section condition~\eqref{eq:sectioncond}, this theory is strictly equivalent to type II or 11d supergravity, and an explicit dictionary between the ExFT fields and the ones from the supergravity theories has been constructed in ref.~\cite{Hohm:2013vpa}.

    \subsection{Field Equations} \label{sec:ExFTfieldequations}
    We list in this section the field equations of each field of the theory, which are 
    derived from the Lagrangian~\eqref{eq:E6lag}.

    \paragraph{Two-form tensors \boldmath ${\cal B}_{\mu\nu\,M}$}
    \begin{equation} \label{eq:ExFTeqB}
        0 = d^{MKL}\partial_L \bigg(\sqrt{\vert \g \vert}\,{\cal M}_{MN} {\cal F}^{\mu \nu}{}^N + \frac{\sqrt{10}}{6}\,\varepsilon^{\mu\nu\rho\sigma\tau}\,{\cal H}_{\rho\sigma\tau}{}_M \bigg).
    \end{equation}

    \paragraph{Vector fields \boldmath ${\cal A}_{\mu}{}^{M}$}
    \begin{equation}
        \begin{aligned}
            0 &= {\cal D}_{\nu}\left({\cal F}^{\mu\nu\,N}{\cal M}_{NM}\right) - 2\,{\cal D}_{\nu}\left(\g^{\rho[\mu}e_{a}{}^{\nu]}\partial_{M}e_{\rho}{}^{a}\right) + 2\, e_{a}{}^{\mu}e_{b}{}^{\nu}\,\partial_{M}\omega_{\nu}{}^{ab}  + \frac{1}{12}\,{\cal D}^{\mu}{\cal M}^{PQ}\partial_{M}{\cal M}_{PQ} \\[5pt]
            &\quad - \frac{1}{\sqrt{\vert \g \vert}}\,\partial_{N}\left(\sqrt{\vert \g \vert}\,{\cal D}^{\mu}{\cal M}^{PN}{\cal M}_{PM} \right) + \dfrac{\sqrt{10}}{8}\,\frac{1}{\sqrt{\vert \g \vert}}\,\varepsilon^{\mu\nu\rho\sigma\tau}\,d_{MNK}\,{\cal F}_{\tau\nu}{}^{K}{\cal F}_{\rho\sigma}{}^{N},
        \end{aligned}
    \end{equation}
    where $\omega_{\nu}{}^{ab}$ is the spin connection associated to the fünfbein $e_{\mu}{}^{a}$.

    \paragraph{Metric \boldmath $\g_{\mu\nu}$}
    \begin{equation} \label{eq:ExFTEinstein}
        \begin{aligned}
            0 &= \widehat{R}_{\mu\nu}-\frac12\,\g_{\mu\nu}\,\widehat{R} - \frac1{24}\,{\cal J}_\mu{}^{M}{}_{N} {\cal J}_\nu{}^{N}{}_M + \frac1{48}\,\g_{\mu\nu}\,{\cal J}_\rho{}^{M}{}_{N} {\cal J}^\rho{}^{N}{}_M \\
            & \quad -\frac12\,{\cal M}_{MN}\,{\cal F}_{\mu\rho}{}^M {\cal F}_{\nu}{}^{\rho}{}^N +\frac18\,\g_{\mu\nu}\,{\cal M}_{MN}\,{\cal F}_{\rho\sigma}{}^M {\cal F}^{\rho\sigma}{}^N \\
            & \quad -\frac{1}{48}\,\g_{\mu\nu}\,{\cal M}^{MN}\partial_M{\cal M}^{KL}\,\partial_N{\cal M}_{KL} + \frac{1}{4} \,\g_{\mu\nu}\,{\cal M}^{MN}\partial_M{\cal M}^{KL}\partial_L{\cal M}_{NK} \\
            & \quad +\frac{1}{2}\,\g_{\mu\nu}\, \partial_M \partial_N {\cal M}^{MN} +\frac{1}{2}\,\g_{\mu\nu}\,\g^{-1}\partial_M\g\,\partial_N{\cal M}^{MN} -\frac{1}{2}\,  \partial_M {\cal M}^{MN}\partial_N \g_{\mu\nu} \\
            & \quad -\frac{3}{8} \,\g_{\mu\nu}\, \g^{-2} {\cal M}^{MN} \partial_M \g \partial_N \g +\frac{1}{2} \,\g_{\mu\nu}\, \g^{-1}\, {\cal M}^{MN}\partial_N \partial_M \g \\
            & \quad -\frac{1}{4}\, \g^{-1}\, \partial_M \g\, {\cal M}^{MN}\partial_N \g_{\mu\nu} -\frac{1}{2}\,  {\cal M}^{MN} \partial_M \partial_N \g_{\mu\nu}  +\frac{1}{2}\,{\cal M}^{MN} \g^{\sigma\tau} \partial_M \g_{\sigma\mu} \partial_N \g_{\tau\nu} \\
            & \quad -\frac{1}{8}\,\g_{\mu\nu}\,{\cal M}^{MN}\partial_Mg^{\sigma\tau}\partial_N \g_{\sigma\tau}, 
        \end{aligned}
    \end{equation}
    with the scalar currents
    \begin{equation} \label{eq:ExFTcurrents}
        {\cal J}_\mu{}^M{}_N = {\cal M}^{MK} {\cal D}_\mu {\cal M}_{KN}.
    \end{equation}

    \paragraph{Scalar fields \boldmath ${\cal M}_{MN}$}
    The variation of the Lagrangian~\eqref{eq:E6lag} with respect to the generalized metric ${\cal M}_{MN}$ gives
    \begin{equation}
        \delta_{\cal M}\mathscr{L} = \sqrt{\vert \g\vert}\, \delta {\cal M}^{MN}\,{\cal K}_{MN},
    \end{equation}
    where
    \begin{equation}
        \begin{aligned}
           {\cal K}_{MN} &= -\frac{1}{12}\,{\cal D}_{\mu}{\cal J}^{\mu\, P}{}_{N}\,{\cal M}_{PM} + \frac{1}{4}\,{\cal M}_{MP}{\cal F}_{\mu\nu}{}^{P}{\cal M}_{NQ}{\cal F}^{\mu\nu}{}^{Q} \\
           &\quad + \frac{1}{24}\,\partial_{M}{\cal M}^{PQ}\partial_{N}{\cal M}_{PQ} - \frac{1}{\sqrt{\vert \g\vert}}\, \frac{1}{24}\,\partial_{P}\Big(\sqrt{\vert \g\vert}\,{\cal M}^{PQ}\partial_{Q}{\cal M}_{MN}\Big)\\
           &\quad + \frac{1}{\sqrt{\vert \g\vert}}\,\frac{1}{24}\,{\cal M}_{MP}{\cal M}_{NQ}\,\partial_{K}\Big(\sqrt{\vert \g\vert}\,{\cal M}^{KL}\partial_{L}{\cal M}^{PQ}\Big) - \frac{1}{2}\,\partial_{M}{\cal M}^{PQ}\partial_{Q}{\cal M}_{NP} \\
           &\quad + \frac{1}{\sqrt{\vert \g\vert}}\,\frac{1}{2}\,\partial_{P}\Big(\sqrt{\vert \g\vert}\,{\cal M}^{PQ}\partial_{N}{\cal M}_{QM}\Big) -\frac{1}{\sqrt{\vert \g\vert}}\, \frac{1}{2}\,{\cal M}_{MP}{\cal M}_{NQ}\,\partial_{K}\Big(\sqrt{\vert \g\vert}\,{\cal M}^{PL}\partial_{L}{\cal M}^{QK}\Big)\\
           &\quad + \frac{1}{2}\,\g^{-2}\,\partial_{M}\g\,\partial_{N}\g + \frac{1}{4}\,\partial_{M}\g_{\mu\nu}\,\partial_{N}\g^{\mu\nu} - \frac{1}{2}\,\g^{-1}\,\partial_{M}\partial_{N}\g.
        \end{aligned}
    \end{equation}
    As the scalar field ${\cal M}_{MN}$ parametrizes the coset space ${\rm E}_{6(6)}/{\rm USp}(8)$, it is a constrained field and ${\cal K}_{MN} = 0$ are not the field equations of the generalized metric. One first has to project onto the coset, using the projector~\cite{Berman:2019izh,Berman:2020tqn}
    \begin{equation} \label{eq:cosetproj}
        \mathbb{P}_{{\rm coset}\,MN}{}^{KL} = {\cal M}_{MP}\,\mathbb{P}_{N}{}^{P}{}_{Q}{}^{(K}\,{\cal M}^{L)Q},
    \end{equation}
    with $\mathbb{P}_{M}{}^{K}{}_{N}{}^{L}$ the projector on the adjoint representation of ${\rm E}_{6(6)}$ defined in~\eqref{eq:projadj}. The field equations for the scalar fields are then
    \begin{equation}
        0 = \mathbb{P}_{{\rm coset}\,MN}{}^{KL}\,{\cal K}_{KL}.
    \end{equation}

    \subsection{Generalized Scherk-Schwarz Reductions} \label{sec:genSSred}
    One of the main achievements of ${\rm E}_{6(6)}$ ExFT is to allow efficient constructions of consistent truncations from ExFT (or equivalently type II and 11d supergravities) down to five-dimensional maximal gauged supergravity~\cite{Hohm:2014qga,Lee:2014mla}. The ExFT fields~\eqref{eq:ExFTfields} are parametrized as functions of the five-dimensional ones,
    \begin{equation} \label{eq:5dfields}
        \Big\{\bg{g}_{\mu\nu}, \bg{M}_{\underline{MN}}, \bg{A}_{\mu}{}^{\underline{M}}, \bg{B}_{\mu\nu\,\underline{M}}\Big\},
    \end{equation}
    with a generalized Scherk-Schwarz ansatz of the form
    \begin{equation} \label{eq:bgreductionansatz}
    	\begin{aligned}
    		\g_{\mu\nu}(x,Y) & = \rho^{-2}(Y)\,\bg{g}_{\mu\nu}(x) ,\\
    		{\cal M}_{MN}(x,Y) & = U_{M}{}^{\underline{M}}(Y)U_{N}{}^{\underline{N}}(Y)\,\bg{M}_{\underline{MN}}(x),\\
    		{\cal A}_{\mu}{}^{M}(x,Y) & = \rho^{-1}(Y)(U^{-1})_{\underline{M}}{}^{M}(Y)\,\bg{A}_{\mu}{}^{\underline{M}}(x),\\
    		{\cal B}_{\mu\nu,M}(x,Y) & = \rho^{-2}(Y)U_{M}{}^{\underline{M}}(Y)\,\bg{B}_{\mu\nu,\underline{M}}(x).
    	\end{aligned}
    \end{equation}
    Similarly, the ansatz for the gauge parameters is 
    \begin{equation} \label{eq:bgreductionansatzgauge}
        \begin{aligned}
            \Lambda^{M}(x,Y) & =  \rho^{-1}(Y)(U^{-1})_{\underline{M}}{}^{M}(Y)\,\bg{\Lambda}^{\underline{M}}(x),\\
            \Xi_{\mu,M}(x,Y) & =  \rho^{-2}(Y)U_{M}{}^{\underline{M}}(Y)\,\bg{\Xi}_{\mu\,\underline{M}}(x),\\
            \xi^{\mu}(x,Y) &= \bg{\xi}^{\mu}(x),
        \end{aligned}
    \end{equation}
    with the five-dimensional gauge parameters $\Big\{\bg{\Lambda}^{\underline{M}}, \bg{\Xi}_{\mu\,\underline{M}}, \bg{\xi}^\mu\Big\}$. The ${\rm E}_{6(6)}$ group valued twist matrix $U_{M}{}^{\underline{M}}(Y)$ and the scale factor $\rho(Y)$ encompass the $Y$ dependence of the ExFT fields. The truncation to the five-dimensional fields~\eqref{eq:5dfields} is then consistent provided that the tensor
    \begin{equation} \label{eq:embeddingtensor}
        X_{\underline{MN}}{}^{\underline{K}} = 6\,\mathbb{P}_{\underline{L}}{}^{\underline{P}}{}_{\underline{N}}{}^{\underline{K}}\,\Gamma_{\underline{PM}}{}^{\underline{L}} - \frac{3}{2}\,\mathbb{P}_{\underline{M}}{}^{\underline{L}}{}_{\underline{N}}{}^{\underline{K}}\,\Gamma_{\underline{PL}}{}^{\underline{P}} - \Gamma_{\underline{MN}}{}^{\underline{K}}
    \end{equation}
    is a constant. In~\eqref{eq:embeddingtensor} we defined the currents
    \begin{equation} \label{eq:Gamma}
        \Gamma_{\underline{M}\underline{N}}{}^{\underline{K}} =      (U^{-1})_{\underline{N}}{}^L\,\dd_{\underline{M}} U_L{}^{\underline{K}} = \Gamma_{\underline{M}}{}^\alpha t_{\alpha\,\underline{N}}{}^{\underline{K}},\quad  {\rm where} \quad \dd_{\underline{M}} \equiv \rho^{-1}(U^{-1})_{\underline{M}}{}^N\partial_N\equiv {\cal U}_{\underline{M}}{}^N\partial_N.
    \end{equation}
    
    The condition~\eqref{eq:embeddingtensor} ensures that the $Y$ dependence factors out in the ExFT field equations of section~\ref{sec:ExFTfieldequations} when expressed with the ansatz~\eqref{eq:bgreductionansatz}. These field equations then factor into products of the twist matrix and the field equations of five-dimensional supergravity. We will come back to this point in section~\ref{sec:5dsugra}. The tensor $X_{\underline{MN}}{}^{\underline{K}}$ is called the embedding tensor, it belongs to the representation~$\boldsymbol{351}$ of ${\rm E}_{6(6)}$ and defines the gauging of the five-dimensional theory~\cite{deWit:2004nw}. Using its expression~\eqref{eq:embeddingtensor} and the section constraint~\eqref{eq:sectioncond}, one can show that the embedding tensor satisfies the quadratic constraint of maximal five-dimensional supergravity:
    \begin{equation} \label{eq:QC}
        X_{\underline{MK}}{}^{\underline{P}}X_{\underline{NP}}{}^{\underline{L}} - X_{\underline{NK}}{}^{\underline{P}}X_{\underline{MP}}{}^{\underline{L}} = - X_{\underline{MN}}{}^{\underline{P}}X_{\underline{PK}}{}^{\underline{L}}.
    \end{equation}
    This quadratic constraint can equivalently be written as~\cite{deWit:2004nw}:
    \begin{equation} \label{eq:QCZ}
        Z^{\underline{MN}} X_{\underline{NK}}{}^{\underline{L}} = 0 \quad \Longleftrightarrow \quad  X_{\underline{MN}}{}^{[\underline{K}}Z^{\underline{L}]\underline{N}} = 0,
    \end{equation}
    in terms of the constant antisymmetric tensor
    \begin{equation} \label{eq:Ztensor}
    	Z^{\underline{MN}} = 2\,d^{\underline{MPQ}}\,X_{\underline{PQ}}{}^{\underline{N}}.
    \end{equation}   
    The embedding tensor~\eqref{eq:embeddingtensor} generically includes a gauging of the trombone symmetry~\cite{LeDiffon:2008sh}. As a consequence, the five-dimensional theory does not feature an action. In the following, we will restrict ourselves to backgrounds that do not feature any gauging of the trombone symmetry, which implies
	\begin{equation}\label{eq:GammaTr}
		\Gamma_{\underline{PM}}{}^{\underline{P}} = -4\,\rho^{-1}\,\dd_{\uM}\rho.
	\end{equation}

    Internal spaces admitting a twist matrix and a scale factor satisfying the consistency equation~\eqref{eq:embeddingtensor} are called generalized Leibniz parallelizable spaces. Indeed, the consistency condition can be written as~\cite{Lee:2014mla}
    \begin{equation} \label{eq:genpara}
    	\mathbb{L}_{{\cal U}_{\underline{M}}}\,{\cal U}_{\underline{N}} = X_{\underline{MN}}{}^{\underline{K}}\,{\cal U}_{\underline{K}},
    \end{equation}
    with $\, {\cal U}_{\underline{M}}{}^{M} = \rho^{-1}(U^{-1})_{\underline{M}}{}^{M}$. This generalized parallelisability in particular implies that the operator $\dd_{\underline{M}}$ introduced in~\eqref{eq:Gamma} satisfy the algebra
    \begin{equation} \label{eq:commd}
        [\dd_{\underline{M}},\dd_{\underline{N}}] = X_{\underline{MN}}{}^{\underline{K}}\,\dd_{\underline{K}}.
    \end{equation}
    Note that this relation constrains both the antisymmetric and symmetric parts  of the embedding tensor, $X_{[\underline{MN}]}{}^{\underline{K}}$ and $X_{(\underline{MN})}{}^{\underline{K}}$ respectively. The constraint on the symmetric part can equivalently be expressed as
    \begin{equation}
        Z^{\underline{MN}}\,\dd_{\underline{N}} = 0\,,
        \label{eq:IZ1}
    \end{equation}
    as we show explicitly in appendix~\ref{app:Formulas}.
    As a consequence, the flattened derivatives $\dd_{\underline{M}}$ also satisfy the section condition~\eqref{eq:sectioncond}:
    \begin{equation} 
        d^{\underline{MNP}} \,\dd_{\underline{M}} \otimes \dd_{\underline{N}} = 0\,.
        \label{eq:IZ2}
    \end{equation}

    \subsection[\texorpdfstring{$D=5$}{D=5} Gauged Supergravity]{\texorpdfstring{\boldmath $D=5$}{D=5} Gauged Supergravity} \label{sec:5dsugra}

    The equations governing the five-dimensional supergravity theory can be obtained from plugging the ans\"atze~\eqref{eq:bgreductionansatz} and~\eqref{eq:bgreductionansatzgauge} in the ExFT expressions of section~\ref{sec:ExFTrecap} and~\ref{sec:ExFTfieldequations}. For example, for ExFT vectors and co-vectors of weights $\lambda$ with ansätze
    \begin{equation}
        \begin{aligned}
            {\cal V}^{M}(x,Y) &= \rho^{-3\lambda}(Y)(U^{-1})_{\underline{M}}{}^{M}(Y)\,\bg{V}^{\underline{M}}(x), \\
            {\cal W}_{M}(x,Y) &= \rho^{-3\lambda}(Y)U_{M}{}^{\underline{M}}(Y)\,\bg{W}_{\underline{M}}(x),
        \end{aligned}
    \end{equation}
    the covariant derivatives defined from~\eqref{eq:genLie} and~\eqref{eq:ExFTcovder} take the form
    \begin{equation}
        \begin{aligned}
            {\cal D}_{\mu}{\cal V}^{M}(x,Y) &= \rho^{-3\lambda}(Y)(U^{-1})_{\underline{M}}{}^{M}(Y)\,\bg{D}_\mu \bg{V}^{\underline{M}}(x), \\
            {\cal D}_{\mu}{\cal W}_{M}(x,Y) &= \rho^{-3\lambda}(Y)U_{M}{}^{\underline{M}}(Y)\,\bg{D}_\mu\bg{W}_{\underline{M}}(x),
        \end{aligned}
    \end{equation}
    with the five-dimensional background covariant derivatives
    \begin{equation}
        \begin{aligned}
            \bg{D}_\mu \bg{V}^{\underline{M}} &= \partial_\mu \bg{V}^{\underline{M}} - \bg{A}_\mu{}^{\underline{P}}  \,X_{\underline{P}\underline{N}}{}^{\underline{M}}\,\bg{V}^{\underline{N}},\\
            \bg{D}_\mu \bg{W}_{\underline{M}} &= \partial_\mu \bg{W}_{\underline{M}} + \bg{A}_\mu{}^{\underline{P}}  \,X_{\underline{P}\underline{M}}{}^{\underline{N}}\,\bg{W}_{\underline{N}}.
        \end{aligned}
    \end{equation}
 The covariant derivatives inherit the factorized form of the generalized Scherk-Schwarz ansatz thanks to the consistency conditions~\eqref{eq:embeddingtensor} and~\eqref{eq:genpara}.
    This is true for all ExFT quantities like field strengths and field equations, in the following we thus give only the expressions of the five-dimensional field strengths and field equations.

    The background two-form field strengths of the vector fields $\bg{A}_{\mu}{}^{\underline{M}}$ following from~\eqref{eq:ExFTF} is
    \begin{equation} \label{eq:5dF}
        \bg{F}_{\mu\nu}{}^{\underline{M}}= 2\,\partial_{[\mu}\bg{A}_{\nu]}{}^{\underline{M}}-\bg{A}_{[\mu}{}^{\underline{K}}\bg{A}_{\nu]}{}^{\underline{L}}\,X_{\underline{KL}}{}^{\underline{M}} -Z^{\underline{MN}}\,\bg{B}_{\mu\nu\,\underline{N}}.
    \end{equation}
    The three-form background field strengths of the tensor fields $\bg{B}_{\mu\nu\,\underline{M}}$ can be defined through the Bianchi identities~\eqref{eq:ExFTbianchi} and their five-dimensional descendants
    \begin{equation}
        3\,\bg{D}_{[\mu} \bg{F}_{\nu\rho]}{}^{\underline{M}} = -Z^{\underline{MN}}\,\bg{H}_{\mu\nu\rho\,\underline{N}}.
    \end{equation}
    $\bg{H}_{\mu\nu\rho\,\underline{N}}$ is then known only under projection with the tensor $Z^{MN}$ of~\eqref{eq:Ztensor}, which will be sufficient in the following\footnote{This projection is the five-dimensional descendant of the projection with $d^{MNK}\partial_{K}$ in ExFT.}. Finally, the five-dimensional scalar currents following from~\eqref{eq:ExFTcurrents} are
    \begin{equation} \label{eq:5dcurrent}
        \bg{J}_\mu{}^{\underline{M}}{}_{\underline{N}} 
        =\bg{M}^{\underline{MK}}\,\bg{D}_\mu \bg{M}_{\underline{KN}}.
    \end{equation}

    \paragraph{}
    
    We list in the following the five-dimensional background equations, which descend from the field equations of section~\ref{sec:ExFTfieldequations}. Thanks to the consistency of the truncation of the ExFT fields~\eqref{eq:ExFTfields} to the five-dimensional ones~\eqref{eq:5dfields}, every solution of the five-dimensional background equations below gives a solution of the ExFT field equations of section~\ref{sec:ExFTfieldequations} through the generalized Scherk-Schwarz ansatz~\eqref{eq:bgreductionansatz}. As ExFT is equivalent to type II or 11d supergravity, this will also define solutions of these theories.

    \paragraph{Two-form tensors \boldmath $\bg{B}_{\mu\nu\,\underline{M}}$}
    \begin{equation}
         0 = \bg{E}^{\mu\nu\,\underline{M}} = Z^{\underline{ML}} \bigg[\frac{\sqrt{10}}{6}\,\bg{\epsilon}^{\mu\nu\rho\sigma\tau}\,\bg{H}_{\rho\sigma\tau\,\underline{L}}+ \bg{M}_{\underline{LN}}\,\bg{F}^{\mu\nu\,\underline{N}}\bigg]\,,
         \label{TensorBkg}
    \end{equation}
    where $\bg{\epsilon}^{\mu\nu\rho\sigma\tau}$ is the background Levi-Civita tensor.

    \paragraph{Vector fields \boldmath $\bg{A}_{\mu}{}^{\underline{M}}$}
    \begin{equation}    \label{VectorBkg}
        0 = \bg{E}^\mu{}_{\underline{M}}= \bg{D}_{\nu}\big(\bg{F}^{\mu \nu\,\uN}{}\bg{M}_{\uN\underline{M}} \big)+  \dfrac{1}{6} X_{\underline{MP}}{}^{\underline{K}} \bg{J}^{\mu \, \underline{P}}{}_{\underline{K}}  +     \frac{\sqrt{10}}{8} \,\bg{\epsilon}^{\mu\nu\rho\sigma\tau} d_{\underline{MNK}}\, \bg{F}_{\tau \nu}{}^{\underline{K}}  \bg{F}_{\rho \sigma}{}^{\underline{N}},
    \end{equation}
    where here and in the following the derivative $\bg{D}_{\nu}$ is covariant both with respect to the background gauge field $\bg{A}_{\mu}{}^{\underline{M}}$ and the background metric $\bg{g}_{\mu\nu}$. For example:
    \begin{equation}
        \bg{D}_{\rho}\bg{F}_{\mu \nu}{}^{\underline{M}} = \bg{\nabla}_{\rho}\bg{F}_{\mu \nu}{}^{\underline{M}} - \bg{A}_\rho{}^{\underline{P}}  \,X_{\underline{P}\underline{N}}{}^{\underline{M}}\,\bg{F}_{\mu \nu}{}^{\underline{N}},
    \end{equation}
    with $\bg{\nabla}_{\mu}$ the covariant derivative with respect to the background metric $\bg{g}_{\mu\nu}$.

    \paragraph{Metric \boldmath $\bg{g}_{\mu\nu}$}
    \begin{equation}
        \begin{aligned}
            0 = \bg{E}_{\mu\nu} &=
            \bg{R}_{\mu\nu}-\frac12\,\bg{g}_{\mu\nu}\,\bg{R} 
            - \frac1{24}\,\bg{J}_\mu{}^{\uM}{}_{\uN} \bg{J}_\nu{}^{\uN}{}_\uM
            +\frac1{48}\,\bg{g}_{\mu\nu}\,\bg{J}_\rho{}^{\uM}{}_{\uN} \bg{J}^\rho{}^{\uN}{}_\uM \label{MetricBkg}\\
            &{}
            -\frac12\,\bg{M}_{\uM\uN}\,\bg{F}_{\mu\rho}{}^\uM \bg{F}_{\nu}{}^{\rho}{}^\uN  
            +\frac18\,\bg{g}_{\mu\nu}\,
            \bg{M}_{\uM\uN}\,\bg{F}_{\rho\sigma}{}^\uM \bg{F}^{\rho\sigma}{}^\uN  
            +\frac12\,\bg{g}_{\mu\nu}\,\bg{V},
        \end{aligned}
    \end{equation}
    with the supergravity potential
    \begin{equation} \label{eq:sugraV}
    \bg{V} =
    \frac{1}{12}\,\bg{M}^{\underline{MN}}X_{\underline{MP}}{}^{\underline{R}}
    \left(X_{\underline{NR}}{}^{\underline{P}} +\frac15\,X_{\underline{NT}}{}^{\underline{S}} \,\bg{M}^{\underline{PT}}\bg{M}_{\underline{RS}}
    \right).
    \end{equation}

    \paragraph{Scalar fields \boldmath $\bg{M}_{\underline{MN}}$}
    \begin{equation} \label{eq:fieldeqbgscal}
        \begin{aligned}
            0=\bg{E}_{\underline{MN}} &=  -\frac{1}{12}\,\bg{M}_{\underline{MK}}\bg{D}_{\mu}\bg{J}^{\mu\,\underline{K}}{}_{\underline{N}} \\
            &\quad+ \bg{\mathbb{P}}_{{\rm coset}\,\underline{MN}}{}^{\underline{KL}} \, \bigg[ \frac{1}{4}\,\bg{M}_{\underline{KP}}\bg{ F}_{\mu\nu}{}^{\underline{P}}\bg{M}_{\underline{LQ}}\bg{ F}^{\mu\nu\,\underline{Q}} -\frac{1}{12}\,X_{\underline{KP}}{}^{\underline{Q}}X_{\underline{LQ}}{}^{\underline{P}} \\
            & \qquad\qquad\qquad\qquad -\frac{1}{60}\,X_{\underline{KP}}{}^{\underline{R}}X_{\underline{LQ}}{}^{\underline{S}}\bg{M}^{\underline{PQ}}\bg{M}_{\underline{RS}} -\frac{1}{60}\,X_{\underline{PK}}{}^{\underline{R}}X_{\underline{QL}}{}^{\underline{S}}\bg{M}^{\underline{PQ}}\bg{M}_{\underline{RS}}\\
            & \qquad\qquad\qquad\qquad+\frac{1}{60}\,\bg{M}_{\underline{KU}}\bg{M}_{\underline{LV}}\,X_{\underline{PR}}{}^{\underline{U}}X_{\underline{QS}}{}^{\underline{V}}\bg{M}^{\underline{PQ}}\bg{M}^{\underline{RS}}\bigg],
        \end{aligned}
    \end{equation}
    where $\bg{\mathbb{P}}_{{\rm coset}\,\underline{MN}}{}^{\underline{KL}}$ is the projector~\eqref{eq:cosetproj} computed using the background scalars $\bg{M}_{\underline{MN}}$.

    \paragraph{}
    The previous field equations are invariant under the gauge transformations
    \begin{equation} \label{eq:backgroundgaugetransfo}
        \begin{aligned}
            \bg{\delta} \bg{g}_{\mu\nu} &= \bg{\xi}^{\rho} \bg{\nabla}_{\rho} \bg{g}_{\mu\nu} + 2\,\bg{\nabla}_{(\mu}\bg{\xi}^{\rho} \bg{g}_{\nu)\rho},\\[7pt]
            \bg{\delta} \bg{M}_{\underline{MN}} &= \bg{\xi}^\mu \bg{D}_{\mu} \bg{M}_{\underline{MN}} -2\bg{\Lambda}^{\underline{P}} X_{\underline{P}(\underline{M}}{}^{\underline{Q}} \, \bg{M}_{\underline{N}) \underline{Q}}, \\[7pt]
            \bg{\delta} \bg{A}_{\mu}{}^{\underline{M}} &= \bg{\xi}^{\nu}\bg{F}_{\nu\mu}{}^{\underline{M}} + \bg{D}_{\mu}\bg{\Lambda}^{\underline{M}} + Z^{\underline{MN}}\,\bg{\Xi}_{\mu\,N}, \\[7pt]
            \bg{\delta}\bg{B}_{\mu\nu\,\underline{M}}  &=  2\,\bg{D}_{[\mu}\bg{\Xi}_{\nu]\,\underline{M}} +d_{\underline{MKL}}\bg{\Lambda}^{\underline{K}}\bg{F}_{\mu\nu}{}^{\underline{L}} + \frac{1}{2\,\sqrt{10}}\,\bg{\xi}^{\rho}\,\bg{\epsilon}_{\mu\nu\rho\sigma\tau}\, \bg{F}^{\sigma\tau\,\underline{N}}\,\bg{M}_{\underline{MN}}  - d_{\underline{MKL}}\,\bg{A}_{[\mu}{}^{\underline{K}}\, \delta \bg{A}_{\nu]}{}^{\underline{L}},
        \end{aligned}
    \end{equation}
    which are deduced from their ExFT analogues~\eqref{eq:internalgaugetransfo} and~\eqref{eq:externalgaugetransfo} using the ansätze~\eqref{eq:bgreductionansatz} and~\eqref{eq:bgreductionansatzgauge}.

\paragraph{}
In sections~\ref{sec:homoyopytransfer} and~\ref{sec:BHbackground} we will mostly discuss applications with backgrounds in the five-dimensional ${\rm SO}(6)$ gauged supergravity of ref.~\cite{Gunaydin:1985cu}, which is a consistent truncation of type IIB supergravity on~$S^{5}$~\cite{Lee:2014mla,Hohm:2014qga,Baguet:2015sma}. The twist matrix $U_{M}{}^{\underline{M}}$ and the scale factor $\rho$ leading to this background in the ExFT formalism can be found in ref.~\cite{Baguet:2015sma}. The tensors defining the theory are most nicely written in the ${\rm SL}(6)\times {\rm SL}(2)$ basis of ${\rm E}_{6(6)}$, in which the fundamental representation $\mathbf{27}$ splits as follows:
\begin{equation}
    \begin{aligned}
       {\rm E}_{6(6)} &\ \supset\ {\rm SL}(6)\times {\rm SL}(2) \\
       \mathbf{27} & \longrightarrow (15,1) \oplus (6',2) \\
       V^{\underline{M}} & \longrightarrow \big\{V^{ab}, V_{a\alpha}\big\},
    \end{aligned}
\end{equation}
with $a,b\in\llbracket1,6\rrbracket$ and $\alpha\in\{1,2\}$. The tensors $d^{\underline{MNP}}$, $X_{\underline{MN}}{}^{\underline{P}}$ and $Z^{\underline{MN}}$ then have the following non-vanishing components:
\begin{equation}
    \begin{cases}
        \displaystyle d^{ab}{}_{c\alpha\,d\beta} = \frac{1}{\sqrt{5}}\,\delta_{cd}{}^{ab}\,\varepsilon_{\alpha\beta},\\[7pt]
        \displaystyle d^{ab\,cd\,ef} = \frac{1}{\sqrt{80}}\,\varepsilon^{abcdef},
    \end{cases}\quad
    \begin{cases}
        \displaystyle X_{ab\,cd}{}^{ef} = 2\sqrt{2}\,\delta_{[a}{}^{[e}\delta_{b][c}\delta_{d]}{}^{f]},\\[7pt]
        \displaystyle X_{ab}{}^{c\alpha}{}_{d\beta} = -\sqrt{2}\,\delta_{[a}{}^{c}\delta_{b]d}\delta_{\alpha}{}^{\beta},
    \end{cases}\quad
    Z_{a\alpha\,b\beta} = \sqrt{10}\,\varepsilon_{\alpha\beta}\,\delta_{ab}.
\end{equation}

\section{Perturbation Theory around Generalized Scherk-Schwarz Backgrounds} \label{sec:pertubtheory}
Thanks to the consistent truncation reviewed in section~\ref{sec:genSSred}, every five-dimensional background solving the field equations of section~\ref{sec:5dsugra} will give a solution in type II or 11d supergravity. We now turn to the analysis of linear perturbations around this solution in higher dimensions. These perturbations organize into infinite towers of Kaluza-Klein modes, with the five-dimensional fields~\eqref{eq:5dfields} sitting at the lowest level of the tower. They can be captured in the ExFT framework using the generalized Scherk-Schwarz ansatz~\eqref{eq:bgreductionansatz}~\cite{Malek:2019eaz,Malek:2020yue}. The fluctuations
\begin{equation} \label{eq:fluctfield}
    \Big\{h_{\mu\nu}, m_{\underline{MN}}, a_{\mu}{}^{\underline{M}}, b_{\mu\nu\,\underline{M}}\Big\},
\end{equation}
which all depend both on the external and internal coordinates $x^{\mu}$ and $Y^{M}$, are introduced directly in the ansatz~\eqref{eq:bgreductionansatz}, with all the ExFT tensorial structure factored out:
\begin{equation} \label{eq:reductionansatz}
    \begin{aligned}
        \g_{\mu\nu}(x,Y) & =  \rho^{-2}(Y)\,\Big(\bg{g}_{\mu\nu}(x) + h_{\mu\nu}(x,Y)\Big),\\
        \M_{MN}(x,Y) & = U_{M}{}^{\underline{M}}(Y)U_{N}{}^{\underline{N}}(Y)\,\Big(\bg{M}_{\underline{MN}}(x) + m_{\underline{MN}}(x,Y)\Big),\\
        {\cal A}_{\mu}{}^{M}(x,Y) & =  \rho^{-1}(Y)(U^{-1})_{\underline{M}}{}^{M}(Y)\,\Big(\bg{A}_{\mu}{}^{\underline{M}}(x) + a_{\mu}{}^{\underline{M}}(x,Y)\Big),\\
        {\cal B}_{\mu\nu\,M}(x,Y) & =  \rho^{-2}(Y)U_{M}{}^{\underline{M}}(Y)\,\Big(\bg{B}_{\mu\nu\,\underline{M}}(x) + b_{\mu\nu\,\underline{M}}(x,Y)\Big),
    \end{aligned}
\end{equation}
and for the gauge parameters
\begin{equation} \label{eq:reductionansatzgauge}
    \begin{aligned}
        \Lambda^{M}(x,Y) & =  \rho^{-1}(Y)(U^{-1})_{\underline{M}}{}^{M}(Y)\,\Big(\bg{\Lambda}^{\underline{M}}(x) + \lambda^{\underline{M}}(x,Y)\Big),\\
        \Xi_{\mu\,M}(x,Y) & =  \rho^{-2}(Y)U_{M}{}^{\underline{M}}(Y)\,\Big(\bg{\Xi}_{\mu\,\underline{M}}(x) + \xi_{\mu\,\underline{M}}(x,Y)\Big),\\
        \xi^{\mu}(x,Y) &= \bg{\xi}^{\mu}(x) + \zeta^{\mu}(x,Y).
    \end{aligned}
\end{equation}
The scale factor $\rho$ and the twist matrix $U$ define a consistent truncation, \textit{i.e.} they satisfy the consistency equation~\eqref{eq:genpara}. The analysis of these linear perturbations has so far been restricted to ${\rm AdS}_{5}$ backgrounds for which the background fields $\bg{A}_{\mu}{}^{\underline{M}}$ and $\bg{B}_{\mu\nu\,\underline{M}}$ are vanishing and $\bg{M}_{\underline{MN}}$ is constant. We generalize in the following the analysis to general five-dimensional backgrounds satisfying the field equations of section~\ref{sec:5dsugra}.

    \subsection{Background Field Expansion} \label{sec:fluctuations}
    Let us begin by computing the expansion of the ExFT fields with these new ansätze. The computation is analogous to the one described in section~\ref{sec:5dsugra}. Upon linearisation with respect to the fields~\eqref{eq:fluctfield}, the field strength~\eqref{eq:ExFTF} yields
    \begin{equation}
        \rho\,U_M{}^{\underline{M}}\,{\cal F}_{\mu\nu}{}^{M} = \bg{F}_{\mu\nu}{}^{\underline{M}} + f_{\mu\nu}{}^{\underline{M}},
    \end{equation}
    where the background field strength $\bg{F}_{\mu\nu}{}^{\underline{M}}$ is given in~\eqref{eq:5dF} and the fluctuation is
    \begin{equation} \label{eq:fluctF}
        f_{\mu\nu}{}^{\underline{M}} = 2\,\bg{D}_{[\mu}a_{\nu]}{}^{\underline{M}}-{\cal Z}^{\underline{MN}}\,\tilde{b}_{\mu\nu\,\underline{N}}.
    \end{equation}
    In this equation, the background covariant derivative is defined as
    \begin{equation} \label{eq:fluctDa}
        \bg{D}_{\mu}a_{\nu}{}^{\underline{M}} = \nabla_{\mu}a_{\nu}{}^{\underline{M}}-\bg{A}_{\mu}{}^{\underline{K}}\left(X_{\underline{KL}}{}^{\underline{M}}+\delta_{\underline{L}}{}^{\underline{M}}\,\dd_{\underline{K}}\right) a_{\nu}{}^{\underline{L}},
    \end{equation}
    and the antisymmetric tensor $Z^{\underline{MN}}$ of~\eqref{eq:Ztensor} is promoted to the operator
    \begin{equation} \label{eq:Zoperator}
        {\cal Z}^{\underline{MN}} = Z^{\underline{MN}} -10\,d^{\underline{MNP}}\,\dd_{\underline{P}}.
    \end{equation}		
    We have also defined the modified two-form fluctuation
    \begin{equation}
        \tilde{b}_{\mu\nu\,\underline{M}}=b_{\mu\nu\,\underline{M}}+d_{\underline{MKL}}\,\bg{A}_{[\mu}{}^{\underline{K}}\,a_{\nu]}{}^{\underline{L}}.
    \end{equation}
    It is this combination of the fluctuations $a_{\mu}{}^{\underline{M}}$ and $b_{\mu\nu\,\underline{M}}$ that is the most natural to define the field strengths, the gauge transformations and the field equations.

    The three-form field strength ${\cal H}_{\mu\nu\rho\,M}$ can again be computed from the Bianchi identity~\eqref{eq:ExFTbianchi} giving, up to linear order in the fluctuations~\eqref{eq:fluctfield},
    \begin{equation} \label{eq:LinF}
        3\,\rho\,U_M{}^{\underline{M}}\,{\cal D}_{[\mu} \left(\rho^{-1} (U^{-1})_{\underline{N}}{}^M {\cal F}_{\nu\rho]}{}^{\underline{N}}(x,Y)\right) = 3\,\bg{D}_{[\mu} \bg{F}_{\nu\rho]}{}^{\underline{M}}-{\cal Z}^{\underline{MN}} h_{\mu\nu\rho\,\underline{N}},
    \end{equation}
    with the linearized field strength
    \begin{equation}
        h_{\mu\nu\rho\,\underline{M}} = 3\,\bg{D}_{[\mu} \tilde{b}_{\nu\rho]\,\underline{M}} -     3\,d_{\underline{M}\underline{K}\underline{L}}\,\bg{F}_{\nu\rho} {}^{\underline{K}} \,a_\mu{}^{\underline{L}},
    \end{equation}
    and the background covariant derivative
    \begin{equation} \label{eq:fluctDb}
        \bg{D}_{\mu}\tilde{b}_{\nu\rho\,\underline{N}} = \nabla_{\mu}\tilde{b}_{\nu\rho\,\underline{N}}    				-\bg{A}_{\mu}{}^{\underline{K}}\left(-X_{\underline{KN}}{}^{\underline{L}}+\delta_{\underline{N}}{}^{\underline{L}}\,\dd_{\underline{K}}\right) \tilde{b}_{\nu\rho\,\underline{L}}.     
    \end{equation}

    Finally, the expansion of the scalar current~\eqref{eq:ExFTcurrents} to first order is
    \begin{equation}
        U_M{}^{\underline{M}} \, (U^{-1})_{\underline{N}}{}^N\,{\cal J}_\mu{}^M{}_N = \bg{J}_\mu{}^{\underline{M}}{}_{\underline{N}} + j_\mu{}^{\underline{M}}{}_{\underline{N}},
    \end{equation}
    with the background current~\eqref{eq:5dcurrent} and its perturbation
    \begin{equation} \label{eq:scalar-current}
        j_\mu{}^{\underline{M}}{}_{\underline{N}} = -\bg{M}^{\underline{MK}}\,m_{\underline{KL}}\,\bg{J}_{\mu}{}^{\underline{L}}{}_{\underline{N}} + \bg{M}^{\underline{MK}}\,\Big(\bg{D}_\mu m_{\underline{KN}} +\Pi_{\underline{KN},\underline{P}}\,a_\mu{}^{\underline{P}} \Big).
    \end{equation}
    Here, we have used the background covariant derivative
    \begin{equation} \label{eq:fluctDm}
        \bg{D}_\mu m_{\underline{KN}} = \partial_\mu m_{\underline{KN}} + \bg{A}_\mu{}^{\underline{L}} X_{\underline{LK}}{}^{\underline{P}}m_{\underline{PN}} + \bg{A}_\mu{}^{\underline{L}} X_{\underline{LN}}{}^{\underline{P}}m_{\underline{PK}} - \bg{A}_\mu{}^{\underline{L}} \dd_{\underline{L}} m_{\underline{KN}},
    \end{equation}
    and defined the operator
    \begin{equation}
        \Pi_{\underline{KN},\underline{P}} = 2\,X_{\underline{P}(\underline{K}}{}^{\underline{L}} \,\bg{M}_{\underline{N})\underline{L}} - 12\,\bg{M}_{\underline{L}(\underline{K}}\,\mathbb{P}_{\underline{N})}{}^{\underline{L}}{}_{\underline{P}}{}^{\underline{Q}}{} \,\dd_{\underline{Q}},
        \label{PiOp}
    \end{equation}
    which interpolates between scalar and vector fields.

    Note that as all fluctuations depend on the internal coordinates $Y^{\underline{M}}$, their background covariant derivative include a term $- \bg{A}_\mu{}^{\underline{K}}\,\dd_{\underline{K}}$, \textit{i.e.} 
    \begin{equation}
        \bg{D}_{\mu} = \nabla_{\mu} + A_{\mu}{}^{\underline{K}}X_{\underline{K}} - \bg{A}_\mu{}^{\underline{K}}\,\dd_{\underline{K}}\,,
        \label{eq:covD}
    \end{equation}
    see~\eqref{eq:fluctDa}, \eqref{eq:fluctDb} and~\eqref{eq:fluctDm} for examples. These derivatives satisfy the following commutation rules:
    \begin{equation}
        \big[\bg{D}_{\mu},\bg{D}_{\nu}\big] = \big[\bg{\nabla}_{\mu},\bg{\nabla}_{\nu}\big] + \bg{F}_{\mu\nu}{}^{\underline{K}}X_{\underline{K}} - \bg{F}_{\mu\nu}{}^{\underline{K}}\,\dd_{\underline{K}}.
    \end{equation}

\subsection{Gauge Transformations}
We now turn to the gauge and external diffeomorphism transformations of the fluctuations~\eqref{eq:fluctfield}. They transform both under the background gauge transformations and external diffeomorphisms of parameters $\bg{\Lambda}^{\underline{M}}(x), \bg{\Xi}_{\mu\,\underline{M}}(x)$ and $\bg{\xi}^{\mu}(x)$, and under their perturbations of parameters $\lambda^{\underline{M}}(x,Y)$, $\xi_{\mu\,\underline{M}}(x,Y)$ and $\zeta^{\mu}(x,Y)$. We give in the following these transformation laws, which follow from plugging the ansätze~\eqref{eq:reductionansatz} and~\eqref{eq:reductionansatzgauge} in~\eqref{eq:internalgaugetransfo} and~\eqref{eq:externalgaugetransfo}. For each field, the first line gives the background transformations and the second line the linearized transformations under the fluctuations:
\begin{subequations} \label{eq:alltransfofluct}
    \begin{align}
        \delta h_{\mu\nu} &= \bg{\xi}^\lambda \bg{D}_\lambda h_{\mu\nu} + 2\,\bg{D}_{(\mu}\bg{\xi}^\lambda h_{\nu)\lambda} +\bg{\Lambda}^{\underline{K}}\, \dd_{\underline{K}}\,h_{\mu\nu}\nonumber\\
        &\quad + 2\,\bg{D}_{(\mu}\zeta_{\nu)}+ \frac{2}{3}\,\dd_{\underline{P}}\tilde{\lambda}^{\underline{P}}\, \bg{g}_{\mu\nu},\\[10pt]
        \delta m_{\underline{MN}} &= \bg{\xi}^\mu \bg{D}_{\mu} m_{\underline{MN}} + \bg{\Lambda}^{\underline{K}}\Big(-2 X_{\underline{K}(\underline{M}}{}^{\underline{L}} m_{\underline{N})\underline{L}}+ \delta_{\underline{M}}{}^{\underline{L}} \dd_{\underline{K}}m_{\underline{NL}}\Big)\nonumber\\
        &\quad + \zeta^\mu \bg{D}_\mu \bg{M}_{\underline{MN}} - \Pi_{\underline{MN},\underline{P}}\tilde{\lambda}^{\underline{P}},\\[10pt]
        \delta a_{\mu}{}^{\underline{M}} &= {\cal L}_{\bg{\xi}} a_{\mu}{}^{\underline{M}} + \bg{\Lambda}^{\underline{K}} \Big(X_{\underline{KL}}{}^{\underline{M}} + \delta_{\underline{L}}{}^{\underline{M}} \dd_{\underline{K}}\Big) a_\mu{}^{\underline{L}}\nonumber\\
        &\quad + \zeta^{\nu} \bg{F}_{\nu\mu}{}^{\underline{M}} + \bg{M}^{\underline{MN}}\,\dd_{\underline{N}} \zeta_{\mu} + \bg{D}_\mu \tilde{\lambda}^{\underline{M}}  + \mathcal{Z}^{\underline{MN}}\, \tilde{\xi}_{\mu\,\underline{N}} ,\\[10pt]
        {\cal Z}^{\underline{NM}}\bg{\delta}\tilde{b}_{\mu\nu\,\underline{M}} &= {\cal Z}^{\underline{NM}}\bigg[{\cal L}_{\bg{\xi}}\tilde{b}_{\mu\nu\,\underline{M}} +\bg{\Lambda}^{\underline{K}}\Big(-X_{\underline{KM}}{}^{\underline{L}} + \delta_{\underline{M}}{}^{\underline{L}} \dd_{\underline{K}}\Big) \tilde{b}_{\mu\nu\,\underline{L}} \nonumber \\
        &\qquad + \frac{1}{2\,\sqrt{10}}\,\zeta^\rho\bg{\epsilon}_{\mu\nu\rho\sigma\tau}\, \bg{F}^{\sigma\tau\,\uN}{}\bg{M}_{\uN\underline{M}} + 2\,\bg{D}_{[\mu} \tilde{\xi}_{\nu]\underline{M}} + d_{\underline{MKL}}\, \tilde{\lambda}^{\underline{K}} \bg{F}_{\mu\nu}{}^{\underline{L}}  \bigg],
    \end{align}
\end{subequations}
where we have defined the modified gauge parameters fluctuations
\begin{equation}
    \begin{aligned}
        \tilde{\lambda}^{\underline{M}} &= \lambda^{\underline{M}} - \bg{\xi}^{\mu}a_{\mu}{}^{\underline{M}}, \\
        \tilde{\xi}_{\mu\,\underline{M}} &= \xi_{\mu\,\underline{M}} + d_{\underline{MNP}}\,\bg{\Lambda}^{\underline{N}}\,a_{\mu}{}^{\underline{P}} - \bg{\xi}^{\nu}\,\tilde{b}_{\nu\mu\,\underline{M}}.
    \end{aligned}
\end{equation}
We also used the covariantized Lie derivative ${\cal L}_{\bg{\xi}}$, which for example acts on $a_{\mu}{}^{\underline{M}}$ as
\begin{equation}
    {\cal L}_{\bg{\xi}} a_{\mu}{}^{\underline{M}} = \bg{\xi}^{\nu}\bg{D}_{\nu}a_{\mu}{}^{\underline{M}} + \bg{D}_{\mu}\bg{\xi}^{\nu}a_{\nu}{}^{\underline{M}}.
\end{equation}
In~\eqref{eq:alltransfofluct} and in the following, five-dimensional indices $\mu,\nu\dots$ are lowered and raised using the background metric $\bg{g}_{\mu\nu}$ and its inverse. Note that the transformations of $\tilde{b}_{\mu\nu\,\underline{M}}$ in~\eqref{eq:alltransfofluct} are defined under projection with the operator ${\cal Z}^{\underline{MN}}$ of~\eqref{eq:Zoperator} (which is the equivalent to the projection with $d^{MNK}\partial_{K}$ in ExFT), as it will always appear in this form (see~\eqref{eq:ExFTF} and~\eqref{eq:ExFTeqB}).

\paragraph{}
One can deduce from~\eqref{eq:alltransfofluct} the gauge transformations of the field strengths and scalar current, that we list now for later convenience:
\begin{subequations}
    \begin{align}
        \delta f_{\mu\nu}{}^{\underline{M}} &= {\cal L}_{\bg{\xi}}f_{\mu\nu}{}^{\underline{M}} + \bg{\Lambda}^{\underline{K}}\Big(X_{\underline{K}\underline{N}}{}^{\underline{M}} +\delta_{\underline{N}}{}^{\underline{M}}\,\dd_{\underline{K}}\Big)f_{\mu\nu}{}^{\underline{N}} \nonumber\\
        & \quad - \bg{F}_{\mu\nu}{}^{\underline{K}}\bigg(-X_{\underline{LK}}{}^{\underline{M}}+6\, \,\mathbb{P}_{\underline{L}}{}^{\underline{P}}{}_{\underline{K}}{}^{\underline{M}}  \, \dd_{\underline{P}} -\frac{1}{3}\,\delta_{\underline{K}}{}^{\underline{M}}\,\dd_{\underline{L}}  \bigg) \tilde{\lambda}^{\underline{L}}\\
        & \quad + \bg{\cal L}_\zeta  \bg{F}_{\mu\nu}{}^{\underline{M}} + 2\,\bg{D}_{[\mu} \Big( \bg{M}^{\underline{MN}} \dd_{\underline{N}}\zeta_{\nu]}\Big) + \frac{\sqrt{10}}{2}\, \dd_{\underline{K}} \zeta^\rho\,\bg{\epsilon}_{\mu\nu\rho\sigma\tau}\, {d}^{\underline{MNK}}  \bg{F}^{\sigma\tau\,\underline{P}}{} \bg{M}_{\underline{PN}}, \nonumber \\[10pt]
        {\cal Z}^{\underline{PM}}\,\delta h_{\mu\nu\rho\,\underline{M}} &= {\cal Z}^{\underline{PM}} \bigg[{\cal L}_{\bg{\xi}}h_{\mu\nu\rho\,\underline{M}} - \bg{\Lambda}^{\underline{K}}\,X_{\underline{K}\underline{M}}{}^{\underline{N}}\,   h_{\mu\nu\rho\,\underline{N}} + \bg{\Lambda}^{\underline{K}}\,\dd_{\underline{K}}\,  h_{\mu\nu\rho\,\underline{M}}\nonumber\\
        &\qquad + \bg{H}_{\mu\nu\rho\,\underline{K}} \bigg(-X_{\underline{LM}}{}^{\underline{K}}+6\,\mathbb{P}_{\underline{L}}{}^{\underline{P}}{}_{\underline{M}}{}^{\underline{K}}{} \,\dd_{\underline{P}} + \frac{2}{3}\, \delta_{\underline{M}}{}^{\underline{K}}\,\dd_{\underline{L}} \bigg) \tilde{\lambda}^{\underline{L}} \label{eq:h3-gauge} \\
        & \qquad + \frac{1}{2\sqrt{10}}\,{\cal L}_{\zeta}\Big(\bg{\epsilon}_{\mu\nu\rho\sigma\tau}\bg{F}^{\sigma\tau\,\underline{N}}\bg{M}_{\underline{NM}}\Big) - 3\,d_{\underline{MKL}}\bg{F}_{[\mu\nu}{}^{\underline{K}} \bg{M}^{\underline{LN}}\,\dd_{\underline{N}}\zeta^{\sigma}\bg{g}_{\rho]\sigma} \nonumber\\
        &\qquad - \frac{1}{6\sqrt{10}}\,\zeta^{\sigma} \bg{\epsilon}_{\mu\nu\rho\sigma\tau} X_{\underline{MK}}{}^{\underline{L}}\bg{J}^{\tau\,\underline{K}}{}_{\underline{L}} \bigg], \nonumber \\[10pt]
        \delta j_{\mu}{}^{\underline{M}}{}_{\underline{N}} &= {\cal L}_{\bg{\xi}}j_{\mu}{}^{\underline{M}}{}_{\underline{N}} + \bg{\Lambda}^{\underline{K}}\bigg(X_{\underline{KL}}{}^{\underline{M}}\,\delta_{\underline{N}}{}^{\underline{P}} - X_{\underline{K}\underline{N}}{}^{\underline{P}}\,\delta_{\underline{L}}{}^{\underline{M}} + \delta_{\underline{L}}{}^{\underline{M}}\delta_{\underline{N}}{}^{\underline{P}}\,\dd_{\underline{K}}\bigg) j_\mu{}^{\underline{L}}{}_{\underline{P}}\nonumber\\
        &\quad + \bg{\mathcal{L}}_{\zeta} \bg{J}_\mu{}^{\underline{M}}{}_{\underline{N}} \,+2\,\bg{M}^{\underline{MK}} \bg{M}^{\underline{PQ}} X_{\underline{P}(\underline{K}}{}^{\underline{L}} \bg{M}_{\underline{N})\underline{L}} \dd_{\underline{Q}} \zeta_\mu \\
        &\quad + 12\,\bg{M}^{\underline{MK}} \mathbb{P}_{\underline{P}}{}^{\underline{R}}{}_{(\underline{K}}{}^{\underline{L}}{} \bg{M}_{\underline{N})\underline{L}} \bigg(\bg{F}_{\mu \nu}{}^{\underline{P}}\dd_{\underline{R}} \zeta^\nu - \bg{M}^{\underline{PQ}} \dd_{\underline{R}}\dd_{\underline{Q}} \zeta_\mu \bigg)\nonumber\\
        & \quad + \bg{J}_\mu{}^{\underline{K}}{}_{\underline{L}} \bigg(X_{\underline{PK}}{}^{\underline{M}}\,\delta_{\underline{N}}{}^{\underline{L}} - 6\,\delta_{\underline{N}}{}^{\underline{L}}\,\mathbb{P}_{\underline{K}}{}^{\underline{M}}{}_{\underline{P}}{}^{\underline{Q}}\,\dd_{\underline{Q}} - X_{\underline{PN}}{}^{\underline{L}}\,\delta_{\underline{K}}{}^{\underline{M}} + 6\,\delta_{\underline{K}}{}^{\underline{M}}\,\mathbb{P}_{\underline{N}}{}^{\underline{L}}{}_{\underline{P}}{}^{\underline{Q}}\,\dd_{\underline{Q}}\bigg)\tilde{\lambda}^{\underline{P}}.\nonumber
    \end{align}
\end{subequations}

\subsection{Fluctuation Equations}
\label{subsec:fluctuationEqs}

The field equations of the higher-dimensional fluctuations~\eqref{eq:fluctfield} around the generic five-dimensional background~\eqref{eq:5dfields} follow from expanding to linear order the ExFT field equations of section~\ref{sec:ExFTfieldequations} when evaluated on the generalized Scherk-Schwarz ansatz~\eqref{eq:reductionansatz}. We display in the following the resulting linearized field equations. By construction, all field equations are manifestly invariant under the gauge transformations~\eqref{eq:alltransfofluct}, provided that the background fields~\eqref{eq:5dfields} are solutions of the five-dimensional field equations of section~\ref{sec:5dsugra}.

\paragraph{Two-form tensors \boldmath $\tilde{b}_{\mu\nu\,\underline{M}}$}

\begin{equation}
         \label{eq:FlucTensors}
   \begin{aligned}
        0 = {\cal Y}^{\mu\nu\,\underline{M}} &= {\cal Z}^{\underline{LM}} \Bigg[\frac{\sqrt{10}}{6}\,\bg{\epsilon}^{\mu\nu\rho\sigma\tau}\,h_{\rho\sigma\tau,\underline{M}}+\bg{M}_{\underline{MN}} f^{\mu\nu\,\underline{N}}+\bg{F}^{\mu\nu}{}^{\underline{N}}\, m_{\underline{NM}}\\
		& \qquad\qquad +2\,\bg{M}_{\underline{MN}}\,\bg{F}^{\rho[\mu} {}^{\underline{N}}\,h^{\nu]}{}_{\rho} + \frac{1}{2}\,\bg{M}_{\underline{MN}}\bg{F}^{\mu\nu} {}^{\underline{N}}\,h^\rho{}_\rho\Bigg].
    \end{aligned}
\end{equation}
The first two terms give the standard field equations of massive two-form fields, covariantized with respect to the background. The other terms are couplings to the scalar and spin-2 fields, that vanish for backgrounds with vanishing field strength $\bg{F}_{\mu\nu}{}^{\underline{M}}$. The variation of this equation under the external and internal gauge transformations~\eqref{eq:alltransfofluct} is
\begin{equation}
    \begin{aligned}
        \delta_{\lambda,\zeta} {\cal Y}^{\mu\nu\,\underline{M}} &= \bg{E}^{\mu\nu\,\underline{L}}  \bigg(- X_{\underline{KL}}{}^{\underline{M}}\,+6\,\mathbb{P}_{\underline{L}}{}^{\underline{M}}{}_{\underline{K}}{}^{\underline{P}}{}\,\dd_{\underline{P}}+\frac{2}{3}\,\delta_\uM{}^\uL\,\dd_{\underline{K}}\bigg) \tilde{\lambda}^\uK,
    \end{aligned}
\end{equation}
where the formula~\eqref{eq:usefulformula} has been used to show that the variation w.r.t $\zeta^{\mu}$ vanishes. Hence, the linearized field equations for the two-forms are gauge invariant if the background field equation~\eqref{TensorBkg} is satisfied.

\paragraph{Vector fields \boldmath $a_{\mu}{}^{\underline{M}}$}

\begin{equation} \label{eq:vectoreomfluct}
    \begin{aligned}
        0 = \mathcal{Y}^\mu{}_{\underline{M}} &=   \bg{M}_{\underline{MN}}\, \bg{D}_{\nu}  f^{\mu\nu\,\underline{N}}  +    \bg{D}_{\nu} \bg{{F}}^{\mu \nu}{}^{\underline{N}} \; m_{\underline{MN}}  +    \frac{1}{6} \,X_{\underline{MK}}{}^{\underline{L}} j^{\mu \, \underline{K}}{}_{\underline{L}} \\
         & \quad    
         +\bg{J}_\nu{}^{\uK}{}_{\uL}\,\bg{M}_{\underline{MK}}\,f^{\mu\nu\,\underline{L}}  + \bg{{F}}^{\mu \nu}{}^{\underline{L}}\,\bg{M}_{\underline{MK}} \,  j_\nu{}^{\underline{K}}{}_{\underline{L}}   +\bg{{F}}^{\mu \nu}{}^{\underline{L}}\,\bg{J}_{\nu}{}^{\underline{K}}{}_{\underline{L}}\,m_{\underline{MK}} \\
         &\quad + \frac{\sqrt{10}}{4} \,\bg{\epsilon}^{\mu\nu\rho\sigma\tau} d_{\underline{MKL}} \,\bg{{F}}_{\tau \nu}{}^{\underline{K}} f_{\rho\sigma}{}^{\underline{L}} - \frac{\sqrt{10}}{16} h^\lambda{}_\lambda \,\bg{\epsilon}^{\mu\nu\rho\sigma\tau} d_{\underline{MKL}} \,\bg{{F}}_{\tau \nu}{}^{\underline{K}} \bg{{F}}_{\rho\sigma}{}^{\underline{L}}\\
        & \quad + \bg{D}_{\nu}   \bigg(\dfrac{1}{2}   h^{\rho}{}_{\rho}\, \delta^{[\nu}{}_{\sigma}   -2h^{[\nu}{}_{\sigma}\bigg)\, \bg{{F}}^{\mu] \sigma}{}^{\underline{N}}\bg{M}_{\underline{MN}}\\
         & \quad  -2h^{[\nu}{}_{\sigma}\,  \bg{D}_{\nu}  \bg{{F}}^{\mu] \sigma}{}^{\underline{N}}\bg{M}_{\underline{MN}}  -2h^{[\nu}{}_{\sigma}\, \bg{{F}}^{\mu] \sigma}{}^{\underline{L}}\, \bg{J}_\nu{}^{\uK}{}_{\uL}\,\bg{M}_{\underline{MK}}         -   \frac{1}{6} \,X_{\underline{MK}}{}^{\underline{L}}  \bg{J}^{\nu \, \underline{K}}{}_{\underline{L}}\,  h^{\mu}{}_{\nu} \\
         & \quad + \dd_{\underline{M}} \bigg(  \bg{D}_\nu   h^{\nu\mu} - \bg{D}^{\mu}  h_{\nu}{}^\nu +\frac83\, \dd_{\underline{N}}  {a}^{\mu}{}^{\underline{N}} \bigg)    +   \bg{J}^{\nu \; \underline{N}}{}_{\underline{M}} \, \dd_{\underline{N}} \,\bigg(\dfrac{1}{2}h^{\rho}{}_{\rho}\, \delta^{\mu}{}_{\nu} - h^{\mu}{}_{\nu}  \bigg)  \\
         &\quad + \dd_{\underline{N}}   j^{\mu \, \underline{N}}{}_{\underline{M}}  - \frac{1}{12}\,\bg{J}^{\mu}{}^{\underline{K}}{}_{\underline{L}} \bg{M}^{\underline{LP}} \, \dd_{\underline{M}} m_{\underline{KP}}  \\
         &\quad +\bg{F}^{\mu \nu\,\uN}{}\bg{M}_{\uM\underline{K}}  \bigg( - X_{\underline{LN}}{}^{\underline{K}} \;+6\, \mathbb{P}_{\underline{N}}{}^{\underline{K}}{}_{\underline{L}}{}^{\underline{Q}} \,  \dd_{\underline{Q}} -\frac23\, \delta^\uK{}_{\uN}   \, \dd_{\underline{L}} \bigg) a_\nu{}^{\underline{L}}      
    \end{aligned}
\end{equation}
The first term is the standard one from Yang-Mills equations. Note that many of the couplings are due to the non-trivial five-dimensional background. Couplings to $h_{\mu\nu}$ like the ones in the sixth lines are involved in the spin-2 Higgs mechanism. 

The variation of the vectors field equations~\eqref{eq:vectoreomfluct} under linearized internal and external diffeomorphisms~\eqref{eq:alltransfofluct} is
\begin{equation}
    \begin{aligned}
      \delta_{\lambda,\zeta} \mathcal{Y}^\mu{}_{\underline{M}} &= \bg{E}^\mu{}_{\underline{L}} \, \bigg(-X_{\underline{KM}}{}^{\underline{L}} + 6\,\mathbb{P}_{\underline{M}}{}^{\underline{L}}{}_{\underline{K}}{}^{\underline{P}} \,\dd_{\underline{P}} +\dfrac{2}{3} \,\delta_{\underline{M}}{}^{\underline{L}}\,\dd_{\underline{K}} \bigg) \tilde{\lambda}^{\underline{K}}\\
      &\quad + \bg{\cal L}_\zeta \bg{E}^\mu{}_{\underline{M}}     -12 \bg{E}_{\underline{MK}} \bg{M}^{\underline{KL}}\,\dd_{\underline{L}} \zeta^\mu + \bg{E}^{\mu\nu\,\underline{N}}  \,\dd_{\underline{P}} \zeta_{\nu} \bg{M}^{\underline{PQ}} d_{\underline{MNQ}} \\
      & \quad +2\,\bg{E}^{\mu\nu}{} \dd_{\underline{M}} \zeta_\nu  - \frac{2}{3} \,  \bg{E}^{\nu}{}_\nu \,\dd_{\underline{M}} \zeta^\mu.  
    \end{aligned}
\end{equation}
The field equations~\eqref{eq:vectoreomfluct} are then invariant if the five-dimensional field equations are satisfied. Let us note that all five-dimensional equations are needed, and not only the one of the vectors: the linearized external diffeomorphisms parametrized by $\zeta^{\mu}$ mix the different terms in the action.

\paragraph{Metric \boldmath $h_{\mu\nu}$}
\begin{equation}
    \begin{aligned}
     0 = {\cal Y}_{\mu\nu} &= -\frac12\,h_{\mu\nu}\,\bg{R}
    +\frac12\,\bg{g}_{\mu\nu}\,h^{\rho\sigma}\,\bg{R}_{\rho\sigma}
    -\frac12\,\bg{D}^\rho \bg{D}_{\rho} h_{\mu\nu}
    -\frac12\,\bg{D}_{(\mu} 
     \bg{D}_{\nu)} h_{\rho}{}^{\rho}
    +\bg{D}_\rho 
     \bg{D}_{(\mu} h_{\nu)}{}^\rho\\
    & \quad +\frac12\, \bg{g}_{\mu\nu}\,
    \,\bg{D}^\rho  \bg{D}_{\rho} h_{\sigma}{}^\sigma
    -\frac12\,\bg{g}_{\mu\nu}\,
    \bg{D}^\rho 
     \bg{D}_{\sigma} h_{\rho}{}^\sigma
    +\dd_{\underline{M}} \bg{D}_{(\mu}   {a}_{\nu)}{}^{\underline{M}}
    -\bg{g}_{\mu\nu} \,\dd_{\underline{M}} \bg{D}^\rho  {a}_{\rho}{}^{\underline{M}}\\
    & \quad -\dfrac{1}{12}\,\bg{J}_{(\mu}{}^{\underline{M}}{}_{\underline{N}} \; j_{\nu)}{}^{\underline{N}}{}_{\underline{M}}+\dfrac{1}{48}h_{\mu \nu} \bg{J}_{\rho}{}^{\underline{M}}{}_{\underline{N}} \; \bg{J}^{\rho}{}^{\underline{N}}{}_{\underline{M}} - \dfrac{1}{48}\,\bg{g}_{\mu \nu} h^{\rho \sigma} \bg{J}_{\rho}{}^{\underline{M}}{}_{\underline{N}} \; \bg{J}_{\sigma}{}^{\underline{N}}{}_{\underline{M}}+ \dfrac{1}{24}\,\bg{g}_{\mu \nu}  \bg{J}_{\rho}{}^{\underline{M}}{}_{\underline{N}} \; j^{\rho}{}^{\underline{N}}{}_{\underline{M}} \\
    & \quad -\dfrac{1}{2}\,m_{\underline{MN}}\bg{F}_{\mu\rho}{}^{\underline{M}}\bg{F}_{\nu}{}^{\rho}{}^{\underline{N}} +\dfrac{1}{2}\,\bg{M}_{\underline{MN}} h^{\rho \sigma}\bg{F}_{\mu\rho}{}^{\underline{M}}\bg{F}_{\nu \sigma}{}^{\underline{N}} - \bg{M}_{\underline{MN}} \bg{F}_{(\mu}{}^{\rho}{}^{\underline{M}}f_{\nu )  \rho}{}^{\underline{N}} \\
    & \quad + \dfrac{1}{8}\,h_{\mu \nu} \bg{M}_{\underline{MN}}\bg{F}_{\rho \sigma}{}^{\underline{M}}\bg{F}^{ \rho \sigma}{}^{\underline{N}} - \dfrac{1}{4}\,\bg{g}_{\mu \nu} \bg{M}_{\underline{MN}} h _{\tau}{}^{\rho} \bg{F}_{\rho \sigma}{}^{\underline{M}}\bg{F}^{\tau\sigma}{}^{\underline{N}} + \dfrac{1}{4}\,\bg{g}_{\mu \nu} \bg{M}_{\underline{MN}}\bg{F}_{\rho \sigma}{}^{\underline{M}}f^{ \rho \sigma}{}^{\underline{N}} \\
    & \quad  +\frac{1}{2}\,h_{\mu\nu}\,\bg{V} + \frac{1}{24}\,\bg{g}_{\mu \nu}\,m^{\underline{MN}}\,M_{\underline{MK}}\,D_{\mu}J^{\mu\,\underline{K}}{}_{\underline{N}} - \frac{1}{12}\,\bg{g}_{\mu \nu}\, X_{\underline{ML}}{}^{\underline{P}} \bg{M}^{\underline{MN}} \bg{M}^{\underline{KL}} \,\dd_{\underline{N}} m _{\underline{PK}} \\
    & \quad - \frac{1}{2}\,\bg{g}_{\mu \nu}\, \dd_{\underline{M}} \dd_{\underline{N}}\,m^{\underline{MN}} + \frac{1}{2}\,\bg{g}_{\mu \nu}\, \bg{M}^{\underline{MN}}\,\dd_{\underline{M}} \dd_{\underline{N}} h^{\rho}{}_{\rho} -\frac{1}{2}\,\bg{M}^{\underline{MN}}\, \dd_{\underline{M}} \dd_{\underline{N}} h_{\mu \nu},
    \end{aligned}
\end{equation}
with $m^{\underline{MN}}=\bg{M}^{\uM \uK}\bg{M}^{\uN \uL} m_{\uK \uL}$, and where the background scalar field equations~\eqref{eq:fieldeqbgscal} have been used. The variation of these equations under linearized diffeomorphisms is
\begin{equation}
    \begin{aligned}
       \delta_{\lambda,\zeta} \mathcal{Y}_{\mu\nu}  &= \bg{\cal L}_\zeta \bg{E}_{\mu\nu}   - \bg{M}^{\underline{MN}}\, \dd_{\uM}\zeta_{(\mu} \,\bg{E}_{\nu) \underline{N}},
    \end{aligned}
\end{equation}
which vanishes on background field equations.

\paragraph{Scalar fields \boldmath $m_{\underline{MN}}$}
\begin{equation}
    \begin{aligned}
        0 = {\cal Y}_{\underline{MN}} &= -\frac{1}{12}\,\bg{M}_{\underline{K}(\underline{M}}\,\bg{D}_{\mu}j^{\mu\,\underline{K}}{}_{\underline{N})} +\frac{1}{12}\,M_{(0)}^{2}{}_{\underline{MN}}{}^{\underline{KL}}\,m_{\underline{KL}}\\
        &\quad -\frac{1}{12}\,m_{\underline{P}(\underline{M}}\bg{D}_{\mu}\bg{J}^{\mu\,\underline{P}}{}_{\underline{N})} - \frac{1}{12}\,\bg{M}_{\underline{P}(\underline{M}}\mathbb{P}_{\underline{N})}{}^{\underline{P}}{}_{\underline{Q}}{}^{[\underline{K}}\bg{M}^{\underline{L}]\underline{Q}} \,m_{\underline{KR}} \bg{D}_{\mu}\bg{J}^{\mu\,\underline{R}}{}_{\underline{L}}\\
        &\quad + \frac{1}{4}\,\bg{\mathbb{P}}_{{\rm coset}\,\underline{MN}}{}^{\underline{KL}} \, m_{\underline{KP}}\bg{F}_{\mu\nu}{}^{\underline{P}}\bg{M}_{\underline{LQ}}\bg{F}^{\mu\nu\,\underline{Q}} + \frac{1}{4}\, m_{\underline{P}(\underline{M}}\mathbb{P}_{\underline{N})}{}^{\underline{P}}{}_{\underline{Q}}{}^{\underline{K}} \bg{M}_{\underline{KR}}\,\bg{F}_{\mu\nu}{}^{\underline{R}}\bg{F}^{\mu\nu\,\underline{Q}}\\
        &\quad - \frac{1}{12}\, \bg{M}_{\underline{K}(\underline{M}}\mathbb{P}_{\underline{N})}{}^{\underline{K}}{}_{\underline{R}}{}^{\underline{L}}\, \bigg[- \bg{J}^{\mu\,\underline{P}}{}_{\underline{L}}\,\Big(X_{\underline{QP}}{}^{\underline{R}} - 6\,\mathbb{P}_{\underline{P}}{}^{\underline{R}}{}_{\underline{Q}}{}^{\underline{S}}\,\dd_{\underline{S}}\Big)\,a_{\mu}{}^{\underline{Q}} \\
        &\qquad\qquad\qquad\qquad\qquad\ \ + \bg{J}^{\mu\,\underline{R}}{}_{\underline{P}}\,\Big(X_{\underline{QL}}{}^{\underline{P}} - 6\,\mathbb{P}_{\underline{L}}{}^{\underline{P}}{}_{\underline{Q}}{}^{\underline{S}}\,\dd_{\underline{S}}\Big)\,a_{\mu}{}^{\underline{Q}}\bigg] \\
        &\quad + \frac{1}{12}\, \bg{M}_{\underline{KM}}\bg{J}^{\mu\,\underline{K}}{}_{\underline{N}}\,\dd_{\underline{P}}a_{\mu}{}^{\underline{P}} + \frac{1}{2}\, \bg{\mathbb{P}}_{{\rm coset}\,\underline{MN}}{}^{\underline{KL}} \bg{M}_{\underline{KP}}\bg{M}_{\underline{LQ}}\,\bg{F}^{\mu\nu\,\underline{P}}\,\big(2\,\bg{D}_{\mu}a_{\nu}{}^{\underline{Q}} - {\cal Z}^{\underline{QR}}\,\tilde{b}_{\mu\nu,\underline{R}}\big) \\
        &\quad + \frac{1}{12}\,\bg{M}_{\underline{K}(\underline{M}}\bg{D}_{\mu}\Big(h^{\mu\nu}\bg{J}_{\nu}{}^{\underline{K}}{}_{\underline{N})}\Big) - \frac{1}{24}\, \bg{M}_{\underline{KM}}\bg{J}_{\mu}{}^{\underline{K}}{}_{\underline{N}}\,\bg{D}^{\mu}h_{\rho}{}^{\rho} + \frac{1}{24}\,\bg{M}^{\underline{PQ}}\, \Pi_{\underline{MN},\underline{P}}\,\dd_{\underline{Q}}h_{\mu}{}^{\mu} \\
        &\quad - \frac{1}{2}\,\bg{\mathbb{P}}_{{\rm coset}\,\underline{MN}}{}^{\underline{KL}} \, \bg{M}_{\underline{KP}}\bg{ F}_{\mu\rho}{}^{\underline{P}}\bg{M}_{\underline{LQ}}\bg{F}_{\nu}{}^{\rho\,\underline{Q}} \, h^{\mu\nu},
    \end{aligned}
    \label{ScEq}
\end{equation}
with the scalar mass operator~\cite{Malek:2020yue}
\begin{equation}
\begin{aligned}
    M_{(0)}^{2}{}_{\underline{MN}}{}^{\underline{KL}} &= \bg{\mathbb{P}}_{{\rm coset}\,\underline{MN}}{}^{\underline{XY}} \, \bigg[ \,X_{\underline{XP}}{}^{\underline{Q}}X_{\underline{RQ}}{}^{\underline{P}}\bg{M}^{\underline{RK}} + \frac{1}{5}\,X_{\underline{PR}}{}^{\underline{U}}X_{\underline{QS}}{}^{\underline{K}}\bg{M}^{\underline{PQ}}\bg{M}^{\underline{RS}}\bg{M}_{\underline{XU}}\\
        &\qquad\qquad\qquad\quad\ + \frac{1}{5}\,\Big(X_{\underline{XP}}{}^{\underline{R}}X_{\underline{UQ}}{}^{\underline{S}}+X_{\underline{PX}}{}^{\underline{R}}X_{\underline{QU}}{}^{\underline{S}}\Big)\bg{M}^{\underline{PQ}}\bg{M}_{\underline{RS}}\bg{M}^{\underline{UK}} \\
        &\qquad\qquad\qquad\quad\  +2\,X_{\underline{PX}}{}^{\underline{K}}\bg{M}^{\underline{PQ}}\,\dd_{\underline{Q}}- 2\,X_{\underline{PR}}{}^{\underline{U}}\bg{M}^{\underline{PQ}}\bg{M}^{\underline{RK}}\bg{M}_{\underline{UX}}\,\dd_{\underline{Q}}\bigg] \delta_{\underline{Y}}{}^{\underline{L}}\\
        & +  \bg{\mathbb{P}}_{{\rm coset}\,\underline{MN}}{}^{\underline{XY}} \, \bigg[ - \frac{1}{5}\,\Big(X_{\underline{XP}}{}^{\underline{K}}X_{\underline{YQ}}{}^{\underline{L}}+X_{\underline{PX}}{}^{\underline{K}}X_{\underline{QY}}{}^{\underline{L}}\Big)\bg{M}^{\underline{PQ}} \\
        &\qquad\qquad\qquad\quad\  + \frac{1}{5}\,\Big(X_{\underline{XP}}{}^{\underline{R}}X_{\underline{YQ}}{}^{\underline{S}}+X_{\underline{PX}}{}^{\underline{R}}X_{\underline{QY}}{}^{\underline{S}}\Big)\bg{M}_{\underline{RS}}\bg{M}^{\underline{PK}}\bg{M}^{\underline{QL}}\\
        &\qquad\qquad\qquad\quad\ - \frac{1}{5}\,\Big(X_{\underline{PR}}{}^{\underline{U}}X_{\underline{QS}}{}^{\underline{V}}+X_{\underline{RP}}{}^{\underline{U}}X_{\underline{SQ}}{}^{\underline{V}}\Big)\bg{M}^{\underline{RS}}\bg{M}_{\underline{UX}}\bg{M}_{\underline{VY}}\bg{M}^{\underline{PK}}\bg{M}^{\underline{QL}} \\
        &\qquad\qquad\qquad\quad\ + 2\, X_{\underline{XP}}{}^{\underline{L}}\bg{M}^{\underline{PK}}\,\dd_{\underline{Y}} - 2\, X_{\underline{PX}}{}^{\underline{R}}\bg{M}^{\underline{PK}}\bg{M}^{\underline{UL}}\bg{M}_{\underline{RY}}\,\dd_{\underline{U}}\, \bigg]\\
        &\quad -  \delta_\uM{}^\uK \delta_\uN{}^\uL \bg{M}^{\underline{PQ}}\,\dd_{\underline{P}}\dd_{\underline{Q}} + 12\,\bg{\mathbb{P}}_{{\rm coset}\,\underline{MN}}{}^{\underline{QL}} \,\bg{M}^{\underline{PK}}\,\dd_{\underline{P}}\dd_{\underline{Q}},
        \end{aligned}
\end{equation}
with the following variation under external diffeomorphisms:
\begin{equation}
    \begin{aligned}
        \delta_{\lambda,\zeta} {\cal Y}_{\underline{MN}} &= \bg{E}_{\underline{KL}} \, \bigg(-2\,X_{\underline{P}(\underline{M}}{}^{\underline{K}}\,\delta_{\underline{N})}{}^{\underline{L}} + 12\,\delta_{(\underline{M}}{}^{\underline{K}}\,\mathbb{P}_{\underline{N})}{}^{\underline{L}}{}_{\underline{P}}{}^{\underline{Q}} \,\dd_{\underline{Q}} - \frac{2}{3} \,\delta_{(\underline{M}}{}^{\underline{K}}\,\delta_{\underline{N})}{}^{\underline{L}}\,\dd_{\underline{P}} \bigg) \tilde{\lambda}^{\underline{P}} \\
        &\quad + \bg{\cal L}_\zeta \bg{E}_{\underline{MN}} + \bg{M}_{\uK(\uM} \mathbb{P}_{\underline{N})}{}^{\underline{K}}{}_{\underline{R}}{}^{\underline{L}}\,   \bg{M}^{\uR \uP} \; \bg{E}^\mu{}_\uP \; \dd_\uL \zeta_\mu.
    \end{aligned}
\end{equation}
This variation vanishes for a background satisfying the five-dimensional equations of motion.

\paragraph{}
We will now discuss in section~\ref{sec:homoyopytransfer} and~\ref{sec:BHbackground} two different applications of the general framework developed in section~\ref{sec:pertubtheory}. These sections will not use the ExFT framework of section~\ref{sec:ExFTrecap}, so there will be no need to distinguish the ExFT internal indices $M$ from the ones of five-dimensional supergravity $\underline{M}$. Hence, in order to lighten the notations, from now on we will write the ${\rm E}_{6(6)}$~indices of five-dimensional supergravity without the under line, e.g. $M_{MN}$ instead of $M_{\underline{MN}}$. We will as well remove the tilde from the modified fields $\tilde{b}_{\mu\nu\,\underline{M}}$ and gauge parameters $\tilde{\lambda}^{\underline{M}}$ and $\tilde{\xi}_{\mu\,\underline{M}}$.

\section{Higgs Mechanism via Homotopy transfer} \label{sec:homoyopytransfer}

In this section we display the Higgs mechanism by means of homotopy transfer for a special class of backgrounds. 
We restrict our study to backgrounds where
one-form and two-form gauge fields are zero, and the background generalized (internal) metric is 
constant. Note, however, that this still describes  a large class of backgrounds including AdS$_5\times S^5$. 
We encode these data  in a chain complex of fields, gauge parameters and other quantities defining the linearized field theory, 
and then describe the reorganization of the gauge redundant fields into massive physical (and hence gauge invariant) fields by homotopy transfer. This reorganization  
defines the Higgs mechanism.

\subsection{Restricted Backgrounds}

We begin by specifying the field content and the equations of motion on the restricted backgrounds. 
Those backgrounds are defined by a general five-dimensional external metric 
$\bg{g}_{\mu\nu}(x)$ and a constant internal generalized metric $\bg{M}_{MN}$.
We recall that even though 
$\bg{M}_{MN}$ is constant, via the Scherk-Schwarz embedding formulas it still gives rise to a generally curved internal 
geometry such as the sphere $S^5$. The following background quantities are assumed to vanish:  
\begin{equation}
    \bg{F}_{\mu\nu}{}^{M} = 0\,, \quad  \bg{J}_\mu{}^{M}{}_{N} =0\,, \quad \bg{B}_{\mu\nu, \, M} =0\,.
\end{equation}

For this truncation 
the gauge transformations~\eqref{eq:alltransfofluct} under the fluctuations $(\lambda^ {M},\zeta_{\mu})$ reduce to 
\begin{equation}\label{gaugeVAR} 
	\begin{split}
		\delta h_{\mu\nu} &= 2\,\nabla_{(\mu} \zeta_{\nu)} + \frac{2}{3}\, \dd_{M} \lambda^M \bg{g}_{\mu\nu}\,, \\
		\delta m_{MN} &= -\Pi_{MN,K} \lambda^{K}\,,   \\
		\delta a_{\mu}{}^{M} &= \partial_{\mu} \lambda^{M} +  \mathcal{Z}^{MN} \xi_{\mu N}  + \dd^{M} \zeta_{\mu} \,, \\
		{\cal Z}^{NM}\delta b_{\mu\nu \, M} &= {\cal Z}^{NM}\,2\,\partial_{[\mu} \xi_{\nu] M}\,, 
	\end{split}
\end{equation}
where we recall that $\nabla_{\mu}=\partial_{\mu}+\Gamma_{\mu} $ is the (background) diffeomorphism invariant derivative, 
and $\Pi_{MN,K} $ and $\mathcal{Z}^{MN}$ are the operators defined in section~\ref{sec:fluctuations}, c.f.~(\ref{eq:Zoperator}) and (\ref{PiOp}), which obey 
 \be\label{Nilpotentcyrel} 
  \Pi_{MN,K}\mathcal{Z}^{KP}=0\;, \qquad \mathcal{Z}^{MN} \dd_{N}=0= \dd_{N}\mathcal{Z}^{NM}\,, 
 \ee
where the first equation is shown in appendix~\ref{app:Formulas}, and the second equation follows from~(\ref{eq:IZ1}) and~(\ref{eq:IZ2}).
Moreover, we use the external and internal background metrics to raise and lower indices: 
\be
	\zeta_{\mu} := \bg{g}_{\mu\nu} \zeta^{\nu} \;, \qquad 
	\dd^{M} := \bg{M}^{MN} \dd_{N}. 
\ee
As a consequence of the identities  (\ref{Nilpotentcyrel}) we have the following gauge for gauge 
transformations: 
 \be
  \lambda^M = -{\cal Z}^{MN}\chi_N\;, \qquad \xi_{\mu M}=\partial_{\mu}\chi_M\;, 
 \ee
where $\chi_M$ is arbitrary, i.e., the gauge transformations of 
fields with respect to these special parameters are zero.

Furthermore, the background field equations of section~\ref{sec:5dsugra} reduce to 
\begin{align}
    0 = \bg{E}_{\mu\nu} &=\bg{R}_{\mu\nu}-\frac12\,\bg{g}_{\mu\nu}\,\bg{R} +\frac{1}{2}\,\bg{g}_{\mu\nu}\,\bg{V},\\[7pt]
    0 = \bg{E}_{MN} &=  -\frac{1}{12}\bg{\mathbb{P}}_{{\rm coset}\,MN}{}^{KL} \, \bigg[ \,X_{KP}{}^{Q}X_{LQ}{}^{P}  -\frac{1}{5}\,\bg{M}_{KU}\bg{M}_{LV}\,X_{PR}{}^{U}X_{QS}{}^{V}\bg{M}^{PQ}\bg{M}^{RS} \nonumber\\
    & \qquad\qquad\qquad\quad +\frac{1}{5}\,X_{KP}{}^{R}X_{LQ}{}^{S}\bg{M}^{PQ}\bg{M}_{RS} +\frac{1}{5}\,X_{PK}{}^{R}X_{QL}{}^{S}\bg{M}^{PQ}\bg{M}_{RS} \bigg]\,,
\end{align}
while  the ones of the 2-forms and vectors are trivial: $\bg{E}^{\mu\nu\,M}\equiv 0 \equiv {\bg{E}}^\mu{}_{M}$. In the following we assume that the backgrounds are on-shell so that 
the above  equations are obeyed. 

We next turn to the linearized (first order) field equations for the fluctuations around a background. 
To this end we need the linearized field strengths and currents of section~\ref{sec:fluctuations}, which reduce to 
\begin{align}
    f_{\mu\nu}{}^{M} &= 2\,\partial_{[\mu}a_{\nu]}{}^{M}-{\cal Z}^{MN}\,b_{\mu\nu\,N}\,,\\
    h_{\mu\nu\rho\,M} &= 3\,\partial_{[\mu} {b}_{\nu\rho]\,M}\,,\\
    j_\mu{}^{M}{}_{N}  &= \bg{M}^{MK}\,\Big(\partial_\mu m_{KN} +\Pi_{KN,P}\,a_\mu{}^{P}\Big)\,.
\end{align}
The equations of motion, at linear order in the fluctuations, (\ref{eq:FlucTensors})--(\ref{ScEq}) then 
read
\begin{align}
    0 = {\cal Y}^{\mu\nu M} &= {\cal Z}^{MN}\,\bigg(
    \frac{\sqrt{10}}{6}\,\bg{\epsilon}^{\mu\nu\rho\sigma\tau}\,h_{\rho\sigma\tau,\, N} +\bg{M}_{NK} f^{\mu\nu\, K}\bigg), 
    \label{fgkgbkg}
    \\[7pt]%
    %%%%%%%%
    0 =  \mathcal{Y}^\mu{}_{M} &=   
    \bg{M}_{MN}\, \bg{\nabla}_{\nu}  f^{\mu\nu\,N} +  \frac{1}{6} \,X_{MP}{}^{K} j^{\mu \, P}{}_{K} + \dd_{N}   j^{\mu \, N}{}_{M}  +\frac83\,  \dd_{M} \dd_{N}  {a}^{\mu}{}^{N}\nonumber\\
    &+ \dd_{M} \Big(  \bg{\nabla}_\nu   h^{\nu\mu} - \bg{\nabla}^{\mu}  h_{\nu}{}^\nu  \Big), \label{eq:eomvecAdS}
    \\[7pt]
 0={\cal Y}^{MN} &= 
    -\frac{1}{12}\, \,\bg{M}^{K(M}\,\bg{\nabla}_{\mu}j^{\mu\,N)}{}_{K} +\frac{1}{12}\,M_{(0)}^{2\,MN,KL}\,m_{KL}+ \frac{1}{24}\, \Pi^ {MN,P}\,\dd_{P}h_{\mu}{}^{\mu},\\[7pt]
    %%%
   0={\cal Y}^{\mu\nu}
    &= -\frac12\,h^{\mu\nu}\,\bg{R}
    +\frac12\,\bg{g}^{\mu\nu}\,h^{\rho\sigma}\,\bg{R}_{\rho\sigma}
    -\frac12\,\bg{\nabla}^\rho  \bg{\nabla}_{\rho} h^{\mu\nu}
    -\frac12\,\bg{\nabla}^{(\mu} 
     \bg{\nabla}^{\nu)} h_{\rho}{}^{\rho}
    +\bg{\nabla}_\rho 
     \bg{\nabla}^{(\mu} h^{\nu)\rho} \nonumber\\
    & +\frac12\, \bg{g}^{\mu\nu}\,
    \,\bg{\nabla}^\rho  \bg{\nabla}_{\rho} h_{\sigma}{}^\sigma
    -\frac12\,\bg{g}^{\mu\nu}\,
    \bg{\nabla}^\rho 
     \bg{\nabla}_{\sigma} h_{\rho}{}^\sigma
    +\frac{1}{2}\,h^{\mu\nu}\,\bg{V} +\dd_{M} \bg{\nabla}^{(\mu} {a}^{\nu)M}\nonumber\\
    &
    -\bg{g}^{\mu\nu} \,\dd_{M} \bg{\nabla}^\rho  {a}_{\rho}{}^{M} 
    - \frac{1}{12}\,\bg{g}^{\mu \nu}\, X_{{ML}}{}^{{P}} \bg{M}^{{MN}} \bg{M}^{{KL}} \,\dd_{{N}} m _{{PK}} 
     - \frac{1}{2}\,\bg{g}^{\mu \nu}\, \dd_{{M}} \dd_{{N}}\,m^{{MN}}
    % \frac{1}{2}\, \bg{g}^{\mu\nu} m^{KL} \bg{E}_{KL} 
   %+ \frac{1}{12}\,\bg{g}^{\mu \nu}\, \Pi^{\dagger\, N,PK}\dd_{N} m _{PK} 
    \nonumber\\
    &+ \frac{1}{2}\,\bg{g}^{\mu \nu}\, \bg{M}^{MN} \dd_{M} \dd_{N} h^{\rho}{}_{\rho} -\frac{1}{2}\,\bg{M}^{MN}\, \dd_{M} \dd_{N} h^{\mu \nu} \label{fgkgbkg2} \, ,
\end{align}
where we recalled the notation ${\cal Y}$ for the specific functions of the fields that define these linearized field equations.  

One may verify that these equations are invariant under internal and external generalized diffeomorphisms provided 
the background field equations are obeyed, see section~\ref{sec:fluctuations}.

\subsection{Chain complex of gauge algebra and fields}

We now  describe the sub-complex of the gauge algebra and fields. More precisely, this complex includes the 
gauge parameters, the linearized gauge transformations of fields, and the gauge-for-gauge parameters that yield 
vanishing gauge transformations.
There will almost certainly be further spaces of gauge-for-gauge-for-gauge parameters, etc., but in this paper we truncate here.
This truncation is sufficient to describe the homotopy transfer to physical fields.

The chain complex will be defined in terms of three elementary differentials. The first essentially consists of the familiar de Rham differential and its adjoint. 
Since in the following we also need the basic assumptions of Hodge theory we will briefly recall these here, and then define the three differentials. 
(See \cite{Arvanitakis:2021ecw} for a brief introduction to Hodge theory in relation to homotopy transfer.) 

\medskip

\paragraph{Differentials and Hodge Theory}
We consider the spaces of differential $p$-forms $\Omega^p$ on a compact Riemannian manifold, which form the 
de Rham complex displayed in the figure \ref{fig:deRham}, with d$^2=0$.
 \begin{figure}[t]
	\begin{centering}
		\begin{tikzpicture}
		    \node at (1,0) {$0$};
		    \draw[->] (1.5,0)--(2.5,0);
		    %\node at (2,0.3) {$\text{d}$};
			\node at (3,0) {$\Omega^{0}$}; 
			\draw[->] (3.5,0)--(4.5,0);
			\node at (4,0.3) {$\text{d}$};
			\node at (5,0) {$\Omega^1$}; 
			\draw[->] (5.5,0)--(6.5,0);
			\node at (6,0.3) {$\text{d}$};
			\node at (7,0) {$\Omega^2$}; 
			\draw[->] (7.5,0)--(8.5,0);
		    \node at (8,0.3) {$\text{d}$};
			\node at (9,0){$\hdots$};
			%------------------------------
			%\draw[->] (2.5,-0.3) to [bend left=45] (1.5,-0.3);
			%\node at (2,-1) {d$^\dagger$};
			\draw[->] (8.5,-0.3) to [bend left=45] (7.5,-0.3);
			\node at (8,-1) {d$^\dagger$};
			%---------------------------------
			\draw[->] (4.5,-0.3) to [bend left=45] (3.5,-0.3);
			\node at (4,-1) {d$^\dagger$};
			\draw[->] (6.5,-0.3) to [bend left=45] (5.5,-0.3);
			\node at (6,-1) {d$^\dagger$};
			%---------------------------------
		\end{tikzpicture}
		\caption{de Rham chain complex.}
		\label{fig:deRham}
	\end{centering}
\end{figure}
We have also indicated the adjoint d$^{\dagger}$, which is defined with respect to the inner product 
on $p$-forms that exists thanks to the Riemannian metric. More precisely,  defining the inner product as 
\be\label{innerproductFORMS}  
\langle a,b\rangle = \int\, a \wedge \star b\,, 
\ee
where $\star$ is the Hodge operator, the adjoint is defined via 
\be\label{adjointd}
\langle \text{d} a , b \rangle = \langle a, \text{d}^\dagger b \rangle\;, 
\ee
and also obeys $({\rm d}^{\dagger})^2=0$. Harmonic forms are now defined as the zero modes of the Laplace 
operator 
 \be\label{Laplaceian} 
  \Delta:= \text{d}\text{d}^\dagger+\text{d}^\dagger {\rm d}\;. 
 \ee 
The core statement we will need is that a $p$-form $a$ is 
harmonic if and only if  it is closed and co-closed, i.e, if and only if $\text{d}a=\text{d}^\dagger a=0$.
One direction is obvious: a closed and co-closed form is 
harmonic by definition (\ref{Laplaceian}). Conversely, suppose that the form $a$ is harmonic. 
We then have $\Delta a=0$ and hence, using again the definition (\ref{Laplaceian}) together with (\ref{adjointd}), 
\be
0= \langle \Delta a, a \rangle =  \langle \text{d}^\dagger a, \text{d}^\dagger a \rangle + \langle \text{d} a, \text{d} a \rangle \;. 
\ee
Since the metric is Riemannian, the inner product (\ref{innerproductFORMS}) 
 is positive definite, and we conclude that $\text{d}^\dagger a = \text{d} a=0$, 
 i.e., that $a$ is closed and co-closed, completing the proof. Furthermore, we have the Hodge decomposition of 
 a general $p$-form as 
  \be
   a= {\rm d}\beta +{\rm d}^{\dagger}\gamma +a_0\;, 
  \ee
where $\beta\in \Omega^{p-1}$, $\gamma\in \Omega^{p+1}$ and $a_0$ is harmonic, $\Delta a_0=0$.   
This decomposition is unique (up to exact shifts of $\beta$ and co-exact shifts of $\gamma$), because by (\ref{adjointd})
the three terms live in orthogonal subspaces with respect to the inner product (\ref{innerproductFORMS}).
Indeed, suppose that $a={\rm d}\beta$ is harmonic, then by the above it is closed (which is trivial in this case) 
and co-closed, hence ${\rm d}^{\dagger}a=0$, hence 
 \be
  \langle a, a\rangle = \langle {\rm d}\beta, a\rangle  
  = \langle \beta, {\rm d}^{\dagger} a\rangle =  0\;. 
 \ee
Therefore, by the positive-definiteness of the norm, $a={\rm d}\beta=0$. The analogous argument implies  that a harmonic 
co-exact form must be zero. This in turn implies that  
the projector onto the zero modes of the Laplace operator must annihilate  
any exact or co-exact form.

 \medskip
  In the following we will use the generalization of the above to the three operators  $\dd$, $\Pi$ and ${\cal Z}$ from the previous section, see equations~\eqref{eq:Gamma}, \eqref{PiOp} and~\eqref{eq:Zoperator}.  
 Because of the form of the gauge transformations displayed later in equation \eqref{gaugeVAR223}, we consider the operator $\dd_M$  that takes the divergence of vectors as the analogue of the de Rham differential. Moreover, $\dd^{M}$ with one index raised by means of the background metric $\bg{M}^{MN}$, which can be thought of as its  adjoint, maps scalars to vectors.

\begin{figure}[t]
	\begin{centering}
		\begin{tikzpicture}
			\node at (2,0) {one-forms}; 
			\draw[->] (3.0,0)--(4.2,0);
			\node at (3.6,0.35) {$\frak{D}$};

			\node at (5,0) {vectors}; 
			\draw[->] (5.8,0)--(7.0,0);
			\node at (6.4,0.35) {$\frak{D}$};

			\node at (8.5,0) {scalar metrics}; 

			%---------------------------------
			\draw[->] (4.5,-0.3) to [bend left=45] (2.8,-0.3);
			\node at (3.65,-1.05) {$\frak{D}^{\dagger}$};

			\draw[->] (7.6,-0.3) to [bend left=45] (5.5,-0.3);
			\node at (6.55,-1.05) {$\frak{D}^{\dagger}$};
			%---------------------------------

			\node at (5,-2.2) {scalars};

			\draw[->] (4.85,-1.8)--(4.85,-0.45);
			\node at (4.45,-1.1) {$\dd^{\dagger}$};

			\draw[->] (5.15,-0.45)--(5.15,-1.8);
			\node at (5.65,-1.1) {$\dd$};
		\end{tikzpicture}
		\caption{Bi-complex underlying gauge algebra.}
        \label{fig:2DchainComplex}
	\end{centering}
\end{figure}
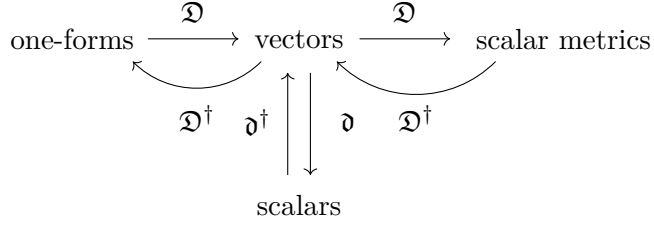

 Similarly, we assume that $\Pi$ has an adjoint $\Pi^{\dagger}$ and 
 ${\cal Z}$ has an adjoint ${\cal Z}^{\dagger}$. More precisely, let us define the inner products by means of the background metric: 
  \be 
   \begin{split}
       \big\langle f_1, f_2\big\rangle_{\rm scalars} \ &:= \ \int dy\, f_{1} f_{2} \;,\\ 
       \big\langle \omega_1, \omega_2\big\rangle_{\rm one-forms}\  &:= \ \int dy\, \bg{M}^{MN} \omega_{1M} \omega_{2N} \;, \\
        \big\langle a_1, a_2\big\rangle_{\rm vectors} \ &:= \  \int dy\, \bg{M}_{MN} a_{1}{}^{M} a_{2}{}^{N} \;,\\ 
    \big\langle m_1, m_2\big\rangle_{\text{scalar metrics}}\  &:= \ \int dy\, \bg{M}^{MK} \bg{M}^{NL}\,  m_{1 MN} m_{2 KL} \;. 
   \end{split}  
  \ee 
  Introducing  then the operators 
 \be 
  \begin{split}
   \frak{D}&: \;\text{one-forms}  \longrightarrow  \text{vectors}\;,  \qquad\qquad \; (\frak{D}\omega)^M\ := \  {\cal Z}^{MN} \omega_N\;, \\
   \frak{D}&: \;\text{vectors}  \longrightarrow  \text{scalar metrics}\;,  \qquad (\frak{D} a)_{MN} \ := \ \Pi_{MN,K} a^K\;, \\
   \dd&: \;\text{vectors}  \longrightarrow  \text{scalars}\;,  \qquad\qquad\qquad   \;\;(\dd a) \ := \  \dd_M a^M\;, 
  \end{split} 
 \ee 
and the adjoint ones in the other direction, 
 \be
  (\dd^{\dagger}  f)^M \ := \  \dd^M f\;, \qquad (\frak{D}^{\dagger} m)^{M} := \Pi^{\dagger M,KL}m_{KL}\;, \qquad
  (\frak{D}^{\dagger}a)_M := {\cal Z}^{\dagger}_{MN} a^N\;, 
 \ee
we indeed have\footnote{Note the sign convention here differs from (\ref{adjointd}) but is more convenient in the present context.}
 \be\label{adjointssss} 
 \begin{split} 
    \big\langle f, \dd a \big\rangle_{\rm scalars} \ &= \ -\big\langle \dd^\dagger  f, a \big\rangle_{\rm vectors} \;, \\
    \big\langle m, \frak{D}a  \big\rangle_{\text{scalar metrics} }  \ &= \ - \big\langle \frak{D}^{\dagger} m, a  \big\rangle_{\text{vectors} } \;, \\
    \big\langle a, \frak{D}\omega  \big\rangle_{\text{vectors} }  \ &= \ - \big\langle \frak{D}^{\dagger} a, \omega  \big\rangle_{\text{one-forms } }\;.  
\end{split} 
 \ee
 The notation with  $\frak{D}$ and $\dd$ is motivated by the fact that the sequence 
 $(\text{one-forms}  \longrightarrow  \text{vectors} \longrightarrow \text{scalar metrics})$ actually forms a chain complex 
 with $\frak{D}^2=0$, which follows from the first equation in (\ref{Nilpotentcyrel}). In contrast, $\dd$ does not 
 form a chain complex with $\frak{D}$; rather, $\frak{D}$ and $\dd$  should be viewed as forming a bi-complex with a double grading 
 as displayed in figure \ref{fig:2DchainComplex}.

 \medskip
 
 As an illustration, let us determine the explicit form of ${\cal Z}^{\dagger}$: 
  \be
  \begin{split}
   \big\langle a, \frak{D}\omega\big\rangle_{\rm vectors} &=  
   \int dy\,\bg{M}_{MN} a^{M} (\frak{D}\omega)^N = 
   \int dy \, a_N {\cal Z}^{NK}\omega_K \\
   &=  \int dy \, a_N \big(Z^{NK}-10\,d^{NKL}\dd_L \big)\omega_K\\
   &= -\int dy\,\omega_K\big(Z^{KN}-10\,d^{KNL}\dd_L\big)a_N\\
   &= - \big\langle \frak{D}^{\dagger} a, \omega  \big\rangle_{\text{one-forms } }\;, 
  \end{split} 
  \ee
 and hence 
  \be\label{Zdagger} 
   {\cal Z}^{\dagger}_{MN} = Z_{MN}-10\,d_{MNK}\dd^K\;, 
  \ee
where we used that $Z^{MN}$ is antisymmetric, we integrated by parts,  and we used that $d^{MNK}$
is symmetric. Thus, ${\cal Z}^{\dagger}$ is obtained from ${\cal Z}$ by lowering the indices. 
With a similar analysis  one obtains  
\be\label{explicitP
Idagger} 
\Pi^{\dagger\, K,MN}= - \bg{M}^{KP} \bg{M}^{L(M} \bigg(2X_{PL}{}^{N)} + 12 \mathbb{P}_{L}{}^{N)}{}_{P}{}^{R}\, 
\dd_R \bigg) \;.
\ee
 \medskip

 In analogy to (\ref{Laplaceian}) we can now define three `Laplace-type' operators, which upon expanding into harmonics 
are related  to the spin-2, vector and tensor mass matrices: 
\be\label{ThreeLaplacians} 
\begin{split}
	\Delta &:= \dd^{M} \dd_{M}\;, 	\\
	(\Delta_V)^M{}_N &:=  \Pi^{\dagger\, M,PQ} \Pi_{PQ,N}\;, \\
	(\Delta_T)_M{}^N &:=\mathcal{Z}^\dagger_{MP} \mathcal{Z}^{PN} \;. 
\end{split}
\ee
These operators each have zero modes and non-zero modes. 
Denoting the projection to zero modes of $\Delta$ by $[\cdot ]$, 
every function $F$ can be split as 
\be 
F=[F]+{\underline{F}} \text{ ,\quad } \Delta[F]=0\;. 
\ee 
On non-zero modes $\Delta$ is invertible, and we denote the inverse 
on this subspace by $K$, extending its definition to the full space by setting it to zero on zero modes.  
$\Delta K$ then defines the projector onto the non-zero modes:
\be 
K\Delta F= \Delta K F= F -[F] = \underline{F}\;. 
\ee 
Similarly we define 
\be 
A^M = [A^M]_V + 
\underline{A^M}_V
\qquad \qquad B_M=[B_M]_T + \underline{B_M}_T\;, 
\ee
where 
\be\label{nonzeromodeprojectors}  
\begin{split}
&(K_V)^M{}_N (\Delta_V)^N{}_P A^P = 
(\Delta_V)^M{}_N (K_V)^N{}_P A^P= \underline{A^M}_V\;,  \\
&(K_T)_M{}^N (\Delta_T)_N{}^P B_P = 
(\Delta_T)_M{}^N (K_T)_N{}^P B_P= \underline{B_M}_T\;, 
\end{split}
\ee 
and 
\be 
(\Delta_V)^M{}_N [A^N]_V=0 \;, \qquad (\Delta_T)_M{}^N [B_N]_T=0\;. 
\ee 
The perhaps somewhat tortuous looking notation used here is consistent with our previous paper  \cite{Eloy:2025ebd}
treating the torus having 
one Laplacian.  
Note that with these definitions we have, for instance, 
$\langle \Delta_Va, b\rangle_{\rm vectors}  
= \langle a, \Delta_V b\rangle_{\rm vectors}$, hence 
$(\Delta_V)^{\dagger}=\Delta_V$, and therefore also  
$(K_V)^{\dagger}=K_V $ and similarly $(\Delta_T)^{\dagger}=\Delta_T$ and 
$(K_T)^{\dagger}=K_T$. 

Moreover, we will use that, in complete analogy to the harmonic analysis of differential forms on a compact manifold 
reviewed above,  a field that is a zero mode of any of the Laplacians is annihilated by the corresponding differential and its 
adjoint. This follows by the same argument. For instance, if $a^M$ is a zero mode of $\Delta_V$ we have 
 \be
  0 = \big\langle \Delta_V a,a\big\rangle_{\text{vectors}} = \big\langle \Pi^{\dagger}\Pi a,a\big\rangle_{\text{vectors}} 
  =-\big\langle \Pi a,\Pi a\big\rangle_{\text{scalar metrics}} \;, 
 \ee
and hence $\Pi a =0$ or $\Pi_{MN,K}a^K=0$ as a consequence of the positive definiteness of the inner product. 
Similarly,  ${\cal Z}$ annihilates any $\Delta_T$ zero mode, $\Pi^{\dagger}$ maps 
to the non-zero modes of $\Delta_V$, and  ${\cal Z}^{\dagger}$ 
maps to the non-zero modes of $\Delta_T$.  
Finally, in analogy to (\ref{Nilpotentcyrel}) we then also have 
 \be\label{Nilpotentcyrel2} 
  \mathcal{Z}^{\dagger}_{MN} \Pi^{\dagger {N,KL}} =0\;, \qquad \mathcal{Z}^{\dagger}_{MN} \dd^{N}=0\;. 
 \ee
The first relation expresses $(\frak{D}^{\dagger})^2=0$  and follows from $\frak{D}^2=0$ using (\ref{adjointssss}): 
$\langle (\frak{D}^{\dagger})^2m, \omega\rangle_{\text{one-forms}}
=\langle m, \frak{D}^2\omega\rangle_{\text{scalar metrics}}=0$. 
Similarly, the second relation expresses $\frak{D}^{\dagger}\circ \dd^\dagger=0$ on scalars, which follows from $\dd\circ \frak{D}=0$ 
on one-forms, which in turn follows from the second equation in (\ref{Nilpotentcyrel}). 

The relations (\ref{Nilpotentcyrel})
and (\ref{Nilpotentcyrel2}) have important consequences for the projectors onto zero and non-zero modes that we now derive. Specifically, since we can lower and raise indices with $M^{MN}$ we  can apply the zero mode projector of $V$ to a non-zero mode 
w.r.t.~$T$, for instance. 
Leaving the index positions implicit, (\ref{Zdagger}) can then be written as ${\cal Z}={\cal Z}^{\dagger}$, 
and hence we also have  
${\cal Z}\Pi^{\dagger}=0=\Pi {\cal Z}^{\dagger}$ and thus for the Laplacians (\ref{ThreeLaplacians}) 
 \be\label{DeltaREL} 
  \Delta_T\Delta_V = 0 = \Delta_V\Delta_T\;. 
 \ee
Acting then with a $V$ zero mode projection onto a $T$ non-zero mode we compute  with (\ref{nonzeromodeprojectors}) 
for a generic field $B=(B_M)$ 
 \be\label{TrivialVproj} 
  \big[\underline{B}_T\big]_V
  = \big[\Delta_TK_T B\big]_V
  = ({\bf 1}-K_V\Delta_V)(\Delta_TK_TB) = 
  \Delta_TK_TB
  = \underline{B}_T\;, 
 \ee
where we used (\ref{DeltaREL}) 
in the third equality.
Thus, on a $T$ non-zero mode the $V$ zero mode projector acts as the identity and can hence be dropped. 

\medskip 

%\noindent{\textit{Gauge algebra and chain complex:}}\\
\paragraph{Gauge algebra and chain complex}

We now write  the gauge structure in terms of these differentials.  
To this end we use that in the above nomenclature $h_{\mu\nu}$ and its gauge parameters $\zeta_{\mu}$ are scalars, 
 $m_{MN}$ is a scalar metric, $a_{\mu}{}^{M}$ and $\lambda^M$ are vectors, and finally $b_{\mu\nu M}$ and $\xi_{\mu M}$ are 
 one-forms. The gauge transformations (\ref{gaugeVAR}) can then be written as
\begin{equation}\label{gaugeVAR223} 
	\begin{split}
		\delta h_{\mu\nu} \ &= \ 2\, \nabla_{(\mu} \zeta_{\nu)} + \dfrac{2}{3}\, (\dd \lambda)\, \bg{g}_{\mu\nu}\,, \\
		\delta m  \ &= \ -\frak{D} \lambda  \,,   \\
		\delta a_{\mu} \  &= \ \partial_{\mu} \lambda +  \frak{D} \xi_{\mu}  + \dd ^{\dagger}   \zeta_{\mu} \,, \\
		\delta b_{\mu\nu} &= 2\, \partial_{[\mu} \xi_{\nu] }\,,
	\end{split}
\end{equation}
where we recall that strictly speaking the gauge transformations of $b_{\mu\nu M}$ are not complete since this field only appears under ${\cal Z}^{MN}$ projection 
in the equations of motion. Hence, $b_{\mu\nu M}$ can be shifted by any term that is annihilated by ${\cal Z}^{MN}$, 
which can be interpreted as a new gauge redundancy. One can characterize this redundancy quite precisely 
by using the notions introduced above: there is an 
additional gauge symmetry with parameters $\omega_{\mu\nu M}$: 
 \be\label{newgaugeonb} 
  \delta b_{\mu\nu M} = 2\,\partial_{[\mu}\xi_{\nu]M}  + \omega_{\mu\nu M}\;, \qquad \omega_{\mu\nu M}\equiv [\omega_{\mu\nu M}]_{T}\;, 
 \ee
where the new parameter carries only zero modes with respect to $\Delta_T$. By our above assumptions 
it then indeed drops out under ${\cal Z}^{MN}$. This new gauge symmetry naturally corresponds to a new 3-form gauge field 
that we know  must be there in the full non-linear off-shell theory as the `covariantly constrained' fields 
that generally appear among the $(n-2)$-forms in $n$ external dimensions \cite{Hohm:2014fxa,Hohm:2019wql}. In our  case they read 
$c_{\mu\nu\rho M}=[c_{\mu\nu\rho M}]_T$ 
and also carry only $\Delta_T$ zero modes and hence obey ${\cal Z}^{MN} c_{\mu\nu\rho N}=0$. These  fields do not show up 
at the present truncation, but since their gauge parameters act on $b_{\mu\nu M}$ we have to  keep these new gauge parameters.

We are now ready to define the gauge chain complex, which is displayed in figure \ref{fig:homotopy transfer} together with the homotopy maps to be defined below. 
\begin{figure}[b]
	\begin{centering}
		\begin{tikzpicture}
			\node at (3,0) {$X_{-2}$}; 
			\draw[->] (3.5,0)--(4.5,0);
			\node at (4,0.3) {$\partial_{-2}$};
			\node at (5,0) {$X_{-1}$}; 
			\draw[->] (5.5,0)--(6.5,0);
			\node at (6,0.3) {$\partial_{-1}$};
			\node at (7,0) {$X_0$}; 
			%------------------------------

            \draw[->] (4.5,-0.3) to [bend left=45] (3.5,-0.3);
			\node at (4,-1) {$\frak{h}_{-1}$};
            \draw[->] (6.5,-0.3) to [bend left=45] (5.5,-0.3);
			\node at (6,-1) {$\frak{h}_0$};
			%---------------------------------
			\node at (3,-0.5) {$ \chi$};
			\node at (5,-0.5) {$\Lambda$};
			\node at (7,-0.5) {$\Phi$};
		\end{tikzpicture}
		\caption{Truncated chain complex.}
			\label{fig:homotopy transfer}
	\end{centering}
\end{figure}
The space in degree $-2$ consists of gauge-for-gauge parameters, which we collectively denote by 
\be \label{eq:X-2}
  \chi= \begin{pmatrix} \chi_M \\  \chi_{\mu M} \end{pmatrix} 
  \in X_{-2} 
  \;, \qquad  \chi_{\mu M}\equiv [ \chi_{\mu M}]_{T}\;. 
 \ee
The space in degree $-1$ consists of gauge parameters collectively denoted by 
 \be \label{eq:X-1}
   \Lambda= \begin{pmatrix}  \zeta_\mu   \\   \lambda^M  \\    \xi_{\mu\,M}  \\   \omega_{\mu\nu M} \end{pmatrix} 
   \in X_{-1} \;, 
  \ee
and the space in degree $0$ consists of fields collectively denoted by 
 \be \label{eq:X0}
  \Phi= \begin{pmatrix}  h_{\mu\nu} \\  m_{MN} \\   a_\mu{}^M \\  b_{\mu\nu\,M} \end{pmatrix} \in X_0\;. 
 \ee  
The two non-trivial differentials displayed in the figure are then given by 
	\be\label{generalizedDifferentials} 
\begin{split}
	{\partial}_{-2}(\chi) = \begin{pmatrix} 
								0\\[0.5ex]
								%-\mathcal{Z}^{MN} \chi_N
								-(\frak{D}\chi)^{M}\\[0.5ex]  
								\partial_\mu \chi_M-\chi_{\mu M} \\[0.5ex]  
								 2\,\partial_{[\mu}\chi_{\nu]M} 
								\\[0.5ex]  
							\end{pmatrix} 	\;, \qquad 
							{\partial}_{-1}(\Lambda) = \begin{pmatrix} 
	2\,\nabla_{(\mu} \zeta_{\nu)} + \frac{2}{3} (\dd  \lambda) \bg{g}_{\mu\nu}\\[0.5ex]
	-(\frak{D}\lambda)_{MN} % -\Pi_{MN,K} \lambda^{K}
		 \\[0.5ex]  
		 \partial_{\mu} \lambda^{M} +  (\frak{D}\xi_{\mu})^M%\mathcal{Z}^{MN} \xi_{\mu N}  
		 + (\dd^{\dagger} \zeta_{\mu})^M   \\[0.5ex]  
   2\,\partial_{[\mu} \xi_{\nu] M} + \omega_{\mu\nu M} 
   \end{pmatrix}\;. 
\end{split}
\ee
The condition $\partial_{-1}\circ \partial_{-2}=0$ making figure~\ref{fig:homotopy transfer} into a chain complex quickly follows from 
$\frak{D}^2=0$, $\dd\circ \frak{D}=0$ and ${\cal Z}^{MN}\chi_{\mu N}=0$.

\subsection{Homotopy transfer to gauge invariant fields}

In this subsection, we construct the homotopy transfer data using the notation of~\cite{Eloy:2025ebd}. 
\paragraph{Gauge invariant fields} 
We begin by defining new field variables that are close to being gauge invariant and that, by a slight abuse of language, 
we will refer to as gauge invariant fields. To this end we  define field dependent functions into the space of gauge parameters: 
 \be\label{homotopyh0one} 
	\frak{h}_0(\Phi) := \begin{pmatrix} 
								\frak{h}_0(\Phi)_{\mu} \\[0.5ex]
								\frak{h}_0(\Phi)^M \\[0.5ex]  
								\frak{h}_0(\Phi)_{\mu M} \\[0.5ex]  
								\frak{h}_0(\Phi)_{\mu\nu M} 
								\\[0.5ex]  
							\end{pmatrix} \ := \ 
							\begin{pmatrix} 
								K \dd_M \left( a_\mu{}^M + (K_V)^M{}_N\, \Pi^{\dagger\, N,PQ} \partial_\mu m_{PQ} \right) \\[0.5ex]
								-(K_V)^M{}_N\, \Pi^{\dagger\, N,PQ}\, m_{PQ} \\[0.5ex]  
								 (K_T)_M{}^N \mathcal{Z}^\dagger_{NP} \left(a_\mu{}^P
								  - \partial_\mu (\frak{h}_0)^P
	 - \dd^P (\frak{h}_0)_\mu \right) \\[0.5ex]  
								[b_{\mu\nu M}]_T
								\\[0.5ex]  
							\end{pmatrix}  \;. 
 \ee	
 The notation indicates that below this will be interpreted as  homotopy maps in degree zero. 
Note that the last line of the definition is well-defined thanks to the explicit projection onto $\Delta_T$ zero modes. 
An explicit computation determines their gauge transformations as 
\be\label{h0Gauge}  
\begin{split}
	\delta(\frak{h}_{0})_{\mu}&= \underline{\zeta_\mu} + \partial_\mu\big(K\dd_M  [\lambda^M]_V\big)\;, \\
	\delta(\frak{h}_{0})^{M} &= \underline{\lambda^M}_V \;, \\
	\delta(\frak{h}_{0})_{\mu\, M} 
	&=  \underline{\xi_{\mu M}}_T + \partial_\mu\Big((K_T)_M{}^N \mathcal{Z}^\dagger_{NP} [{\lambda}^P]_V\Big)\;, \\
	\delta(\frak{h}_{0})_{\mu\nu \, M} 
	&= \omega_{\mu\nu M} + 2\,\partial_{[\mu}[\xi_{\nu]M}]_T \;. 
\end{split}
\ee 
The structure of these gauge transformations now suggests to define the following fields: 
\be\label{almostgaugeinv} 
\begin{split}
         \widehat{h}_{\mu\nu} &:= h_{\mu\nu}{} - 2\, 
         \nabla_{(\mu} \frak{h}_0(\Phi)_{\nu)} -  \frac{2}{3}\dd_M \frak{h}_0(\Phi)^M \bg{g}_{\mu\nu}\,,\\[7pt]
	\widehat{m}_{MN} &:= m_{MN} +  \Pi_{MN,K}\frak{h}_0(\Phi)^K\,,\\
    &=\mathbb{P}_{MN}{}^{PQ} m_{PQ}\,,\\[7pt]
	\widehat{a}_{\mu}{}^M &:= a_{\mu}{}^M - \partial_\mu \frak{h}_0(\Phi)^M -\mathcal{Z}^{MN} \frak{h}_0(\Phi)_{\mu\,N} - \dd^M \frak{h}_0(\Phi)_\mu\,, \\
    &=\mathcal{P}^M{}_N \big(a_\mu{}^N +  (K_V)^N{}_R \Pi^{\dagger\,R,PQ} \partial_\mu m_{PQ}\big) \,,\\[7pt]
	\widehat{b}_{\mu\nu\,M} &:= b_{\mu\nu\,M} - 2\, \partial_{[\mu} \frak{h}_0(\Phi)_{\nu]\,M} 
	- \frak{h}_0(\Phi)_{\mu\nu M} \,, 
\end{split}
\ee 
with 
\be\label{undertildeproj} 
\mathcal{P}^M{}_N := \delta^M{}_N - \mathcal{Z}^{MK} (K_T)_K{}^L \mathcal{Z}^{\dagger}_{LN}
-\dd^M K \dd_N\;, 
\ee
 and 
 \be\label{Piproj}
 \mathbb{P}_{MN}{}^{KL} := \delta_{(M}{}^{K} \delta_{N)}{}^{L} 
  -\Pi_{MN,P} (K_V)^{P}{}_{Q} \Pi^{\dagger\,Q,KL}\,.
   \ee
   It is straightforward to verify that these operators  are indeed projectors satisfying ${\mathcal{P}}^2={\mathcal{P}}$ and ${\mathbb{P}}^2={\mathbb{P}}$. They implement the simultaneous projection onto the space of vectors that are `divergence-free' w.r.t.~$\dd_M$ and 
   ${\cal Z}^{\dagger}$, and on the space of scalar metrics that are annihilated by $\Pi^\dagger$. Moreover, $\widehat{b}$ carries only $T$ non-zero modes, so that the hatted fields in total are subject to the constraints: 
 \be\label{HattedConstraints} 
 \begin{split}
  &\dd_M \widehat{a}_\mu{}^M = 0 = {\cal Z}^{\dagger}_{MN}\widehat{a}_\mu{}^N\,,\\
  &\Pi^{\dagger\,K,MN} \widehat{m}_{MN} = 0\;, \\
  & \widehat{b}_{\mu\nu M} = \underline{\widehat{b}_{\mu\nu\,M}}_T\;. 
  \end{split}
 \ee 
Their gauge transformations are quickly computed from (\ref{h0Gauge}): 
\be\label{gaugeofgaugeinvariantfields} 
\begin{split}
	\delta \widehat{m}_{MN} &= 0\;, \\[7pt]
    \delta \widehat{a}_{\mu}{}^M 
    &= \partial_\mu [\lambda^M]_V -\mathcal{Z}^{MN} (K_T)_N{}^P \mathcal{Z}^{\dagger}_{PQ} \partial_\mu [{\lambda}^Q]_V
     - \dd^M K \dd_P \partial_\mu [\lambda^P]_V \\
    &=\partial_\mu[\widehat{\lambda}^M]_V\;, \\[7pt]
    \delta 	\widehat{b}_{\mu\nu\,M}  &= 0 \;, \\[7pt]
	\delta \widehat{h}_{\mu\nu}  &= 2\, \nabla_{(\mu} [\zeta_{\nu)}] +\frac{2}{3} \dd_M [\lambda^M]_V \bg{g}_{\mu\nu} 
	- 2\, K  \nabla_{\mu} \nabla_{\nu}\, \dd_M [\lambda^M]_V \;, 
\end{split}
\ee 
where we used in the second line the projector ${\cal P}$ and 
defined $\widehat{\lambda}^M=\mathcal{P}^M{}_N\lambda^N$.  
The gauge transformations (\ref{gaugeofgaugeinvariantfields}) are consistent with the constraints (\ref{HattedConstraints}). 
We infer that the internal metric fluctuations  $m_{MN}$ and the two-forms  $b_{\mu\nu\, M}$ are indeed fully gauge invariant while the spin-2 field  $h_{\mu\nu}$ and the vectors $a_\mu{}^M$ are gauge invariant under the higher mode gauge transformation but transform under the zero modes,
hence defining a notion of gauge covariance. 
Nevertheless, by slight  abuse of language, we will often refer to these hatted  fields as gauge invariant.

\medskip 

\paragraph{Downstairs chain complex} 
We now define a second chain complex of gauge invariant fields and establish homotopy transfer from the above 
`upstairs' chain complex of gauge redundant fields to this `downstairs' chain complex. 

In degree zero we have the fields 
 \be \label{eq:oX0}
\begin{split}
	\down{\Phi} &= \begin{pmatrix} 
	\down{h}_{\mu\nu}  \\[0.5ex]
    \down{m}_{MN} \;,   \quad\Pi^{\dagger\,K,MN} \down{m}_{MN}\equiv 0  \\[0.5ex]
	\down{a}_\mu{}^M \;, \;\; \mathcal{Z}^\dagger_{MN} \down{a}_\mu{}^N\equiv 0 \equiv \dd_M \down{a}_\mu{}^M  \\[0.5ex]  
  \down{b}_{\mu\nu M} \equiv \underline{\down{b}_{\mu\nu M}}_T  \\[0.5ex] 
  \end{pmatrix}\in \down{X}_0 \;, 
\end{split}
\ee 
where in the last  three entries we displayed the constraints that the downstairs fields are born with. 
As suggested by (\ref{gaugeofgaugeinvariantfields}), in the downstairs complex only the respective zero modes 
of two gauge parameters are left: 
\be \label{eq:oX-1}
\begin{split}
	\down{\Lambda} &= \begin{pmatrix} 
	\down{\zeta}_\mu \equiv [\down{\zeta}_\mu]\\[0.5ex]
	 \down{\lambda}^M  \equiv  [\down{\lambda}^M]_V 
   \end{pmatrix} \ \in \ \down{X}_{-1} \;, 
\end{split}
\ee 
with the differential given by 
	\be \label{eq:od-1}
	\down{\partial}_{-1}(\down{\Lambda}) = \begin{pmatrix} 
	2\, \nabla_{(\mu} \down{\zeta}_{\nu)} + \frac{2}{3}\, \dd_{M} \down{\lambda}^M \bg{g}_{\mu\nu}
  -2\, K\nabla_{\mu}\nabla_{\nu}\, \dd_{M} \down{\lambda}^M\\[0.5ex]
	0 \\[0.5ex]
	 \partial_{\mu} \mathcal{P}^M{}_N \down{\lambda}^{N} 
		 \\[0.5ex]  
		0  
   \end{pmatrix}\;. 
\ee
Finally, there is only one gauge-for-gauge parameter left, 
 \be \label{eq:oX-2}
  \down{\chi} = (\down{\chi}_M)\in \down{X}_{-2} \;, 
 \ee
on which the differential acts as 
\be\label{donwstairstrivialgauge} 
	\begin{split}
	\down{\partial}_{-2}(\down{\chi}) &= \begin{pmatrix} 
	0\\[0.5ex]
	 -\mathcal{Z}^{MN}\down{\chi}_N 
	    \end{pmatrix} 	. 
	\end{split}
	\ee
Note that $\down{\chi}_M$ need not be constrained, since 
by the first relation in (\ref{Nilpotentcyrel}) $\down{\partial}_{-2}(\down{\chi})$ is well-defined: 
$\mathcal{Z}^{MN}\down{\chi}_N$ is annihilated by $\Pi$ and hence  belongs to the $V$ zero modes. 
We have  $\down{\partial}_{-1}\circ \down{\partial}_{-2}=0$, which follows with a short 
computation using the form  of the projector (\ref{undertildeproj}).

 \paragraph{Projection and inclusion} We now define the homotopy data: projection, inclusion and homotopy map, 
beginning with the former two. 
The projection  and inclusion on the fields is 
\be\label{degreeZEROPROJ}
\begin{split}
\widehat{\Phi}:= p_0(\Phi)=p_0 \begin{pmatrix}
{h}_{\mu\nu}\\[0.5ex]
	{m}_{MN}\\[0.5ex]  
	{a}_\mu{}^M\\[0.5ex]  
	{b}_{\mu\nu M}%\\[0.5ex]  
	%{h}_{\mu\nu}
\end{pmatrix}&=
 \begin{pmatrix} 
 \widehat{h}_{\mu\nu} \\[0.5ex]
		 \widehat{m}_{MN} \\[0.5ex]
	    \widehat{a}_\mu{}^M  \\[0.5ex]  
		 \widehat{b}_{\mu\nu\,M} % \\[0.5ex]
		%
		  %\widehat{h}_{\mu\nu} 
	\end{pmatrix} \;, \qquad \iota_{0}(\down{\Phi}) = \down{\Phi}\;, 
\end{split}
\ee 
in terms of the hatted fields (\ref{almostgaugeinv}), with the inclusion being trivial. 
	The projection and inclusion on the gauge parameters is
\be\label{degreezeroprojectionandinclusion} 
	\begin{split}
	p_{-1}(\Lambda) = p_{-1} \begin{pmatrix} 
	\zeta_\mu \\[0.5ex]
	 \lambda^M \\[0.5ex]  
   \xi_{\mu M}\\[0.5ex]  
   \omega_{\mu\nu  M}
   \end{pmatrix}= \begin{pmatrix} 
	[\zeta_\mu] \\[0.5ex]
	 [\lambda^M]_V 
   \end{pmatrix} 	, \quad 
   \iota_{-1}(\down{\Lambda}) &= \iota_{-1}  \begin{pmatrix} 
	\down{\zeta_\mu} \\[0.5ex]
	 \down{\lambda}^M 
   \end{pmatrix}= \begin{pmatrix} 
	\down{\zeta_\mu} - \partial_\mu K \dd_M \down{\lambda}^M \\[0.5ex]
	 \down{\lambda}^M \\[0.5ex]  
  -\partial_\mu (K_T)_M{}^N \mathcal{Z}^{\dagger}_{NK} {\down{\lambda}}^K \\[0.5ex] 
  0
   \end{pmatrix} .
	\end{split}
\ee

The non-trivial chain map conditions involving $p_0$ read 
$\down{\partial}_{-1}(p_{-1}(\Lambda))=p_0(\partial_{-1}(\Lambda))= \delta_{\Lambda}\widehat{\Phi}$ 
and follow because we defined the gauge transformations downstairs, i.e.~$\down{\partial}_{-1}$, 
with 
the same formula as the upstairs  gauge transformations of $\widehat{\Phi}$. 
 
To fix the inclusion in degree $-1$ we note that 
the non-trivial chain map condition for the inclusion involving $\iota_0$ and 
$\iota_{-1}$ states that 
 \be\label{RHSChainmapcond} 
 \iota_0(\down{\partial}_{-1}(\down{\Lambda}) ) 
  = \begin{pmatrix} 
  2 \nabla_{(\mu} \down{\zeta}_{\nu)} + \frac{2}{3}\, \dd_{M} \down{\lambda}^M \bg{g}_{\mu\nu}
  -2K\nabla_{\mu}\nabla_{\nu}\, \dd_{M} \down{\lambda}^M\\[0.5ex]
	0 \\[0.5ex]
	 \partial_{\mu} \mathcal{P}^M{}_N{\down{\lambda}}^{N} 
		 \\[0.5ex]  
   \end{pmatrix}
 \ee
should be equal to  
 \be
  {\partial}_{-1}(\iota_{-1}(\down{\Lambda})) = \begin{pmatrix} 
  2 \nabla_{(\mu}\Big( \down{\zeta}_{\nu)}  - \partial_{\nu)} K \dd_M \down{\lambda}^M\Big) + \frac{2}{3}\, \dd_{M} \down{\lambda}^M \bg{g}_{\mu\nu}\\[0.5ex]
	-\Pi_{MN,K} \down{\lambda}^{K}  \\[0.5ex]
	 \partial_{\mu} \down{\lambda}^{M}  
	 +  \mathcal{Z}^{MN}\Big( -\partial_\mu (K_T)_N{}^K \mathcal{Z}^{\dagger}_{KL} {\down{\lambda}}^L\Big)
	 + \dd^{M} \Big(\down{\zeta_\mu} - \partial_\mu K \dd_N \down{\lambda}^N\Big) 
		 \\[0.5ex]  
   \end{pmatrix}\;, 
 \ee 
where we used (\ref{generalizedDifferentials}).  
This is indeed equal to  (\ref{RHSChainmapcond}) since $\Pi_{MN,K} \down{\lambda}^{K} =0$ because $\down{\lambda}^{K}$ 
is born a $V$~zero mode, and  $\dd^{M} \down{\zeta_\mu}=0$ since $\down{\zeta_\mu}$ is born an ordinary $\Delta$ zero mode, 
and finally the remaining terms in the third  line combine into the undertilde projector (\ref{undertildeproj}) that appears 
in the corresponding entry of (\ref{RHSChainmapcond}). 
This shows why we need a non-trivial inclusion in degree $-1$. 
 
Finally, in degree $-2$ we postulate projection and inclusion 
 \be \label{eq:p-2i-2}
  p_{-2} \begin{pmatrix} \chi_M \\ \chi_{\mu M} \end{pmatrix} = \chi_M \;, \qquad 
  \iota_{-2}(\down{\chi}) = \begin{pmatrix} \down{\chi}_M \\ \partial_{\mu}[\down{\chi}_M]_T  \end{pmatrix} \;, 
 \ee 
where the second map is well-defined in its second entry thanks to the explicit projection to $T$ zero modes.  
The chain map condition for projectors indeed holds: 
 \be
  p_{-1}(\partial_{-2}(\chi))=p_{-1}\begin{pmatrix} 
								0\\[0.5ex]
								-\mathcal{Z}^{MN} \chi_N\\[0.5ex]  
								\partial_\mu \chi_M-\chi_{\mu M} \\[0.5ex]  
								 2\partial_{[\mu}\chi_{\nu]M} 
								\\[0.5ex]  
							\end{pmatrix} 
							= \begin{pmatrix} 
	0\\[0.5ex]
	 - [\mathcal{Z}^{MN} \chi_N]_V 
   \end{pmatrix}= \begin{pmatrix} 
	0\\[0.5ex]
	 -\mathcal{Z}^{MN}{\chi}_N 
	    \end{pmatrix} 
  =\down{\partial}_{-2}(p_{-2}(\chi)) 	\;. 
 \ee
For the chain map condition for the inclusion we compute 
 \be
  \iota_{-1} (\down{\partial}_{-2}(\down{\chi}) ) =  \iota_{-1}  \begin{pmatrix} 
	0\\[0.5ex]
	 -\mathcal{Z}^{MN}\down{\chi}_N 
	    \end{pmatrix} 
	    = \begin{pmatrix} 
	  \partial_\mu K \dd_M \mathcal{Z}^{MN}\down{\chi}_N  \\[0.5ex]
	-\mathcal{Z}^{MN}\down{\chi}_N  \\[0.5ex]  
  \partial_\mu (K_T)_M{}^N \mathcal{Z}^{\dagger}_{NK} \mathcal{Z}^{KL}\down{\chi}_L  \\[0.5ex] 
  0
   \end{pmatrix}
   = \begin{pmatrix} 
	 0  \\[0.5ex]
	-\mathcal{Z}^{MN}\down{\chi}_N  \\[0.5ex]  
  \partial_\mu \underline{\down{\chi}_M}_T  \\[0.5ex] 
  0
   \end{pmatrix}\;, 
 \ee
where we used~(\ref{donwstairstrivialgauge}) and~(\ref{degreezeroprojectionandinclusion}). This  is indeed  equal to 
 \be
  {\partial}_{-2}(\iota_{-2}(\down{\chi})) 
  =  {\partial}_{-2}\begin{pmatrix} \down{\chi}_M \\ \partial_{\mu}[\down{\chi}_M]_T   \end{pmatrix}
  = \begin{pmatrix} 
								0\\[0.5ex]
								-\mathcal{Z}^{MN} \down{\chi}_N\\[0.5ex]  
								\partial_\mu \down{\chi}_M-\partial_{\mu}[\down{\chi}_M]_T \\[0.5ex]  
								 0
								\\[0.5ex]  
							\end{pmatrix} , 
 \ee
where we used the first equation in (\ref{generalizedDifferentials}). This completes the verification of the chain map conditions 
$p\circ \partial = \down{\partial}\circ p $ and $\partial\circ \iota = \iota \circ \down{\partial}$ in degrees $-2$, $-1$ and $0$.

\paragraph{Homotopy maps} In order to complete the description of homotopy transfer we need to define 
the homotopy maps $\frak{h}_k: X_k\rightarrow X_{k-1}$. With the data we have we will be able to fix 
$\frak{h}_0$ and $\frak{h}_{-1}$.

 We already defined  $\frak{h}_0$ in (\ref{homotopyh0one}). 
 The homotopy relation in degree zero, 
 \be
  ({\rm id}-\iota p)_{0}(\Phi) = \partial_{-1}(\frak{h}_{0}(\Phi)) +\frak{h}_1(\partial_{0}(\Phi)) = \partial_{-1}(\frak{h}_{0}(\Phi)) \;, 
 \ee
where we assumed $\frak{h}_1=0$, is just a rewriting of $\widehat{\Phi}=\Phi-\partial_{-1}(\frak{h}(\Phi))$ 
that we used in (\ref{almostgaugeinv}) to define the hatted fields. Thus, in degree zero the homotopy relation is satisfied. 

In degree $-1$, on the other hand, we need to satisfy 
  \be
  ({\rm id}-\iota p)_{-1}(\Lambda) = \partial_{-2}(\frak{h}_{-1}(\Lambda)) +\frak{h}_0(\partial_{-1}(\Lambda))\;. 
 \ee
The left-hand side is 
 \be
  ({\rm id}-\iota p)_{-1}(\Lambda) = 
  \begin{pmatrix} 
	\zeta_\mu \\[0.5ex]
	 \lambda^M \\[0.5ex]  
   \xi_{\mu M}\\[0.5ex]  
   \omega_{\mu\nu  M}
   \end{pmatrix}
   - 
    \begin{pmatrix} 
	[\zeta_\mu]-\partial_{\mu}(K\dd_M[\lambda^M]_V) \\[0.5ex]
	 [\lambda^M]_V \\[0.5ex]  
    -\partial_{\mu}((K_T)_M{}^{N} {\mathcal Z}^{\dagger}_{NK} [\lambda^K]_V) \\[0.5ex]  
  0 
   \end{pmatrix}
  = \begin{pmatrix} \underline{\zeta}_{\mu} + \partial_{\mu}(K  \dd_M[\lambda^M]_V) \\ 
  \underline{\lambda^{M}}_V \\ 
  {\xi}_{\mu M} + \partial_{\mu}((K_T)_M{}^N \mathcal{Z}^\dagger_{NK} [{\lambda}^K]_V)\\
   \omega_{\mu\nu  M}
  \end{pmatrix} , 
 \ee
while the second term on the right-hand side can be deduced from (\ref{h0Gauge}):
 \be
  \frak{h}_0(\partial_{-1}(\Lambda)) = \delta_{\Lambda}(\frak{h}_0(\Phi))  
  = \begin{pmatrix} \underline{\zeta_{\mu}} + \partial_{\mu}(K  \dd_M[\lambda^M]_V) \\ 
  \underline{\lambda^{M}}_V \\ 
  \underline{\xi_{\mu M}}_T + \partial_{\mu}((K_T)_M{}^N \mathcal{Z}^\dagger_{NK} [{\lambda}^K]_V)\\
   \omega_{\mu\nu M} + 2\partial_{[\mu}[\xi_{\nu]M}]_T
  \end{pmatrix}. 
 \ee
Thus, 
 \be
  ({\rm id}-\iota_{-1}  p_{-1}    - \frak{h}_0 \partial_{-1} )(\Lambda) 
  =  \begin{pmatrix}  0  \\ 
 0 \\ 
  [{\xi}_{\mu M}]_T \\
  - 2\partial_{[\mu}[\xi_{\nu]M}]_T  
  \end{pmatrix}\;. 
 \ee
By (\ref{generalizedDifferentials}) this ought to be equal to 
\be 
\partial_{-2}(\frak{h}_{-1}(\Lambda)) = \begin{pmatrix} 
								0\\[0.5ex]
								-\mathcal{Z}^{MN} \frak{h}_{-1}(\Lambda)_N\\[0.5ex]  
								\partial_\mu \frak{h}_{-1}(\Lambda)_M-\frak{h}_{-1}(\Lambda)_{\mu M} \\[0.5ex]  
								 2\partial_{[\mu}\frak{h}_{-1}(\Lambda)_{\nu]M} 
								\\[0.5ex]  
							\end{pmatrix} , 
\ee
which works if we set 
 \be \label{eq:h-1}
   \frak{h}_{-1}(\Lambda) = \begin{pmatrix}  \frak{h}_{-1}(\Lambda)_M \\  \frak{h}_{-1}(\Lambda)_{\mu M} \end{pmatrix} 
   = \begin{pmatrix} 0 \\ -[\xi_{\mu M}]_T  \end{pmatrix}  \;.  
 \ee
 
This completes the proof that the upstairs and downstairs chain complexes are homotopy equivalent for the truncation considered.

\subsection{Trait d'union: Homotopy transfer including field equations }

We now extend the above analysis by including spaces in degree $1$, where the equations of motion live, degree $2$, where the Noether identities live, and $3$, where the Noether-for-Noether identities live. Concretely, in degree $1$ we define the space  with generic elements
 \be\label{genericdegreeonedude} 
  {\cal E} \equiv 
  \begin{pmatrix} 
  {\cal E}_{\mu\nu} \\ 
  {\cal E}_{MN} \\
  {\cal E}_{\mu}{}^{M} \\
  {\cal E}_{\mu\nu M}
  \end{pmatrix}  \ \in \ X_1\;, 
 \ee
which is thus isomorphic to the space of fields in degree zero. 
(We can use the internal and external background metrics to freely raise and lower indices). 
The differential mapping from degree $0$ to degree $1$ is given 
by  the tensors ${\cal Y}$ that define  the linearized field equations, c.f.~(\ref{fgkgbkg})--(\ref{fgkgbkg2}): 
 \be \label{differentialZEROOO} 
  \partial_0(\Phi) = \begin{pmatrix} {\cal Y}_{\mu\nu}(\Phi) \\
  {\cal Y}_{MN}(\Phi) \\
  {\cal Y}_{\mu}{}^{M}(\Phi) \\
  {\cal Y}_{\mu\nu M}(\Phi) 
  \end{pmatrix} \ \in \ X_1 \;. 
 \ee
Furthermore, in parallel to the spaces of gauge and gauge-for-gauge parameters we define the spaces in 
degree $2$ and $3$ as follows: 
 \be \label{eq:X2X3}
  {\cal N}=
  \begin{pmatrix} 
   {\cal N}_{\mu} \\
   {\cal N}^M \\ 
   {\cal N}_{\mu M} \\
   {\cal N}_{\mu\nu M} 
  \end{pmatrix} \ \in \ X_2\;, \qquad
  {\cal R} = \begin{pmatrix} 
  {\cal R}_M \\
  {\cal R}_{\mu M} 
  \end{pmatrix} \ \in \ X_3\;, 
 \ee
where ${\cal N}_{\mu\nu M}$ carries only T-zero modes 
(in parallel to the constraint on $\omega_{\mu\nu M}$), 
as does  ${\cal R}_{\mu M}$. 

The differential in degree $1$ is defined in terms of the Noether identities. These in turn can be derived by specializing the general variation of the action with respect to the fields to their gauge variation. 
The general variation 
 can be written in terms of the (partially) index-free notation of the previous subsection as  
 \be
  \delta S = \int \,d^5x \Big( \big\langle 
  \delta h_{\mu\nu} \, ,{\cal Y}^{\mu\nu} \big\rangle_{\rm scalars} 
  + \big\langle \delta m  \,, {\cal Y}\big\rangle_{\text{scalar metrics}}
  + \big\langle \delta a_{\mu} \,, {\cal Y}^{\mu}\big\rangle_{\rm vectors}  
  + \big\langle  \delta b_{\mu\nu  } \,,{\cal Y}^{\mu\nu }_t
  \big\rangle_{\text{one-forms}} \Big) \,, 
 \ee
 where ${\cal Y}^{\mu\nu}_t$ denotes the tensor field equations.  
Specializing to the gauge variation (\ref{gaugeVAR}) (extended by (\ref{newgaugeonb})) then yields: 
 \be\label{Noethercomp} 
  \begin{split}
   0 = \int \,d^5x \Big( &\big\langle 
  	2 \nabla_{\mu} \zeta_{\nu} + \frac{2}{3} (\dd  \lambda) \bg{g}_{\mu\nu} \, ,{\cal Y}^{\mu\nu} \big\rangle_{\rm scalars} 
  - \big\langle \frak{D} \lambda   \,, {\cal Y}\big\rangle_{\text{scalar metrics}} \\
  &+ \big\langle \partial_{\mu} \lambda +  \frak{D} \xi_{\mu}  + \dd^{\dagger}    \zeta_{\mu}  \,, 
  {\cal Y}^{\mu}\big\rangle_{\rm vectors}  
  + \big\langle 2 \partial_{\mu} \xi_{\nu }+\omega_{\mu\nu } \,,{\cal Y}^{\mu\nu }_t
  \big\rangle_{\text{one-forms}} \Big) \;. 
  \end{split}
 \ee  
This  is zero off-shell as a consequence of gauge invariance. 
We recall that here we are using the above notation ${\cal Y}$ collectively for the tensors that define the field equations. 
Integrating by parts we read off the Noether identities 
 \be
  \begin{split}
   0 &\equiv 2\nabla_{\mu} {\cal Y}^{\mu\nu} +\dd {\cal Y}^{\nu} \qquad\qquad\qquad\qquad \quad    (\zeta_{\nu})\;, \\
   0 &\equiv -\frak{D}^{\dagger}{\cal Y}   
   +\frac{2}{3} \dd^{\dagger}  \big(  \bg{g}_{\mu\nu} {\cal Y}^{\mu\nu}  \big)   
   +\nabla_{\mu}{\cal Y}^{\mu}      \qquad\, (\lambda^M)\;, \\
   0 &\equiv\frak{D}^{\dagger} {\cal Y}^{\mu} 
   + 2 \nabla_{\nu} {\cal Y}^{\nu\mu}_t \qquad \qquad 
   \qquad \quad (\xi_{\mu M}) \;,\\
   0&\equiv [{\cal Y}^{\mu\nu}_t]_T \qquad \qquad \qquad
   \qquad \qquad \quad (\omega_{\mu\nu M}) \;, 
  \end{split} 
 \ee
where we indicated in parenthesis the corresponding gauge parameters that imply this Noether identity. Returning to the more explicit index notation 
these read 
 \be\label{EXPlicitREALNoethers} 
  \begin{split}
   0 &\equiv  2\nabla_{\mu} {\cal Y}^{\mu\nu} +\dd^{M} {\cal Y}^{\nu}{}_{M} \;, \\
   0 &\equiv  -\Pi^{\dagger}{}_{M,KL}  {\cal Y}^{KL}   +\frac{2}{3} \dd_M \big(  \bg{g}_{\mu\nu} {\cal Y}^{\mu\nu}  \big)   
   +\nabla_{\mu}{\cal Y}^{\mu}{}_{M}        \;, \\
   0 &\equiv  {\cal Z}^{\dagger MN} {\cal Y}^{\mu}{}_{N} + 2 \nabla_{\nu} {\cal Y}^{\nu\mu M}  \;, \\
   0&\equiv  [{\cal Y}^{\mu\nu M}]_T  \;. 
  \end{split} 
 \ee
These are identities, i.e., they are identically obeyed for 
the explicit expressions ${\cal Y}$ defining the equations of motion. We now postulate that the differential in degree $1$, acting on a general element (\ref{genericdegreeonedude}), 
is of the same structural form: 
\be\label{deg1diff} 
	\begin{split}
	{\partial}_{1}(\mathcal{E}) &= 
   \begin{pmatrix} 
	2\, \nabla_{\nu}  \mathcal{E}^{\mu\nu} + \dd^{M}  \mathcal{E}^\mu{}_M\\[0.5ex]
	 \nabla_{\mu}  \mathcal{E}^{\mu}{}_M + \frac{2}{3}\, \dd_M \mathcal{E}^\mu{}_\mu - \Pi^{\dagger}_{M,KL} \mathcal{E}^{KL}
		 \\[0.5ex]  
		 2\, \nabla_{\nu}  \mathcal{E}^{\nu\mu\,M} + \mathcal{Z}^{\dagger MN} \mathcal{E}^\mu{}_N    \\[0.5ex]  [\mathcal{E}^{\mu\nu\,M}]_T
   \end{pmatrix} 	\ \in \ X_2\;. 
	\end{split}
	\ee
With this definition, the nilpotency of the differential 
in degree zero $\partial_1\circ \partial_0=0$ 
follows from (\ref{differentialZEROOO}) and is just an expression of the Noether identities (\ref{EXPlicitREALNoethers}). 
Requiring nilpotency of the differential 
in degree $1$ quickly yields  
\be\label{deg2diff} 
	\begin{split}
	 {\partial}_{2}(\mathcal{N}) = 
    \begin{pmatrix} 
	{\partial}_{2}(\mathcal{N})^M\\[0.5ex]
	 {\partial}_{2}(\mathcal{N})^{\mu M}
   \end{pmatrix} 	
   = 
   \begin{pmatrix} 
	\nabla_{\mu}  \mathcal{N}^{\mu M} - \mathcal{Z}^{\dagger MN} \mathcal{N}_N\\[0.5ex]
		 \nabla_{\nu}  \mathcal{N}^{\mu\nu\,M} + \frac{1}{2}[\mathcal{N}^{\mu\,M}]_T
   \end{pmatrix}  \ \in \ X_3	\;. 
    \end{split}
	\ee
Indeed, by an explicit computation one verifies $
 0= (\partial_{2}\circ \partial_1)({\cal E})$. 
This completes our presentation of the chain complex up 
to and including spaces in degree $3$.

\paragraph{Downstairs chain complex}
We now describe the projected downstairs chain complex in degrees $1$, $2$ and $3$.  
First, in degree 1, we expect the equations of motion to be subject to the same projections as the fields:
\be\label{genericdegreeonedudefjkf} 
  \down{\cal E} \equiv 
  \begin{pmatrix} 
  \down{\cal E}_{\mu\nu} \\ 
  \down{\cal E}_{MN}\;, \quad 
  \Pi^{\dagger}_{K,MN} \down{{\cal E}}^{MN} = 0\\
  \down{\cal E}_{\mu}{}^{M}\;, \quad  \dd_M \down{\cal E}_{\mu}{}^{M} = 0 = {\cal Z}^{\dagger}_{MN}\down{\cal E}_{\mu}{}^{N} \\
  \down{\cal E}_{\mu\nu M}
  \equiv \underline{\down{\cal E}_{\mu\nu M}}_T
  \end{pmatrix}  \ \in \ \down{X}_1\;, 
 \ee
where we displayed in each component 
the constraints that are  part of the defintion of $\down{X}_1$.  

Next, in analogy to the downstairs gauge chain complex we define 
the space of Noether identities in degree $2$ to consist of 
generic elements 
 \be \label{eq:oX2}
   \down{\cal N}=
   \begin{pmatrix} 
    \down{\cal N}^{\mu}
    \equiv [\down{\cal N}^{\mu}] \\  \down{\cal N}_M 
    \equiv [\down{\cal N}_M]_V
    \end{pmatrix} 
   \ \in \  \down{X}_2\;,
 \ee   
where we displayed again the constraints that are part of the defintion. 
Finally, 
the space of Noether-for-Noether identities in degree $3$ reduces to 
 \be \label{eq:oX3}
  \down{\cal R} = (\down{\cal R}^M) \ \in \ \down{X}_3\;, 
 \ee
without any constraints, as for the corresponding gauge-for-gauge parameter. 

The differentials are defined as follows. In degree zero $\down{\partial}_0$ takes the 
fields, inserts them into the full field equations but then projects them as approppriste for the definition  (\ref{genericdegreeonedudefjkf}) 
of the target space of $\down{\partial}_0$: 
  \be\label{downstairsDIFFZERO} 
  \down{\partial}_0(\down{\Phi}) 
  = \begin{pmatrix} {\cal Y}^{\mu\nu}(\down{\Phi}) \\
  \mathbb{P}^{MN}{}_{KL}  {\cal Y}^{KL}(\down{\Phi} ) \\
  \mathcal{P}_M{}^N {\cal Y}^{\mu }{}_N(\down{\Phi}) \\
  {\cal Y}^{\mu\nu M}(\down{\Phi}) 
  \end{pmatrix} \;, 
 \ee
where the projectors are defined in  (\ref{undertildeproj}) 
and (\ref{Piproj}). Next, in degree $1$ we define the differential as 
 \be \label{eq:op1}
	\begin{split}
	\down{\partial}_{1}(\down{\mathcal{E}}) 
  = 
   \begin{pmatrix} 
	2\, [\nabla_{\nu}  \down{\mathcal{E}}^{\mu\nu}]\\[0.5ex]
	 [   \nabla_{\mu}  \down{\mathcal{E}}^{\mu}{}_M + \frac{2}{3}\, \dd_M \down{\mathcal{E}}^\mu{}_\mu
     -2\dd_MK\nabla_{\mu}\nabla_{\nu}\down{\cal E}^{\mu\nu} ]_V \end{pmatrix} \;. 	
	\end{split}
	\ee
 This  follows from 
the analogous computation of (\ref{Noethercomp}), but using the gauge transformations (\ref{gaugeofgaugeinvariantfields}) of the hatted fields. 
We note that $\down{\partial}_1\circ \down{\partial}_0=0$ 
then follows by a non-trivial computation, using~(\ref{Nilpotentcyrel2}) and  the Noether identities (\ref{EXPlicitREALNoethers})  for  the  ${\cal Y}$. 
Finally, the differential in degree 2 is defined as  
 \be\label{downstairsdegreeTWO} 
	\begin{split}
	 \down{{\partial}}_{2}
	 (\down{\mathcal{N}})^M    = 
	 - \mathcal{Z}^{\dagger MN} \down{\mathcal{N}}_N\;. 
	   \end{split}
	\ee
Nilpotency of the differential, $0= (\down{\partial}_{2}\circ \down{\partial}_1)(\down{\cal E})$,  
quickly follows from (\ref{Nilpotentcyrel2}) and 
the constraints in~(\ref{genericdegreeonedudefjkf}).

\paragraph{Projection and inclusion} We now define the projection and inclusion maps between the upstairs and downstairs chain complexes. In degree zero 
these were already defined in (\ref{degreeZEROPROJ}) as follows:
 \be
   p_0(\Phi) = \widehat{\Phi} \;, \qquad \iota_0(\down{\Phi}) = 
   \down{\Phi}\;, 
 \ee
where $\widehat{\Phi}$ denotes the gauge invariant combinations 
(\ref{almostgaugeinv}) of the fields. 
Next, in degree $1$ we define 
\be\label{degreeONEprojector} 
 p_1({\cal E}) =   
 \begin{pmatrix} {\cal E}^{\mu\nu}   \\  \mathbb{P}^{MN}{}_{KL}{\cal E}^{KL}
  \\ \mathcal{P}_M{}^N {\cal E}^{\mu}{}_{N}  \\ 
  \underline{{\cal E}^{\mu\nu M}}_T  \end{pmatrix} \ \in \ \down{X}_1\;, 
 \ee
using the projectors (\ref{undertildeproj}) 
and (\ref{Piproj}), 
while the inclusion takes the following non-trivial form 
\be\label{degreeONEINclusion} 
  \iota_1(\down{\cal E}) = \iota_1 \begin{pmatrix} \down{\cal E}^{\mu\nu}   \\  \down{\cal E}_{MN} 
  \\ \down{\cal E}_{\mu}{}^{M}  \\ \down{\cal E}^{\mu\nu M}  \end{pmatrix} 
 =  \begin{pmatrix} \down{\cal E}^{\mu\nu}   \\ \down{\cal E}_{MN} +\Pi_{MN,P}(K_V)^P{}_Q \Big(\tfrac{2}{3} 
  \dd^Q {\cal E}^\mu{}_\mu
  + \nabla^\mu \down{\cal E}_{\mu}{}^{Q}
     -2 \dd^Q K\nabla^\mu \nabla^\nu \down{\cal E}_{\mu\nu }  \Big)
  \\  \down{\cal E}_{\mu}{}^{M} 
  -2{\cal Z}^{MK}(K_T)_{K}{}^{L} \nabla^{\nu}\down{\cal E}_{\nu\mu L} - 2 \dd^M K \nabla^\nu \down{\cal E}_{\mu\nu}\\ 
  \down{\cal E}^{\mu\nu M}  \;,  \end{pmatrix} , 
 \ee
which shortly we will see to be necessary in order 
to obey the chain map conditions. 
Next,  in degree $2$ we define projector and inclusion as  
 \be \label{eq:p2i2}
  p_2({\cal N}) = \begin{pmatrix} [{\cal N}^{\mu}] \\ 
  [{\cal N}_M-\dd_MK\nabla_{\mu}{\cal N}^{\mu} -{\cal Z}_{MN} (K_T)^{N}{}_{K}\nabla_{\mu} {\cal N}^{\mu K} ]_V 
  \end{pmatrix} \;, \qquad
  \iota_2\begin{pmatrix} \down{\cal N}^{\mu} \\ \down{\cal N}_M \end{pmatrix}
  =\begin{pmatrix} \down{\cal N}^{\mu} \\ 
  \down{\cal N}_M    \\ 0
  \\0 \end{pmatrix}\;, 
 \ee
where we see that now it is the projection that becomes non-trivial, thanks to the extra terms in the second line. Finally, in degree $3$ we define 
 \be \label{eq:p3i3}
 p_3({\cal R}) = 
 %[{\cal R}^{M} - 2\nabla_{\mu} {\cal %R}^{\mu M}]_V\;, \qquad 
 {\cal R}^{M} - 2\nabla_{\mu} {\cal R}^{\mu M} \;, \qquad 
 \iota_3(\down{\cal R}) = \begin{pmatrix} 
	\down{\cal R}^M\\[0.5ex]
		 0
         %\nabla^{\mu}[{\cal R}^M]_T
   \end{pmatrix} 	\;.  
\ee

The above form of the projection and inclusion maps are such that they define chain maps: they `commute' with the differentials. For instance, 
 for the inclusion maps we need:  
 \be
 \begin{split} 
  \partial_0(\iota_0(\down{\Phi}))
  &= \iota_1(\down{\partial}_0
  (\down{\Phi}))\;, \\
   \partial_1(\iota_1(\down{\cal E})) &= \iota_2(\down{\partial}_1(\down{\cal E})) \;, \\
  \partial_2(\iota_2(\down{\cal N}))&= \iota_3(\down{\partial}_2(\down{\cal N}))\;.  \\
 \end{split} 
 \ee
The left-hand side of the first degree-zero inclusion just 
encodes the original field equations evaluated on the constrained or downstairs fields, but on the right-hand side $\down{\partial}_0(\down{\Phi})$ involves the projections 
in (\ref{downstairsDIFFZERO}). This apparent discrepancy is fixed thanks to the non-trivial $\iota_1$ in 
(\ref{degreeONEINclusion}), 
using the Noether identities 
(\ref{EXPlicitREALNoethers}). 
Similarly, the other relations follow
by explicit computation.

Finally, for the projections we 
demand the chain map conditions 
  \be\label{pchainmap} 
  \begin{split} 
  \down{\partial}_0(p_0(\Phi)) 
  &= 
  p_1(\partial_0(\Phi))\;, \\
  \down{\partial}_1(p_1({\cal E})) &= 
  p_2(\partial_1({\cal E}))\;, \\
  \down{\partial}_2(p_2({\cal N}))
  &=p_3(\partial_2({\cal N}))\;. 
  \end{split} 
 \ee
For instance, in the first relation the right-hand side is given by 
 \be
  p_1(\partial_0(\Phi))
  =\begin{pmatrix} {\cal Y}^{\mu\nu}(\Phi)   \\  \mathbb{P}^{MN}{}_{KL}{\cal Y}^{KL}(\Phi)
  \\\mathcal{P}_M{}^N {\cal Y}^{\mu}{}_{N}(\Phi)  \\ 
  \underline{{\cal Y}^{\mu\nu M}}_T(\Phi)  \end{pmatrix}
  =\begin{pmatrix} {\cal Y}^{\mu\nu}(\widehat\Phi)   \\  \mathbb{P}^{MN}{}_{KL}{\cal Y}^{KL}(\widehat\Phi)
  \\ \mathcal{P}_M{}^N {\cal Y}^{\mu}{}_{N}(\widehat\Phi)  \\ 
  \underline{{\cal Y}^{\mu\nu M}}_T(\widehat\Phi)  \end{pmatrix}\;, 
 \ee
where we used (\ref{differentialZEROOO}) and (\ref{degreeONEprojector}). Moreover, in the second equality we used that the tensors ${\cal Y}$ are gauge invariant, so that the 
difference between $\widehat\Phi$ and $\Phi$, 
which is 
pure gauge by the definition (\ref{almostgaugeinv}) of 
the hatted fields, drops out. 
As $p_0(\Phi)=\widehat{\Phi}$ this equals 
the left-hand side $\down{\partial}_0(p_0(\Phi))$, 
c.f.~(\ref{downstairsDIFFZERO}).  
The other relations follow similarly by 
explicit computation, 
where for the third relation in (\ref{pchainmap}) one has to use that according to (\ref{TrivialVproj}) the $V$ zero mode projection of a $T$ non-zero mode acts trivially and can hence be deleted.

\paragraph{Homotopy maps} In order to complete the homotopy transfer we need the homotopy maps $\frak{h}:X_k\rightarrow X_{k-1}$ so that the homotopy relations are obeyed: 
 \be
  ({\rm id}-\iota p)_k 
  =\frak{h}_{k+1} \circ \partial_k
  +\partial_{k-1}
   \circ \frak{h}_{k}\;. 
 \ee 
Again by exlicit computation one finds 
that this holds for 
 \be \label{eq:h1h2h3}
 \begin{split} 
  \frak{h}_1&=0\;, \\[0.5ex]
   \frak{h}_2({\cal N}) 
  &= \begin{pmatrix} \frak{h}_2({\cal N})_{\mu\nu} \\ 
  \frak{h}_2({\cal N})_{MN} \\ 
  \frak{h}_2({\cal N})_{\mu}{}^{M} \\
  \frak{h}_2({\cal N})_{\mu\nu M} 
  \end{pmatrix}
  = \begin{pmatrix} 0 \\ 
  -\Pi_{MN,K}(K_V)^{K}{}_{L} 
  \big({\cal N}^{L} 
  - \dd^{L}K\nabla^{\mu}{\cal N}_{\mu}
  -{\cal Z}^{LP}(K_T)_{P}{}^{Q} {\cal Z}^{\dagger}_{QR}{\cal N}^{R} \big) \\ 
  \dd^MK{\cal N}_{\mu} +{\cal Z}^{MK} 
  (K_T)_K{}^{L} {\cal N}_{\mu L} \\
  {\cal N}_{\mu\nu M} 
  \end{pmatrix}, \\[0.5ex]
    \frak{h}_3 \begin{pmatrix} {\cal R}^M \\ 
  {\cal R}^{\mu M} \end{pmatrix} 
  &= \begin{pmatrix} \frak{h}_3({\cal R})^{\mu}\\ \frak{h}_3({\cal R})_{M}   \\
  \frak{h}_3({\cal R})^{\mu M}\\
  \frak{h}_3({\cal R})^{\mu\nu M}\end{pmatrix} 
   = \begin{pmatrix} 0\\ 0 \\
  2\,{\cal R}^{\mu M} \\
  0 \end{pmatrix} .
 \end{split}  
 \ee 
Note in particular that the homotopy relation in degree $0$ is still the same as in the pure gauge chain complex, thanks to $\frak{h}_1=0$. 
The homotopy relation can then only be non-trivially checked in degrees $1$ and $2$, as we do not know the differential in degree $3$.

\begin{figure}
    \centering
    \begin{tikzpicture}
       	\foreach \x in {-2,-1,...,3}
    		{\node (X\x) at (2.5*\x,0) {};
    		\draw (X\x) node [right] {$X_{\x}$};}
    
    	\draw (X-2) node [below right = 7pt and 3pt] {\small $\chi$};
    	\draw (X-1) node [below right = 7pt and 3pt] {\small $\Lambda$};
    	\draw (X0) node [below right = 7pt and 3pt] {\small $\Phi$};
    	\draw (X1) node [below right = 7pt and 3pt] {\small ${\cal E}$};
    	\draw (X2) node [below right = 7pt and 3pt] {\small ${\cal N}$};
    	\draw (X3) node [below right = 7pt and 3pt] {\small ${\cal R}$};
    
    	\foreach \x in {-2,-1,...,2}
    		{\draw [-stealth] ($(X\x)+(0.75,0.25)$) arc [start angle=150,end angle=30,x radius=1,y radius=0.35] node [midway, above] {$\partial_{\x}$};}
    
    	\foreach \x in {-1,0,...,3}
    		{\draw [-stealth] ($(X\x)+(-0.05,-0.25)$) arc [start angle=-30,end angle=-150,x radius=1,y radius=0.35] node [midway, below] {$\frak{h}_{\x}$};}
    
    	\foreach \x in {-2,-1,...,3}
    		{\node (oX\x) at (2.5*\x,-3) {};
    		\draw (oX\x) node [right] {$\mathring{X}_{\x}$};}
    
    	\draw (oX-2) node [below right = 7pt and 3pt] {\small $\mathring{\chi}$};
    	\draw (oX-1) node [below right = 7pt and 3pt] {\small $\mathring{\Lambda}$};
    	\draw (oX0) node [below right = 7pt and 3pt] {\small $\mathring{\Phi}$};
    	\draw (oX1) node [below right = 7pt and 3pt] {\small $\mathring{{\cal E}}$};
    	\draw (oX2) node [below right = 7pt and 3pt] {\small $\mathring{{\cal N}}$};
    	\draw (oX3) node [below right = 7pt and 3pt] {\small $\mathring{{\cal R}}$};
    
    	\foreach \x in {-2,-1,...,2}
    		{\draw [-stealth] ($(oX\x)+(0.75,0.05)$) -- +(1.7,0) node [midway, above] {$\mathring{\partial}_{\x}$};}
    
    	\foreach \x in {-2,-1,...,3}
    		{\draw [-stealth] ($(X\x)+(0.15,-1)$) --  ($(oX\x)+(0.15,0.75)$) node [midway, left] {$p_{\x}$};
    		 \draw [stealth-] ($(X\x)+(0.5,-1)$) --  ($(oX\x)+(0.5,0.75)$) node [midway, right] {$\iota_{\x}$};}
    \end{tikzpicture}
    \caption{Homotopy transfer. The definitions of the different objects are given in table~\ref{tab:homotopytransfer}.}
    \label{fig:completecomplex}
\end{figure}
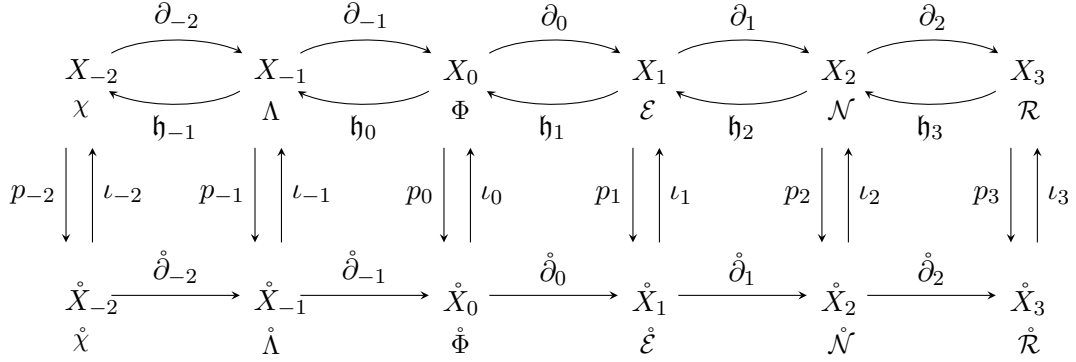

\medskip 
As a help for the reader, we have summarized all the homotopy transfer data in figure~\ref{fig:completecomplex} and table~\ref{tab:homotopytransfer}.

\begin{table}
    \centering
    \begin{tabular}{cccccc|ccccc}
        $X_{-2}$ & $X_{-1}$ & $X_{0}$ & $X_{1}$ & $X_{2}$ & $X_{3}$ & $\partial_{-2}$ & $\partial_{-1}$ & $\partial_{0}$ & $\partial_{1}$ & $\partial_{2}$\\
        \eqref{eq:X-2} & \eqref{eq:X-1} & \eqref{eq:X0}& \eqref{genericdegreeonedude} & \eqref{eq:X2X3} & \eqref{eq:X2X3} & \eqref{generalizedDifferentials} & \eqref{generalizedDifferentials} & \eqref{differentialZEROOO} & \eqref{deg1diff} & \eqref{deg2diff} \\ \hline
        $\mathring{X}_{-2}$ & $\mathring{X}_{-1}$ & $\mathring{X}_{0}$ & $\mathring{X}_{1}$ & $\mathring{X}_{2}$ & $\mathring{X}_{3}$ & $\mathring{\partial}_{-2}$ & $\mathring{\partial}_{-1}$ & $\mathring{\partial}_{0}$ & $\mathring{\partial}_{1}$ & $\mathring{\partial}_{2}$\\
        \eqref{eq:oX-2} & \eqref{eq:oX-1} & \eqref{eq:oX0} & \eqref{genericdegreeonedudefjkf}  & \eqref{eq:oX2} & \eqref{eq:oX3} & \eqref{donwstairstrivialgauge} & \eqref{eq:od-1} & \eqref{downstairsDIFFZERO} & \eqref{eq:op1} & \eqref{downstairsdegreeTWO} \\ \hline
        $p_{-2}$ & $p_{-1}$ & $p_{0}$ & $p_{1}$ & $p_{2}$ & $p_{3}$ & $\frak{h}_{-1}$ & $\frak{h}_{0}$ & $\frak{h}_{1}$ & $\frak{h}_{2}$ & $\frak{h}_{3}$ \\
        \eqref{eq:p-2i-2} & \eqref{degreezeroprojectionandinclusion} & \eqref{degreeZEROPROJ} & \eqref{degreeONEprojector} & \eqref{eq:p2i2} & \eqref{eq:p3i3} & \eqref{eq:h-1} & \eqref{homotopyh0one}  & \eqref{eq:h1h2h3} & \eqref{eq:h1h2h3} & \eqref{eq:h1h2h3} \\ \hline
        $\iota_{-2}$ & $\iota_{-1}$ & $\iota_{0}$ & $\iota_{1}$ & $\iota_{2}$ & $\iota_{3}$ & & & & & \\
        \eqref{eq:p-2i-2} & \eqref{degreezeroprojectionandinclusion} & \eqref{degreeZEROPROJ} & \eqref{degreeONEINclusion} & \eqref{eq:p2i2} & \eqref{eq:p3i3} & & & & &
    \end{tabular}
    \caption{Summary of all quantities defined in the homotopy transfer.}
    \label{tab:homotopytransfer}
\end{table}

\subsection{Kaluza-Klein mass spectrum} 
\label{subsec:KKmass}

We now analyze the equations of motion encoded in (\ref{downstairsDIFFZERO}) in order to determine the Kaluza-Klein spectrum and to derive general expressions for the mass matrices. \medskip

\paragraph{Scalar metric modes}
The simplest case are the scalar metric modes. Passing to gauge invariant variables we first replace each field by its hatted version. Then we act with $\mathbb{P}^{MN}{}_{KL}$ on ${\cal Y}_{KL}=0$, as required by the second entry of (\ref{downstairsDIFFZERO}), which eliminates all $\dd$ exact and $\Pi$ exact terms.  
The final equations read  
 \be 
  0 = 
  \nabla_\mu \nabla^\mu \widehat{m}_{MN}
    -  ({\cal M}_{\rm sm}){}_{MN}{}^{PQ}\,\widehat{m}_{PQ}\;, 
 \ee
where 
\be \label{eq:massmatrixscalarproj}
\begin{split}
    ({\cal M}_{\rm sm}){}_{MN}{}^{KL} &= \mathbb{P}_{MN}{}^{KL} M_{(0)}^{2}{}_{KL}{}^{PQ}\mathbb{P}_{PQ}{}^{KL}\;.
\end{split}
\ee

\paragraph{Vector modes}
The vector equations ${\cal Y}_M{}^{\mu}=0$, given in~\eqref{eq:eomvecAdS},
read explicitly:
 \be\label{recallVECTOReq}  
  0 =   
    \bg{M}_{MN}\, \nabla_{\nu}  f^{\mu\nu\,N} +  \frac{1}{6} \,X_{MP}{}^{K} j^{\mu \, P}{}_{K} + \dd_{N}   j^{\mu \, N}{}_{M}  
    + \dd_{M} \Big(  \nabla_\nu   h^{\nu\mu} - \nabla^{\mu}  h_{\nu}{}^\nu +\tfrac83\,   \dd_{N}  {a}^{\mu}{}^{N} \Big). 
 \ee
We rewrite the terms involving $j^{\mu}$,  which  takes values in the Lie algebra of ${\rm E}_{6(6)}$, so that we can write in terms of the projector onto the adjoint:
\be 
\begin{split} 
 \Pi^{\dagger M,NK} j_{\mu NK} &= 
 -\bg{M}^{ML}\bg{M}^{P(N} \Big(2\,X_{LP}{}^{K)}
 +12\,\mathbb{P}{}_{P}{}^{K)}{}_{L}{}^{Q}\dd_Q\Big) j_{\mu NK} \\
 &= -2\,\bg{M}^{ML} X_{LP}{}^{K} j_{\mu}{}^{P}{}_{K} 
 -12\,\bg{M}^{ML} \mathbb{P}_{P}{}^{K}{}_{L}{}^{Q}\dd_Q 
 j_{\mu}{}^{P}{}_{K} \\
 &= -12\,\bg{M}^{ML}\big(\dd_Q j_{\mu}{}^{Q}{}_{L} 
 +\tfrac{1}{6} X_{LP}{}^{K}  \,j_{\mu}{}^{P}{}_{K}\big) \;, 
\end{split} 
\ee 
where we used from the first to the second line that 
$j_{\mu MN}=\partial_{\mu}m_{MN} + \Pi_{MN,K} a_{\mu}{}^{K}$ 
is symmetric in $MN$ (in addition to being Lie algebra valued once one index is raised by means of $\bg{M}^{MN}$). 
Using this in (\ref{recallVECTOReq}) we have equivalently 
 \be
  0= \nabla_{\nu}  f^{\mu\nu\,M} 
  -\frac{1}{12} \Pi^{\dagger \,M,KL} j^{\mu}{}_{KL} 
  + \dd^{M} \Big(  \nabla_\nu   h^{\nu\mu} - \nabla^{\mu}  h_{\nu}{}^\nu +\tfrac83\,   \dd_{N}  {a}^{\mu}{}^{N} \Big). 
 \ee 
Next, as a consequence of  gauge invariance, 
we can replace each field by its (almost gauge invariant) hatted version, which 
satisfy the constraints of the downstairs complex, 
so that the term $\Pi^{\dagger}\widehat{m}$ drops out. 
Further, we can project the equations by ${\cal P}_{M}{}^{N}$, 
after which the terms under the total $\dd^M$ derivative, 
as well as the term ${\cal Z}^{MN} b_{\mu\nu\,N}$ inside 
$f_{\mu\nu}{}^M$, 
drop out, leading to 
 \be 
 0= {\nabla}_{\nu} \big(\nabla^{\mu} \widehat{a}^{\nu\, M}
 -\nabla^{\nu} \widehat{a}^{\mu\, M}\big)
  -\frac{1}{12}{\cal P}_{P}{}^{M} \Pi^{\dagger \,P,KL}\Pi_{KL,N} \widehat{a}^{\mu\, N}\;. 
 \ee   
 This equation is equivalent to 
 \be
  0={\nabla}_{\nu} \big(\nabla^{\nu} \widehat{a}^{\mu\, M}
 -\nabla^{\mu} \widehat{a}^{\nu\, N}\big) 
 -({\cal M}_{V})^M{}_{N}\, \widehat{a}^{\mu \,N}\;,   
 \ee 
where 
 \be\label{VectorMass} 
 ({\cal M}_{V})^M{}_{N} \ = \  
  - \frac{1}{12}\,  {\mathcal{P}^M}{}_K (\Delta_V)^K{}_L 
  {\mathcal P}^L{}_{N}\;,  
 \ee
and the second projector can be inserted for free as the field on which ${\cal M}_V$ acts is constrained. 
Acting on the above simplified equations of motion with $\nabla_\mu$ we infer the subsidiary condition  $\nabla_{\mu}\widehat{a}^{\mu\, M}=0$
for the non-zero eigenvectors  of ${\cal M}_V$. 
Using this constraint inside the equations of motion and commuting covariant derivatives one obtains 
 \be 
  \Big(\square-\frac{1}{3} V\Big) \widehat{a}_{\mu}{}^{M} 
  - ({\cal M}_V)^M{}_{N}\, \widehat{a}_{\mu}{}^{N}
  =0\;, 
 \ee
where again this only holds for the non-zero eigenvectors of ${\cal M}_V$. 
Here we used the background field equations $R_{\mu\nu} = \frac{1}{2}(R-V)g_{\mu\nu}$, 
which imply
 \be
  R_{\mu\nu}= \frac{1}{3}Vg_{\mu\nu}\;, \quad
  R=\frac{5}{3} V\;, 
 \ee
and we recall that $V$ is constant. (We recall that this does {not} mean that the external spacetime must be AdS$_5$, for the Riemann tensor need not be of the maximally symmetric form. This only means that the background geometry is Einstein.) 
These are the expected Klein-Gordon type equations for a 
massive vector on an Einstein manifold, with mass (\ref{VectorMass}).

The following comment is in order: due to the explicit projectors 
in the definition of the mass matrix (\ref{VectorMass}) any vectors of 
the form 
 \be\label{trivialzeroEigenvectors} 
  u^M = \dd^M\beta + {\cal Z}^{MN} \gamma_N\;, 
 \ee 
for arbitrary $\beta$ and $\gamma$,  are eigenvectors with eigenvalue zero. These should \textit{not} be interpreted as massless vector 
fields. Indeed, the almost gauge invariant vectors are already 
constrained to be divergence-free w.r.t.~to $\dd_M$
and ${\cal Z}^{\dagger}_{MN}$, hence precisely those zero eigenvalues 
must be discarded from the spectrum whose eigenvectors are of the 
form (\ref{trivialzeroEigenvectors}). 
Put differently, only those eigenvalue directions obeying the proper divergence constraints contribute to the physical mass spectrum. 
We have verified with Mathematica at low Kaluza-Klein levels for AdS$_5\times S^5$ 
that \eqref{eq:massmatrixscalarproj} and (\ref{VectorMass}) then lead precisely to the 
correct spectra for the Kaluza-Klein scalar and vector modes.

%\medskip
\paragraph{Tensor modes}
The tensor equations ${\cal Y}^{\mu\nu M}=0$, when restricted to $T$ non-zero modes, are equivalent to 
 \be\label{TensorEQQQ}  
  \frac{\sqrt{10}}{2}\,\varepsilon^{\mu\nu\rho\sigma\tau} 
  \partial_{\rho} \widehat{b}_{\sigma\tau\, M} + 2\,\nabla^{[\mu} \widehat{a}^{\nu]}{}_{M} 
  -{\cal Z}_{MN} \widehat{b}^{\mu\nu \,N } = 0 \;, 
 \ee 
where again due to gauge invariance we can assume all fields 
to be hatted, satisfying the constraints of  the downstairs 
chain complex, even though for convenience we drop the hats. 
Acting on this equation with $\varepsilon_{\mu\nu\lambda\kappa\zeta}\nabla^{\zeta}$ yields 
 \be  
  -\sqrt{10}\, 3! \, \nabla^{\zeta} \nabla_{[\lambda}\widehat{b}_{\kappa\zeta]\,M}
  -\varepsilon_{\mu\nu\lambda\kappa\zeta}{\cal Z}_{MN} \nabla^{\zeta}\widehat{b}^{\mu\nu\,N} = 0 \;. 
 \ee
Using again (\ref{TensorEQQQ}) in the second term we 
can rewrite this as 
 \be\label{ProfTensorEQ}  
  \nabla^{\mu}\big(\nabla_{\mu} \widehat{b}_{\nu\rho\,M} 
  +\nabla_{\nu} \widehat{b}_{\rho\mu \,M}+\nabla_{\rho} \widehat{b}_{\mu\nu\,M}\big)
  +\frac{1}{10} ({\cal Z}^{\dagger}{\cal Z})_M{}^{N} \widehat{b}_{\nu\rho\,N}
  = 0 \;, 
 \ee 
where we used 
${\cal Z}_{MN}\widehat{a}_{\mu}{}^{N}=0$.  
Acting on this equation with $\nabla^{\nu}$ gives $\nabla^{\mu}\widehat{b}_{\mu\nu\,M}=0$ (recalling that we consider only $T$ non-zero modes). 
Using this constraint then in (\ref{ProfTensorEQ}) and commuting covariant derivatives yields 
 \be  
  \big(\square -\tfrac{2}{3}V\big) \widehat{b}_{\mu\nu\,M} 
  +2 \,R_{\mu\rho\nu\sigma} \widehat{b}^{\rho\sigma}{}_{M}  
  - ({\cal M}_T)_M{}^{N} \widehat{b}_{\mu\nu\,N} =0\;, 
 \ee 
where 
 \be
  ({\cal M}_T)_M{}^{N} = 
  - \frac{1}{10} ({\cal Z}^{\dagger}{\cal Z})_M{}^{N}\,. 
  \label{eq:MassTensorAdS}
 \ee

\medskip
\paragraph{Spin-2 modes}
For the spin-2 equations 
(\ref{fgkgbkg2}) we pass again to the gauge invariant variables and use the subsidiary conditions estbalished above. Defining  
 \be\label{DEFSigma}  
  \Sigma_{\mu\nu}:=-\frac{1}{2}\nabla^{\rho}
  \nabla_{\rho}\widehat{h}_{\mu\nu} 
  +\nabla^{\rho}\nabla_{(\mu} \widehat{h}_{\nu)\rho}
  -\frac{1}{2}\nabla_{\mu}\nabla_{\nu} \widehat{h}\;, 
 \ee
theses  spin-2 equations  can then be written as:
 \be\label{simplifiedSPin2} 
  0= \Sigma_{\mu\nu} - \frac{1}{2}\bg{g}_{\mu\nu} \Sigma^{\rho}{}_{\rho}  
  +\frac{1}{2} \bg{g}_{\mu\nu} \,
  R^{\rho\sigma} \widehat{h}_{\rho\sigma} 
  -\frac{1}{2} (R-V) \widehat{h}_{\mu\nu} 
  + \frac{1}{2}\,\bg{g}_{\mu \nu}\, \Delta \widehat{h}^{\rho}{}_{\rho} -\frac{1}{2}\,\Delta \widehat{h}_{\mu \nu} \;.
 \ee
Taking the covariant divergence of the first two terms (which gives zero on flat space) one finds: 
 \be\label{intermediateDIV} 
  \nabla^{\mu}\Sigma_{\mu\nu} - \frac{1}{2} 
  \nabla_{\nu} \Sigma^{\rho}{}_{\rho} 
  = R_{\nu}{}^{\mu}(\nabla^{\rho}\widehat{h}_{\rho\mu} 
  -\frac{1}{2}\nabla_{\mu}\widehat{h}) +\frac{1}{2}\nabla_{\nu}R^{\rho\sigma} \,
  \widehat{h}_{\rho\sigma} \;.
 \ee
This can be simplified by using background field equations, 
which we recall imply $R_{\mu\nu}= \frac{1}{3}Vg_{\mu\nu}$. 
The divergence (\ref{intermediateDIV}) then reads 
  \be\label{intermediateDIV344} 
  \nabla^{\mu}\Sigma_{\mu\nu} - \frac{1}{2} 
  \nabla_{\nu} \Sigma^{\rho}{}_{\rho} 
  = \frac{1}{3}V\Big(\nabla^{\mu}\widehat{h}_{\mu\nu} 
  -\frac{1}{2}\nabla_{\nu}\widehat{h}\Big) \;, 
 \ee
while the spin-2 equations (\ref{simplifiedSPin2}) simplify to 
 \be\label{simplifiedSPin26456} 
  0= \Sigma_{\mu\nu} - \frac{1}{2}\bg{g}_{\mu\nu} \Sigma^{\rho}{}_{\rho}  
  -\frac{1}{3}V\Big(\widehat{h}_{\mu\nu} - \frac{1}{2}g_{\mu\nu} \widehat{h}\Big)
  -\frac{1}{2}\Delta\big(\widehat{h}_{\mu\nu}-g_{\mu\nu} \widehat{h}\big)\;. 
 \ee
Taking the covariant divergence of this equation and using 
(\ref{intermediateDIV344}) we then infer: 
 \be
  0= \frac{1}{2} \Delta(\nabla_{\nu}\widehat{h} 
  -\nabla^{\mu}\widehat{h}_{\mu\nu})\;, 
 \ee
and hence 
 \be\label{intermedsecond} 
\nabla^{\mu}\underline{\widehat{h}}_{\mu\nu} - \nabla_{\nu}\widehat{\underline{h}}=0 
 \ee 
for non-zero modes. This implies that the non-zero mode projection of the trace of (\ref{DEFSigma}) vanishes: 
$\underline{\Sigma}^{\mu}{}_{\mu}=0$. 
Taking then the trace of (\ref{simplifiedSPin26456}) and projecting to the non-zero modes 
yields 
 \be
  \Big(\Delta + \frac{1}{4}V\Big)\widehat{\underline{h}}=0\;. 
 \ee
Generically the operator appearing here  is invertible, hence 
implying 
 \be
  \widehat{\underline{h}}=0\;, 
 \ee
although it might be worth investigating whether there are degenerate points where this operator is non-invertible, which would be related to the so-called partially massless case \cite{Deser:1983mm}. 
Together with~(\ref{intermedsecond}) we thus have 
 \be\label{FinalSUBSID} 
  \nabla^{\mu}\widehat{\underline{h}}_{\mu\nu}=0\;, \qquad
   \widehat{\underline{h}}=0\;, 
 \ee
which are the desired subsidiary conditions. 
Using this in (\ref{DEFSigma}) and commuting covariant derivatives yields 
 \be
  \Sigma_{\mu\nu} = -\frac{1}{2}\nabla^2 
  \widehat{\underline{h}}_{\mu\nu} -R_{\mu\rho\nu\sigma}\widehat{\underline{h}}^{\rho\sigma} 
  +\frac{1}{3}V \widehat{\underline{h}}_{\mu\nu} \;, 
 \ee
which back in the spin-2 equations (\ref{simplifiedSPin26456})  finally gives 
 \be
  (\nabla^2+\Delta) \widehat{\underline{h}}_{\mu\nu} + 2R_{\mu\rho\nu\sigma}\widehat{\underline{h}}^{\rho\sigma} =0\;. 
 \ee
Together with (\ref{FinalSUBSID}) these are precisely the desired massive spin-2 equations on 
a general Einstein background.\footnote{The Riemann tensor term is part of the proper spin-2 Lichnerowicz wave operator as only this combination preserves the 
constraint $ \nabla^{\mu}\underline{h}_{\mu\nu}=0$. } 

%%%%%%%%%%%%%%%%%%%%%%%%%%%%%%%%%%%%%%%
%%%%%%%%%%%%%%%%%%%%%%%%%%%%%%%%%%%%%%%

\section{Mass spectra around AdS$_5$ Black Hole Background} \label{sec:BHbackground}
In this section, we apply the formalism detailed in sections~\ref{sec:ExFT5dsugra} and~\ref{sec:pertubtheory} in order to compute the mass spectrum of fluctuations around AdS$_5$ black hole backgrounds. In contrast to the backgrounds discussed in the previous section, this geometry reduces the AdS$_5\times S^5$ isometry group to a smaller subgroup, allowing for a richer pattern of couplings and symmetry breaking. Specifically, we investigate the near-horizon limit of extremal Kerr-Newman-AdS$_5$ black holes as solutions of IIB supergravity, in which the geometry becomes AdS$_2\times S^3$.
 This extends the analysis of \cite{Ezroura:2024xba} to include the fluctuations corresponding to the full ten-dimensional Kaluza-Klein towers. After reviewing the black hole background in section \ref{NHKNAdSBH}, we discuss in section~\ref{subsec:fluctuation} the stability criteria for fluctuations around the near horizon geometry. We specify the equations of motion around this background in section~\ref{FluEq}. Then, we analyze the linearized fluctuation equations for the various fields. For simplicity, in this paper we restrict to what we call `simple' fields which do not interfere with fields of different $D=5$ spin, even though the near-horizon background generically allows for such couplings at the linearized level. We discuss successively simple scalars, simple tensors, and simple vectors. For each class of fields, we analyze the $D=5$ fluctuation equations from the perspective of the AdS$_2$ factor within the near-horizon geometry, in order to analyze the stability properties of the various modes.
In section~\ref{BFStudy} we summarize and illustrate our findings. In particular, we identify the unstable and stable regions in the two-dimensional parameter space of backgrounds. This also allows a useful comparison with the subsector of modes investigated in \cite{Ezroura:2024xba}.

\subsection{Near-Horizon limit of extremal Kerr-Newman-AdS black holes}
\label{NHKNAdSBH}

We consider the near-horizon geometry of the extremal Kerr-Newman-AdS$_5$ black holes. These are part of the family of black hole solutions constructed in \cite{Chong:2005hr} in minimal $D=5$ supergravity, and, following \cite{Ezroura:2024xba}, we describe their embedding into the maximal $D=5$ supergravity. Via the explicit embedding formulae of \cite{Cvetic:2000nc,Baguet:2015sma}, they may further be uplifted to solutions of IIB supergravity.
Their near-horizon geometry is the fibered product of an AdS$_2$ with a squashed three-sphere $S^3$, preserving an ${\rm SO}(1,2)\times {\rm U}(2)$ group of the AdS$_5$ isometries. The general geometry with these isometries is of the form
\begin{equation}
    {\rm d}s^2 = -A^2(r-r_+)^2{\rm d}t^2+\dfrac{B^2}{(r-r_+)^2}{\rm d}r^2+\dfrac{C^2}{4}(\sigma_1^2+\sigma_2^2) + \dfrac{D^2}{4}\Big(\sigma_3 + E(r-r_+)\,{\rm d}t \Big)^2\, , 
    \label{AdS2NH}
\end{equation}
with constants $A$, $B$, $C$, $D$ and $E$, and the left-invariant forms $\sigma_i$ on $S^3$,
while the near-horizon gauge field takes the general form
\begin{equation}
    \mathcal{A} = -\Big(F(r-r_+){\rm d}t + G \sigma_3\Big)
    \,,
    \label{eq:NHvector}
\end{equation}
with constants $F$, $G$, and field strength  $\mathcal{F} = -F {\rm d}r\wedge {\rm d}t -G\,\sigma_1 \wedge\sigma_2$. 
With the ansatz (\ref{AdS2NH}) and~(\ref{eq:NHvector}), the field equations of five-dimensional minimal supergravity,
\begin{align}
\nabla_\nu {\cal F}^{\mu\nu} \ = \ & \frac14\,\epsilon^{\mu\kappa\lambda\rho\sigma}\,{\cal F}_{\kappa\lambda} {\cal F}_{\rho\sigma}
\,,
\nonumber\\
R_{\mu\nu} \ = \ &
\frac32\,{\cal F}_{\mu}{}^\rho {\cal F}_{\nu\rho} - \frac14\,g_{\mu\nu}\,{\cal F}_{\rho\sigma}{\cal F}^{\rho\sigma} -4\,g_{\mu\nu}
\,,
\label{eq:eom5Dmin}
\end{align}
translate into algebraic relations among the constants $A$, \dots, $G$, which can 
be solved in terms of two parameters, denoted as $a$ and $r_+$ in \cite{Ezroura:2024xba}:
\begin{equation} \label{NHlimitcoeff}
	\begin{aligned}
    A = & \dfrac{2\sqrt{(r_+^2+a^2)(3r_+^2+2a^2+1)}}{r_+^2+2a^2+a^2\sqrt{2r_+^2+2a^2+1}}\,, \quad
    B =  \sqrt{\dfrac{r_+^2+a^2}{4(3r_+^2+2a^2+1)}} \,, \quad
    C =  \sqrt{\dfrac{r_+^2+a^2}{1-a}} \,,\\
    D = & \dfrac{r_+}{(1-a^2)(r_+^2+a^2)}\Big( r_+^2+2a^2+a^2\sqrt{2r_+^2+2a^2+1}\Big)\,, \\
    E = & \dfrac{4a(1-a^2)}{r_+}\dfrac{r_+^2+(2r_+^2+a^2)\sqrt{2r_+^2+2a^2+1}}{(r_+^2+2a^2+a^2\sqrt{2r_+^2+2a^2+1})^2}\,,\\
    F = & \dfrac{2r_+(r_+^2\sqrt{2r_+^2+2a^2+1}-a^2)}{(r_+^2+a^2)(r_+^2+2a^2+a^2\sqrt{2r_+^2+2a^2+1})}\,,\\
    G = & \dfrac{a}{2(1-a^2)(r_+^2+a^2)}(r_+^2 \sqrt{2r_+^2+2a^2+1}-a^2)\,. 
	\end{aligned}
\end{equation}
The geometry then reproduces the
near-horizon limit deduced from the extremal Kerr-Newman AdS case. The physical parameter range is
\begin{equation} \label{eq:BHparam}
    0\le a\le 1\,,\qquad
    r_+^2 \ge \frac14\,\big(\sqrt{1+8a^2}-1\big)\,.
\end{equation}
Supersymmetric solutions correspond to the one-parameter family of solutions defined by
\begin{equation}
r_+^2=a\,(2+a)\,.
\label{eq:NHsusy}
\end{equation}
Let us also note, that the field equations (\ref{eq:eom5Dmin}) support a pure AdS$_5$ solution with $F_{\mu\nu}=0$ and
\begin{equation}
R = -20\,,\quad \Longrightarrow\quad
\ell_{{\rm AdS}_5} = 1
\,,
\label{eq:lAdS5}
\end{equation}
which upon embedding into the maximal $D=5$ theory and subsequent uplift to IIB yields the maximally supersymmetric AdS$_5\times S^5$ solution.

As pointed out in \cite{Ezroura:2024xba}, minimal $D=5$ supergravity can be embedded as a consistent truncation into maximal gauged $D=5$ supergravity with gauge group ${\rm SO}(6)$. I.e.\ the field equations (\ref{eq:eom5Dmin}) are obtained from the maximal $D=5$ field equations summarized in section \ref{sec:5dsugra} above, by breaking the gauge group 
\begin{equation}
{\rm SO}(6) \longrightarrow {\rm U}(3) = {\rm SU}(3) \times {\rm U}(1) \;,\qquad
{\bf 6} \longrightarrow [1,0]^+\oplus[0,1]^-
\,,
\label{eq:breakSO6}
\end{equation}
where we use Dynkin labels to denote the ${\rm SU}(3)$ representations, and superscripts denote the ${\rm U}(1)$ charges.
Truncating the maximal ${\rm SO}(6)$ supergravity to its ${\rm U}(3) $ singlets then yields minimal $D=5$ supergravity with a single one of the 15 physical vector fields surviving according to the breaking
\begin{equation}
{\bf 15} \longrightarrow 
[0,0]^0 \oplus 
[1,0]^{-2}\oplus[0,1]^{+2} \oplus [1.1]^0
\,.
\label{eq:break15}
\end{equation}

In order to embed this background into the ${\rm E}_{6(6)}$ covariant notation of sections~\ref{sec:5dsugra} and~\ref{sec:pertubtheory}, we set all scalar fields to zero, 
\begin{equation}
M_{MN}=\delta_{MN}\,,
\end{equation}
and explicitly embed the background vector (\ref{eq:NHvector}) as
\begin{equation}
    {A}_{\mu}{}^{M} = C^M {\cal A}_{\mu}\,,\qquad
    F_{\mu\nu}{}^M = C^M {\cal F}_{\mu\nu}
    \,,
\end{equation}
into the 27 vector fields ${A}_{\mu}{}^{M}$. The constant vector $C^M$ selects the ${\rm U}(3) $ singlet from (\ref{eq:break15}) and defines the breaking
\begin{equation}
{\rm E}_{6(6)} \longrightarrow {\rm SO}(6) \longrightarrow {\rm U}(3)
\,.
\end{equation}
We choose this vector to be normalized as
\begin{equation}
C^M {\delta}_{MN}\,C^N = 3\,,\qquad
C^KC^MC^N\,d_{KMN} = \frac{6}{\sqrt{10}}\,,
\end{equation}
such that the background field equations~(\ref{VectorBkg}) and~(\ref{MetricBkg}) reduce to the equations of the minimal theory (\ref{eq:eom5Dmin}).
The scalar and two-form background field equations~(\ref{TensorBkg}) and~(\ref{eq:fieldeqbgscal}) are trivially satisfied, 
as there are no ${\rm U}(3)$ singlets in these sectors.
With these conventions, the background covariant derivatives in the $D=5$ theory (\ref{eq:covD}) take the form
\begin{equation}
{D}_\mu {\lambda}^{{M}} = 
\nabla_\mu {\lambda}^{{M}}
-{\cal A}_\mu{}
 \left(C^{{P}} {X}_{{P}{N}}{}^{{M}}\,    \lambda^{{N}}
+
C^{{P}}  \dd_{{P}}\,      \lambda^{{M}} \right) 
= \partial_\mu {\lambda}^{{M}}
- i\,Q{\cal A}_\mu{}
\,{\lambda}^{{M}}
\,,
\label{eq:covDQ}
\end{equation}
where the $Q$'s are integer valued charges. 

Let us also note that the matrix $C^M d_{MKL}$ has eigenvalues
\begin{equation}
    \left\{+\frac{2}{\sqrt{10}}\,,\;\;
    +\frac1{\sqrt{10}}\,,\;\;
    -\frac1{\sqrt{10}}\,
    \right\},
\end{equation}
with multiplicity 1, 12, and 14, respectively.

\subsection{Near-horizon fluctuation equations and stability}
\label{subsec:fluctuation} 
 
In the near-horizon region of the extremal Kerr-Newman-AdS$_5$ black holes,
all fluctuation equations eventually reduce to the Klein-Gordon equation of a massive charged
scalar field that propagates in AdS$_2$ spacetime. 
The two-dimensional line element (in a co-rotating frame) from (\ref{AdS2NH}) is given by
\begin{equation}
    {\rm d}s_2^2 = -A^2(r-r_+)^2 {\rm d} t^2 +\dfrac{B^2}{(r-r_+)^2}{\rm d}r^2\;. 
    \label{eq:AdS2}
\end{equation}
In Poincaré coordinates, $z= \dfrac{1}{r-r_+}$, $\tau = \dfrac{B}{A} t$, this metric reads
\begin{equation}
    {\rm d}s_2^2 = \dfrac{\ell_2^2}{z^2}\,({\rm d}z^2-{\rm d}\tau^2)\, , 
\end{equation}
with the AdS$_2$ length $\ell_{2}=B$. The AdS$_2$ geometry admits a vector field 
\begin{equation}
    {A}_2= -AB(r-r_{+})\,{\rm d}t\,, 
\end{equation}
compatible with the AdS$_2$ isometries, as its field strength reproduces the AdS$_2$ volume form
\begin{equation}
    {F}_2 = {\rm vol}\big({\rm AdS}_{2}\big)\,.
\end{equation}
A free scalar field in the geometry (\ref{eq:AdS2}) is thus characterized by two parameters: its mass $m$ and a U(1) charge $q$,
with field equation
\begin{equation}
\big(\nabla^u - i\,q\,A_2^u \big)
\big(\partial_u - i\,q\,A_{2,u}\big)\,\phi_{(m,q)} 
= m^2\,\phi_{(m,q)}
\,,
\qquad
u=0,1\,.
\label{eq:KG2D}
\end{equation}
For vanishing charge $q=0$, the stability condition of (\ref{eq:KG2D}) is given by the standard two-dimensional Breitenlohner-Freedman (BF) bound \cite{Breitenlohner:1982jf}
\begin{equation}
1+4\,m^2 \ell_2^2 \,\ge\, 0
\,.
\end{equation}
For non-vanishing charge, the stability condition is modified to
\begin{equation}
1+4\,m^2 \ell_2^2- 4\,q^2 \ell_2^4  \,\ge\, 0
\,.
\label{StabCond}
\end{equation}
Indeed, the solution of (\ref{eq:KG2D}) for $\phi_{(m,q)}=\varphi(z)e^{i\omega\tau}$ is given by 
Whittaker functions \cite{Faulkner:2011tm}
\begin{equation}
    \varphi(z) = a_{1}\,W_{iq {A}_{2,\tau},\nu}(-2i\omega z) + a_{2}\,W_{-iq{A}_{2,\tau},\nu}(2i\omega z), \quad {\rm with} \quad \nu=\sqrt{(m\ell_{2})^{2}-q^{2}{A}_{2,\tau}^{2}+\frac{1}{4}}\,,
\end{equation}
and dual to a ${\rm CFT}_{1}$ operator of conformal dimension \cite{Faulkner:2009wj}
\begin{equation}
    \Delta = \frac{1}{2} + \sqrt{(m\ell_{2})^{2}-q^{2}{A}_{2,\tau}^{2}+\frac{1}{4}} \;.
\end{equation}

In the following, we will bring all fluctuation equations into the form (\ref{eq:KG2D}) 
and test stability of the modes according to (\ref{StabCond}).

\subsection{Fluctuation equations}
\label{FluEq}

Let us now turn to the fluctuation equations around this background. We have worked out these equations in section~\ref{subsec:fluctuationEqs} for a general $D=5$ background. Here, we will specify them for the near-horizon geometry (\ref{AdS2NH})--(\ref{NHlimitcoeff}). We then diagonalize the resulting system of equations until each mode satisfies an equation of the type (\ref{eq:KG2D}) whose stability we can then check with (\ref{StabCond}).
As this section is meant as an illustration of the general method rather than a complete analysis, we will make several truncations to the spectrum of fluctuations we consider. A full analysis will be presented elsewhere.

First of all, as a consequence of the smaller isometry group of this background, the $D=5$ fluctuation equations in general mix fields in different $D=5$ Lorentz representations. E.g.\ the fluctuation equation (\ref{eq:FlucTensors}) mixes fluctuations of $D=5$ tensors, vectors, scalars and spin-2. To simplify the analysis, we will restrict to fluctuations that do not couple to fields of other $D=5$ spin. We identify these `simple' fields as fluctuations in representations of the U(3) symmetry group~(\ref{eq:breakSO6}) that only show up once in the full spectrum. Because no other field transforms in the same representation, they have no possible mixing partner, and their field equations are therefore already diagonal. As a second simplification, and since we analyze stability of the fluctuations w.r.t.\ the AdS$_2$ factor in the geometry, we will further truncate the spectrum of fluctuations to singlets under the U(2) isometry group of the five-dimensional metric (\ref{AdS2NH}). In practice, this implies that all fluctuations are parametrized by functions depending only on the AdS$_2$ coordinates $(t,r)$.

To summarize the symmetries of the spectrum: all fluctuations fall into representations of the group
\begin{equation}
\underbrace{{\rm SO}(2,1) \times {\rm U}(2)}_{{\rm SO}(2,4)} \times 
\underbrace{{\rm SU}(3) \times {\rm U}(1)}_{{\rm SO}(6)}\times {\rm SO}(2)
\,,
\label{eq:allGroups}
\end{equation}
of which the first two factors are embedded into the AdS$_5$ isometry group whereas the next two factors are embedded into the ${\rm SO}(6)$ isometry group of the round sphere $S^5$. The last factor of the group (\ref{eq:allGroups}) is the ${\rm SO}(2)\subset {\rm SL}(2)$ duality group of IIB supergravity. The ${\rm SU}(3) \times {\rm U}(1)$ organizes the Kaluza-Klein modes in the compactification from IIB to $D=5$. In turn, the ${\rm U}(2)$ labels the Kaluza-Klein modes in the compactification from $D=5$ to AdS$_2$, by truncating to its singlets we project out all the Kaluza-Klein modes on the squashed $S^3$ in (\ref{AdS2NH}).

In order to find the representation content of the full Kaluza-Klein spectrum, it is most convenient to start from the known spectrum on AdS$_5\times S^5$, summarized in table~\ref{tab:S5}, and to break the ${\rm SO}(2,4) \times {\rm SO}(6)$ representations down to (\ref{eq:allGroups}). We have seen in the general fluctuation equations (\ref{eq:FlucTensors})--(\ref{ScEq}) that for vanishing background scalar fields ${\cal M}_{MN}=\delta_{MN}$ they carry (among other terms) the mass operators derived in section~\ref{subsec:KKmass} for the AdS$_5\times S^5$ background. Since these operators enter the fluctuation equations around the black hole near-horizon geometry, it will be useful to tabulate their eigenvalues as well. These have been derived in \cite{Kim:1985ez,Gunaydin:1984fk} directly from IIB supergravity, and in the present framework in \cite{Malek:2020yue}. Explicitly, they are encoded in the conformal dimensions $\Delta$ given in the first column of table~\ref{tab:S5}, with the masses related as
\begin{align}
& \Delta = 2 + \sqrt{4+M^2 \ell_{{\rm AdS}_5}^2} \quad \mbox{(spin 0 and spin 2)}
\,,\nonumber\\
&  \Delta = 2 + \sqrt{1+M^2 \ell_{{\rm AdS}_5}^2}\quad \mbox{(vectors)}\,,\qquad
\Delta = 2 +\left| M \ell_{{\rm AdS}_5} \right|\quad \mbox{(tensors)}\,.
\end{align}

\begin{table}[bt]
\centering
{\footnotesize
\begin{tabular}{|c|l|} 
    \hline
     $\Delta$ &\\ \hline\hline
     $n+2$ & $[n+2,00]{(0\,0)}$ \\
          $n+3$ & $[n,02]{(0\,0)}\oplus[n,20]{(0\,0)}
\oplus[n+1,00]{(0\,1)}\oplus[n+1,00]{(1\,0)}
\oplus[n,11]{(\frac12\,\frac12)}$ \\
     $n+4$ & $2\!\cdot\![n,00]{(0\,0)}\oplus[n-2,22]{(0\,0)}
+[n-1,02]{(0\,1)}\oplus[n-1,20]{(1\,0)}\oplus2\!\cdot\![n-1,11]{(\frac12\,\frac12)}\oplus[n,00]{(1\,1)}$
\\
     $n+5$ & $[n-2,02]{(0\,0)}\oplus[n-2,20]{(0\,0)}
\oplus[n-1,00]{(0\,1)}\oplus[n-1,00]{(1\,0)}
\oplus[n-2,11]{(\frac12\,\frac12)}$ \\
     $n+6$ & $[n-2,00]{(0\,0)}$ \\
\hline
\end{tabular}
}
\caption{Bosonic spectrum on AdS$_5\times S^5$ in ${\rm SO}(6)\times{\rm SO}(4)$ notation $[n_1,n_2,n_3](j_1\,j_2)$
with ${\rm SO}(6)$ Dynkin labels $n_i$, and $(j_1\,j_2)$ denoting the spins of ${\rm SU}(2)\times {\rm SU}(2)\sim{\rm SO}(4)\subset
{\rm SO}(2,4)$. Specifically, these are AdS$_5$ scalars $(0\,0)$, vectors $(\frac12\,\frac12)$, self-dual tensors $(1\,0)\oplus(0\,1)$ and spin-2 modes $(1\,1)$. For $n=0$, this is the bosonic spectrum of $D=5$ maximal supergravity.}
\label{tab:S5}
\end{table}

As discussed above, we will restrict the analysis here to the `simple' sectors of fields whose ${\rm SU}(3) \times {\rm U}(1)\times {\rm SO}(2)$ representation only appears once in the full Kaluza-Klein spectrum, such that they cannot couple to any other field in their linearized field equations. We will use the notation
\begin{equation}
{\cal R} = [m_1,m_2]_r^Q
\,,
\label{eq:not-rep}
\end{equation} 
to specify their representation content, with ${\rm SU}(3)$ Dynkin labels $m_i$ and $r$ and $Q$ denoting the (integer) charges of ${\rm SO}(2)$ and ${\rm U}(1)$, respectively. Since we restrict to singlets under the ${\rm U}(2)$ factor of (\ref{eq:allGroups}), there is no need to specify its quantum numbers. In the following, we discuss these fields separately in the different $D=5$ spin sectors. We show that in all these sectors, the fluctuation equations eventually reduce to the standard form (\ref{eq:KG2D}) and collect the AdS$_2$ mass and charge values for each mode. The results are summarized in tables \ref{tab:simplecomparison}, \ref{tab:simpletensors}, and \ref{tab:simplevectors} below.

\subsubsection*{Simple scalars}

Upon breaking the spectrum of table~\ref{tab:S5} down to (\ref{eq:allGroups}), we can identify the $D=5$ scalar fields whose representation appears precisely once in the spectrum. At Kaluza-Klein level $n$, we find the following tower of such `simple scalars' 
\begin{equation}
  \bigoplus_{k=0}^{n} 
    [k+2,n-k]^{2k-n-1}_{+} \oplus \,[k,n-k+2]^{2k-n+1}_{-} \oplus [k,n-k]^{2k-n}_{\pm2} 
  \,,
  \label{eq:simp1}
\end{equation}
together with a few additional modes
\begin{equation}
  [0,n+2]^{-n-2}_0 \oplus  [n+2,0]^{n+2}_0
  \oplus
  [n,0]^{n+3}_{+} \oplus [0,n]^{-n-3}_{-} 
  \,.
  \label{eq:simp2}
\end{equation}
For these scalars, the fluctuation equations \eqref{ScEq} simplify to
\begin{equation}
      {D}^\mu {D}_\mu m_{MN} - M_{(0)}^2{}_{MN}{}^{KL} \, m_{KL} -\bg{F}_{\mu\nu}\bg{F}^{\mu\nu}\,\mathbb{Q}_{MN}{}^{KL}\,m_{KL}
 =0
 \,,
 \label{eq:eomsimple}
 \end{equation}
where we have used that the background current vanishes $J_\mu{}^M{}_N=0$, and defined the matrix
\begin{equation}
    \mathbb{Q}_{MN}{}^{KL} = 3\,\bg{\delta}_{P(M} \mathbb{P}_{N)}{}^{P}{}_{Q}{}^{K}{C}^{Q}{C}^{L} + 3\, \delta_{M}{}^{K}\mathbb{P}_{N}{}^{L}{}_{P}{}^{Q} {\delta}_{QR}\,{C}^{R}{C}^{P}\,,
     \label{eq:Paulioperator}
\end{equation}
called Pauli-couplings in \cite{Ezroura:2024xba}.
The covariant derivatives are given by (\ref{eq:covDQ}) with the U(1) charge~$Q$ as indicated in (\ref{eq:not-rep}).
For the simple scalars~(\ref{eq:simp1}) and~(\ref{eq:simp2}), we find that all the corresponding eigenvalues of $\mathbb{Q}_{MN}{}^{KL}$ vanish. The eigenvalues of the mass operator $M_{(0)}^2{}_{MN}{}^{KL}$ correspond to the masses of the corresponding ${\rm SO}(6)$ representations, collected in the second column of table~\ref{tab:simplecomparison}.

\begin{table}[tb]
    \centering
    \hspace*{-1cm}
    \begin{tabular}{|c|c|c|}
    \hline
      ${\cal R} = [m_1,m_2]_r^Q$  &  $\big(M\ell_{{\rm AdS}_{5}}\big)^{2}$ &  Level 0, label from \cite{Ezroura:2024xba}
       \\\hline\hline
      $ [0,n+2]^{-n-2}_0 \oplus  [n+2,0]^{n+2}_0$ & $n^2-4$  &  $t^+:\,  [2,0]^{+2}_0 \oplus [0,2]^{-2}_0  $  \\ \hline
         $\displaystyle \bigoplus_{k=0\vphantom{q}}^{n\vphantom{X}} \, [k,n-k]^{2k-n}_{\pm2} $  & $n(n+4)$ & $\Lambda_\alpha{}^\beta:\,[0,0]^{0}_{\pm2}$  \\ \hline
        $ [n,0]^{n+3}_{+} \oplus [0,n]^{-n-3}_{-} $  & $(n-1)(n+3)$ &  $\varphi^{\textbf{1}}:\,[0,0]_\pm^{\pm3}$ \\
       \hline 
       $\displaystyle \bigoplus_{k=0\vphantom{q}}^{n\vphantom{X}}\, [k+2,n-k]^{2k-n-1}_{+} \oplus \,[k,n-k+2]^{2k-n+1}_{-}$   & $(n-1)(n+3)$ &  $\varphi^{\textbf{6}}:\,
      [2,0]^{-}_{+} \oplus \,[0,2]^{+}_{-}$ \\
 \hline
    \end{tabular}
    \caption{Simple scalars (\ref{eq:simp1}), (\ref{eq:simp2}) with the mass $M$ determined by identifying the ${\rm SO}(6)$ origin of the modes within table~\ref{tab:S5}. Recall, that in our normalization $\ell_{{\rm AdS}_5} = 1$, c.f.\ (\ref{eq:lAdS5}). The level $n=0$ sector of the spectrum has been considered in \cite{Ezroura:2024xba}.}
    \label{tab:simplecomparison}
\end{table}

For scalar fields living only on the AdS$_2$ spacetime, the resulting $D=5$ Klein-Gordon equation (\ref{eq:eomsimple}) reduces to an AdS$_2$ Klein-Gordon equation (\ref{eq:KG2D}). With the background metric (\ref{AdS2NH}) and gauge field (\ref{eq:NHvector}), we find the AdS$_2$ mass and charge given by
\begin{equation}
q=\frac{F-EG}{AB}\,Q\,,\qquad
m^2 = M^2 +\frac{4G^2}{D^2}\,Q^2
\,,
\label{eq:QMscalar}
\end{equation}
in term of the $D=5$ quantities $M$, $Q$ from table~\ref{tab:simplecomparison}, and the constants given by (\ref{NHlimitcoeff}) as functions of 
the parameters $r_+$ and $a$.

\subsubsection*{Simple tensors}

The `simple tensors', i.e.\ tensors in representations that only appear once in the full spectrum are collected in table~\ref{tab:simpletensors}, together with the eigenvalues $M^2$ of the mass operator (\ref{eq:MassTensorAdS}) that can be inferred from the ${\rm SO}(6)$ origin of these modes in table~\ref{tab:S5}. In the near-horizon background, the tensor fluctuation equations (\ref{eq:FlucTensors}) read
\begin{equation}
     {\cal Z}^{LM} \,\Bigg[\frac{\sqrt{10}}{2}\, \bg{\epsilon}^{\mu\nu\rho\sigma\tau}\,\bg{\nabla}_{\rho} b_{\sigma\tau\, M} - \delta_{MN}\,{\cal Z}^{NP}\,b^{\mu\nu}{}_{P}\Bigg]
     = 0\,.
    \label{eq:lineomsimpletensor}
\end{equation}
This is the equation of charged massive tensor fluctuations, with an anti-hermitean mass operator which squares to (\ref{eq:MassTensorAdS}). For a mass eigenmode of this operator, the fluctuation equation reduces to
\begin{equation} \label{eq:eomsimpletensorsinglet}
    3\,D_{[\mu}b_{\nu\rho]} = \frac{iM}{2}\,\bg{\varepsilon}_{\mu\nu\rho\sigma\tau}\,b^{\sigma\tau}\,,
\end{equation}
where again the $D=5$ covariant derivative (\ref{eq:covDQ}) carries the U(1) charge $Q$.

\begin{table}[tb]
\centering
	\begin{tabular}{|c|c|c|}
    \hline
		  ${\cal R} = [m_1,m_2]_r^Q$  & $\big(M\ell_{{\rm AdS}_{5}}\big)^{2}$ &
		   Level 0  \\\hline\hline
			$[0,n-1]^{-n-2}_0 \oplus [n-1,0]^{n+2}_0$ &$(n+2)^2$ & --- \\ 
            \hline
		$[0,n+1]^{-n-1}_+ \oplus [n+1,0]^{n+1}_- $ & $(n+1)^2$  &
		$[0,1]^{-1}_+ \oplus [1,0]^{+1}_- $
		\\
        \hline
	\end{tabular}
	\caption{Simple tensors  with the mass $M$ determined by identifying the ${\rm SO}(6)$ origin of the modes within table~\ref{tab:S5}. Recall, that in our normalization $\ell_{{\rm AdS}_5} = 1$, c.f.\ (\ref{eq:lAdS5}).}
	\label{tab:simpletensors}
\end{table}

It remains, to reduce this equation to AdS$_2$ equations upon truncation the fluctuations to singlets under the ${\rm U}(2)$ of (\ref{eq:allGroups}). This corresponds to a parametrization of the $D=5$ two-form mode as
\begin{equation}
    b = \chi_{(2)} + \chi_{(1)}\wedge\sigma_{3} + \chi_{(0)}\,\sigma_{1}\wedge\sigma_{2}\,,
\end{equation}
with the AdS$_2$  forms
\begin{equation}
    \chi_{(2)} = \chi_{tr}(t,r)\, {\rm d} t\wedge {\rm d} r\,, \quad \chi_{(1)} = \chi_{t}(t,r)\, {\rm d} t+\chi_{r}(t,r)\, {\rm d} r\,, \quad {\rm and} \quad \chi_{(0)} = \chi(t,r)\,.
\end{equation}
With this ansatz, the fluctuation equation (\ref{eq:eomsimpletensorsinglet}) reduces to a set of AdS$_2$ covariant equations, which algebraically determine the forms $\chi_{(1)},  \chi_{(2)}$ as
\begin{equation}
    \begin{aligned}
        \chi_{u} &= \frac{4D^2}{4D^{2}+C^{4}M^{2}}\Big((\bg{\nabla}_u -iqA_{2,u})\chi
        + \frac{i\,MC^{2}}{2D}\,\bg{\epsilon}_{uv}
        (\bg{\nabla}^{v}-iqA_{2}^v)\chi\Big)\,,\\
        \chi_{tr} &= \frac{E}{AB}\,\bg{A}_{2,t}\,\chi_{r} -\frac{8ABG}{C^{2}D}\,\frac{Q}{M}\,\chi
        \,,
    \end{aligned}
    \label{eq:tensors12}
\end{equation}
with the constants from (\ref{NHlimitcoeff}), and the AdS$_2$ charge given by
\begin{equation}
q=\frac{F-EG}{AB}\,Q\,,
\label{eq:Qtensor}
\end{equation}
as in (\ref{eq:QMscalar}).
Plugging equations (\ref{eq:tensors12}) back into the remaining equations 
shows that the remaining mode $\chi_{(0)}$ satisfies the AdS$_2$ Klein-Gordon equation (\ref{eq:KG2D}) with charge (\ref{eq:Qtensor}) and mass 
\begin{equation}
    m^{2} = M^{2}  +\frac{4G^{2}}{D^{2}} \, Q^{2}+ \frac{4D^{2}}{C^{4}}+ \frac{2D\big(F-EG\big)}{ABC^{2}}\,\frac{Q}{M}+\frac{16G^{2}}{C^{4}}\,\frac{Q^{2}}{M^{2}}\,.
\label{eq:Mtensor}
\end{equation}
We note that
the $D=5$ mass $M$ is non-zero for all physical $D=5$ tensor fields (otherwise they are projected out from the spectrum), so the expression is well defined for all tensors.

\subsubsection*{Simple vectors}

We finally turn to the `simple vectors', thus the physical vectors whose representation appears only once in the full physical spectrum, thereby excluding all couplings to other fields.  In table~\ref{tab:simplevectors}, we collect the field content of these vectors together with the eigenvalues $M^2$ of the mass operator~(\ref{VectorMass}) that are inferred from the ${\rm SO}(6)$ origin of these modes in table~\ref{tab:S5}. 
In the general formulas, the simple vectors show up via the field strengths
\begin{equation}
    f_{\mu\nu}{}^M=2\,\bg{D}_{[\mu} a_{\nu]}{}^M
   = 2\,\partial_{[\mu} a_{\nu]}{}^M 
   - i\,Q\,\bg{A}_{[\mu}  a_{\nu]}{}^N\,,
\end{equation}
 and within the currents of the Goldstone scalars 
\begin{equation}
    j_{\mu,MN} = 
    \bg{D}_\mu \varphi_{MN}
    + \Pi_{MN,K}\,a_\mu{}^K
    \,,
\end{equation}
using the $\Pi$ operator defined in (\ref{PiOp}).
The fluctuation equations \eqref{eq:vectoreomfluct} for these simple vectors reduce to
\begin{align}
     0 \ = \  &   
  \delta_{MN}\,\Big( D_{\nu}  f^{\mu\nu\,N}
   +i\,Q\,\bg{F}^{\mu \nu}  \, a_\nu{}^{N} 
  \Big)
          - \frac{\sqrt{10}}{4} \,\bg{\epsilon}^{\mu\kappa\lambda\rho\sigma}\,\bg{{F}}_{\kappa\lambda}{}\, {C}^{K}\,d_{MKL} \, f_{\rho\sigma}{}^{L} 
   \;+ {C}^N m_{MN}\, \bg{\nabla}_{\nu}  \bg{F}^{\mu\nu}           \nonumber\\
         &{}           
          +
            \frac{1}{12}\,
          (\Pi^\dagger)_{M,KL}\, j^{\mu}{}^{KL}
          + \bg{{F}}^{\mu \nu}{}\, {C}^{N} \,  j_\nu{}_{MN}    
         \; .
         \label{eq:flucSimpVec}
\end{align}
After gauging away the Goldstone scalars and observing that the operator $C^N\,\Pi_{MN,P}$ vanishes for simple vectors, we can rewrite this equation as
\begin{equation}  
   \bg{\nabla}_{\nu}  f^{\mu\nu\,M}
           - \frac{1}{4} \,\bg{\epsilon}^{\mu\kappa\lambda\rho\sigma}\,\bg{{F}}_{\kappa\lambda}{} \, f_{\rho\sigma}{}^{M}
  +i\, Q \,\bg{F}^{\mu \nu}  \, a_\nu{}^{M} 
-   ({\cal M}_V)^M{}_{N}\, a^{\mu\,P} 
= 0
         \,,
         \label{eq:flucSimpVecSimp}
\end{equation}
with the positive definite mass operator $ ({\cal M}_V)^M{}_{N}$ from (\ref{VectorMass}) whose eigenvalues $M^2$ are collected in table~\ref{tab:simplevectors}.

\begin{table}[tb]
 \centering
    \hspace*{-1cm}
	\begin{tabular}{|c|c|c|}
    \hline
		${\cal R} = [m_1,m_2]_r^Q$  & $\big(M\ell_{{\rm AdS}_{5}}\big)^{2}$ & Level 0\\\hline\hline
		$[1,n]^{-n-2}_0 \oplus [n,1]^{n+2}_0$ &$n(n+2)$ & 
		$[1,0]^{-2}_0 \oplus [0,1]^{+2}_0$\\
    \hline
		$\displaystyle \bigoplus_{k=0\vphantom{q}}^{n-1\vphantom{X^X}} \,[k,n-k]^{2k-n+3}_{-} 
		\;\;\oplus\; \bigoplus_{k=1\vphantom{q}}^{n\vphantom{X}} \,[k,n-k]^{2k-n-3}_{+} $ & $(n+1)(n+3)$  & --- \\ 
       \hline
	\end{tabular}
        	\caption{Simple vectors with the mass $M$ determined by identifying the ${\rm SO}(6)$ origin of the modes within table~\ref{tab:S5}. Recall, that in our normalization $\ell_{{\rm AdS}_5} = 1$, c.f.\ (\ref{eq:lAdS5}).}
	\label{tab:simplevectors}
    \end{table}

Finally, we perform the reduction around AdS$_2$, retaining the ${\rm U}(2)$ singlets of the $D=5$ vector field by an ansatz
\begin{equation}
a = a_t(t,r) dt + a_r(t,r) dr  + \phi(t,r)\,\Big(\sigma_3+E(r-r_+)\,dt\Big)
\,.
\end{equation}
We further define a scalar $\varphi(t,r)$ by contraction of the two-dimensional field strength
\begin{equation}
\varphi = \epsilon^{uv}\,(\partial_{u} a_{v} - iqA_{2,u}a_v)
\,,
\end{equation}
where the $D=2$ charge $q$ is again related to the $D=5$ charge $Q$ by (\ref{eq:Qtensor}).
Straightforward computation then shows, that the $D=5$ fluctuation equations (\ref{eq:flucSimpVecSimp}) yield a coupled system of AdS$_2$ scalar equations for the scalars $\phi$ and $\varphi$, that can be diagonalized into eigenmodes
\begin{equation}
\phi_\pm = 
C^2 D \Big(ABG\,\varphi +(F-2 EG)\, \phi\Big)+8 ABG^2\, \phi
   \pm
    \sqrt{16 A^2 B^2 C^4 G^2 \,M^2+\left(C^2 DF-8 ABG^2\right)^2}\,\phi
\,,    
\end{equation}
satisfying the AdS$_2$ Klein-Gordon equation (\ref{eq:KG2D}) with charge (\ref{eq:Qtensor}) and masses given by
\begin{align}
    m^2_\pm \ = \ & 
    M^2 +\frac{4G^2}{D^2}\,Q^2
    +\frac{\left(C^2 D E-8 ABG\right)
   \left(C^2 DF-8 ABG^2\right)}{8
   A^2 B^2 C^4 G}
   \nonumber\\
   &{}
\pm\,   \frac{\left(C^2 D E-8 ABG\right)
   \sqrt{16 A^2 B^2 C^4 G^2 \,M^2+\left(C^2 DF-8 ABG^2\right)^2}}{8 A^2
   B^2 C^4 G}
   \,,
\end{align}
in terms of the constants from (\ref{NHlimitcoeff}).
With the explicit parametrization of (\ref{NHlimitcoeff}), the different terms simplify to
\begin{align}
  \frac{\left(C^2 D E-8 ABG\right)
   \left(C^2 DF-8 ABG^2\right)}{8
   A^2 B^2 C^4 G} \ = \ &
   \frac{4}{\sqrt{1+2 (a^2+r_+^2)}\,-1}
   \,,
   \nonumber\\
 \frac{\left(C^2 D E-8 ABG\right)
   \sqrt{16 A^2 B^2 C^4 G^2 \,M^2+\left(C^2 DF-8 ABG^2\right)^2}}{8 A^2
   B^2 C^4 G}
    \ = \ &
     \frac{4\,\sqrt{1+a^2M^2}}{\sqrt{1+2 (a^2+r_+^2)}\,-1}
   \,.
\end{align}

Let us finally note that  for $M=0$, the $D=5$ equation (\ref{eq:flucSimpVecSimp}) still exhibits gauge invariance. Closer inspection shows that in this case the $m_-$ mode can be gauged away, such that the physical spectrum carries a single scalar of mass $m_+$.

\subsection{Summary and stability plots}
\label{BFStudy}

In the previous sections, we have worked out the mass and charge spectra for all `simple' sectors of fields. These include infinitely many fluctuations distributed over all Kaluza-Klein levels $n$. For the fields of different $D=5$ origin, we have derived the explicit formulas which determine their AdS$_2$ masses and charges $\{m, q\}$ in terms of their $D=5$ masses and charges $\{M, Q\}$, and the parameters (\ref{NHlimitcoeff}) of the black hole near-horizon background. For all these fluctuations, we may thus test their stability by computing the charged BF bound (\ref{StabCond}) as a function of the background parameters $(r_+,a)$. Following the example of \cite{Ezroura:2024xba}, we can then plot, for a given fluctuation,  its stability and instability regions in the two-dimensional parameter space. As a useful consistency check of our computations, we also plot in these graphs the curve (\ref{eq:NHsusy}) where the solution becomes supersymmetric and stability is expected by general arguments. Indeed, this curve does not intersect any of the identified instability regions, however it turns out tangent to several of these regions.

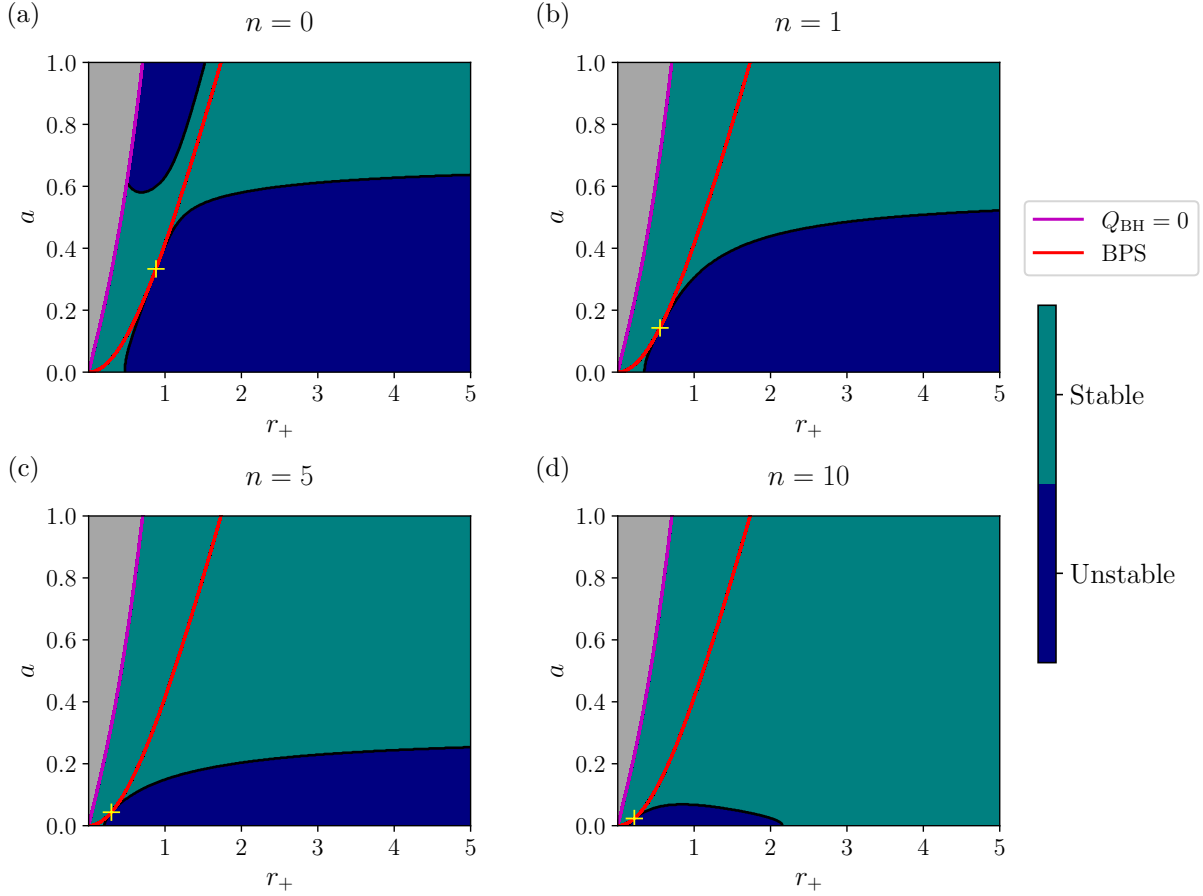
\begin{figure}[t!]
    \centering
    \begin{tikzpicture}
        \draw (0,0) node (n0) {\includegraphics[scale=0.65]{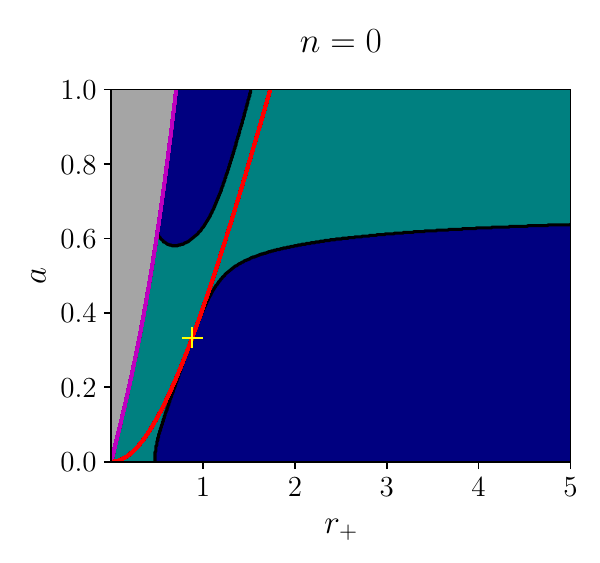}};
        \draw ($(n0.north west)+(0.5,0-.5)$) node {\small (a)};
        
        \draw ($(n0)+(7,0)$) node (n1) {\includegraphics[scale=0.65]{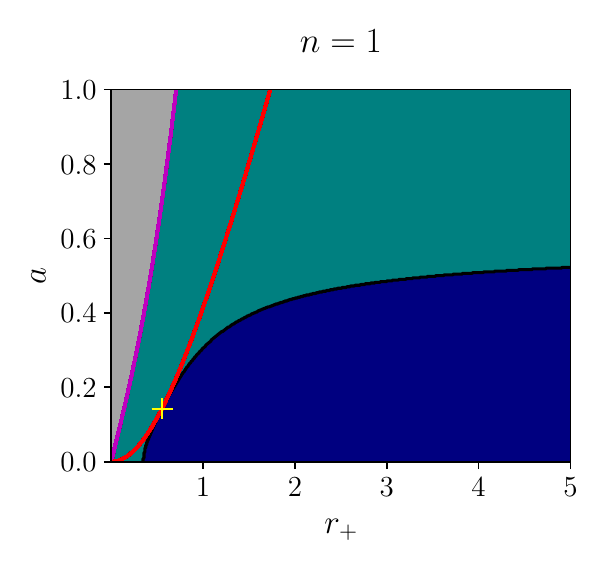}};
        \draw ($(n1.north west)+(0.5,0-.5)$) node {\small (b)};
        
        \draw ($(n0)+(0,-6)$) node (n5) {\includegraphics[scale=0.65]{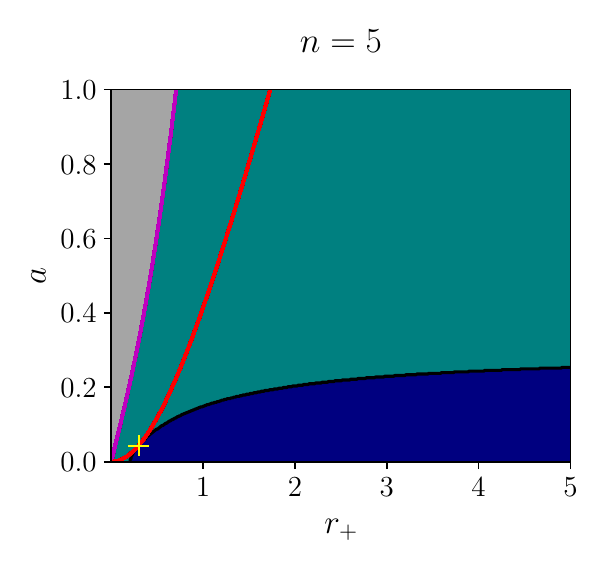}};
        \draw ($(n5.north west)+(0.5,0-.5)$) node {\small (c)};
        
        \draw ($(n0)+(7,-6)$) node (n10) {\includegraphics[scale=0.65]{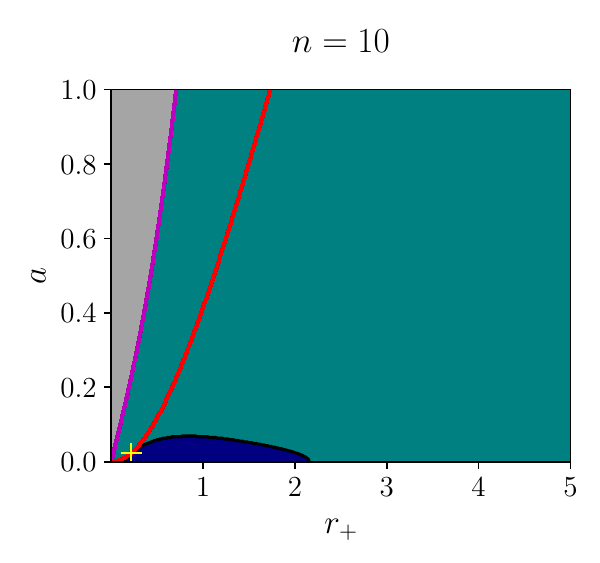}};
        \draw ($(n10.north west)+(0.5,0-.5)$) node {\small (d)};
        
        \draw ($(n0)+(11.4,-0.25)$) node (legend) {\includegraphics[scale=0.75]{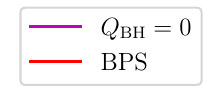}};
        
        \draw ($(n0)+(11.4,-3.5)$) node (n10) {\includegraphics[scale=0.75]{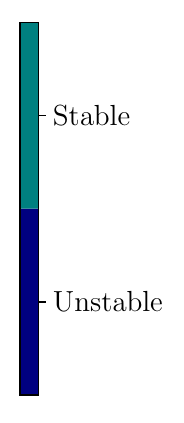}};
    \end{tikzpicture}
    \vspace{-0.6cm}
    \caption{Instability regions for the scalars $t^+$ at Kaluza-Klein levels (a) $n=0$, (b) $n=1$, (c) $n=5$ and (d) $n=10$ in the $(r_{+},a)$ plane. The five-dimensional masses $M$ and charges $Q$ of these modes can be found in table~\ref{tab:simplecomparison}. The BPS line $r_+=\sqrt{a(2+a)}$, in red, is tangent to the BF instability region at the points~\eqref{eq:tangent}, depicted by yellow crosses. The purple line corresponds to the saturation of the left equation in~\eqref{eq:BHparam}. The grey region is unphysical, c.f.~(\ref{eq:BHparam}).}
    \label{fig:Stabilitytp}
\end{figure}

As an illustration, we present these stability plots for some of the fluctuations, focusing on examples with non-trivial instability regions in the parameter space. In particular, we restrict to the sector of simple scalars, since we find that simple vectors and tensors are stable for any level and throughout the entire parameter space.
In figure \ref{fig:Stabilitytp} we show the stability regions for the $t^+$ scalars (see table \ref{tab:simplecomparison}) at various Kaluza-Klein levels. The dark blue region shows the region of the parameter space where the charged BF bound (\ref{StabCond}) is violated ($1+4\,m^2 \ell_{{\rm AdS_2}}^2 < 4\,q^2 \ell_{{\rm AdS_2}}^4$) whereas the BF bound is satisfied in the teal region. The red curve is the supersymmetric (BPS) locus (\ref{eq:NHsusy}) which consistently never crosses into the instability region. The grey region is excluded from the parameter space, c.f.\ (\ref{eq:BHparam}). For $n=0$, we recover the stability analysis of \cite{Ezroura:2024xba}. We have marked by a yellow cross the point in which the BPS locus becomes tangent to the instability region. 

\begin{figure}[t]
    \centering
    \centerline{
    \subfigure{\label{fig:speccomplete}
    \begin{tikzpicture}
        \draw (0,0) node (complete) {\includegraphics[scale=0.7, clip, trim =0 0 3cm 0]{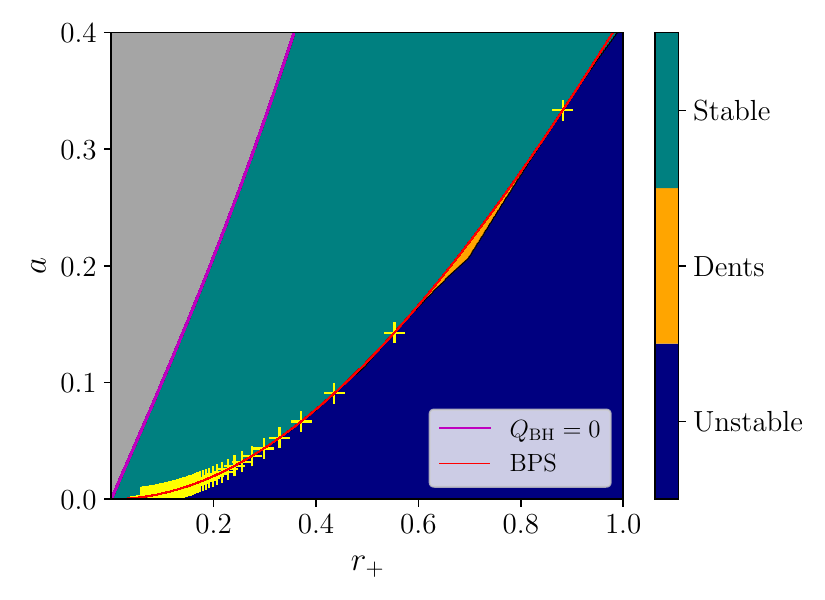}};
    	\draw ($(complete.north west)+(0.5,0)$) node {\small (b)};
    % 	\draw [yellow, fill=yellow] (-2.965,-2.36) circle [radius=1pt];
    % 	\draw [yellow, fill=yellow] (-2.64,-2.36) circle [radius=1pt];
    % 	\draw [yellow, fill=yellow] (-2.965,-0.15) circle [radius=1pt];
    	    
    	\draw (-7.75,0) node (all) {\includegraphics[scale=0.7, clip, trim =0 0 3cm 0]{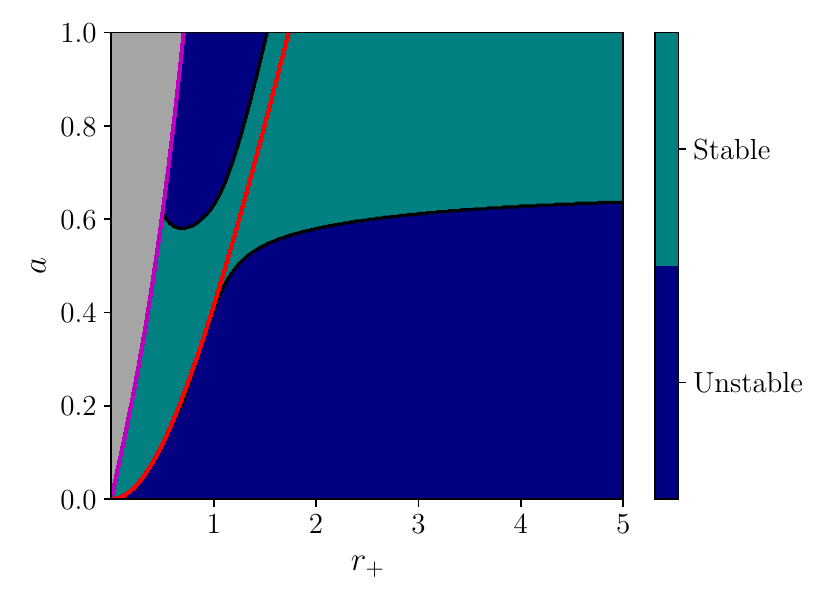}};
    	\draw ($(all.north west)+(0.5,0)$) node {\small (a)};

        \draw [very thick, dashed, gray!20,shift={(-7.75,0)}] (-2.51,-2.36) rectangle (-1.34,-0.15);
    
    	\draw (5,0) node {\includegraphics[scale=0.7, clip, trim =11cm 0 0 0]{plots/Plot_All_Stable_tp_camille.pdf}};
    \end{tikzpicture}
    }
    }
    \subfigure{\label{fig:speczoom}}
    \vspace{-1.2cm}
    \caption{(a) Instability region for the whole Kaluza-Klein tower of $t^{+}$ scalars. (b) Zoom in to the lower left corner (see the dashed rectangle), showing ``stability dents'' below the BPS locus between each tangency points~\eqref{eq:tangent}, drawn as yellow crosses.}
    \label{fig:StabilityAll}
\end{figure}

Perturbative BF stability of course is an all-level statement. In order for the background to be perturbatively stable for given values $(r_+,a)$, all fields at all Kaluza-Klein levels must satisfy the BF bound. 
With growing Kaluza-Klein level $n$, the instability region shrinks (as naively expected), however 
it shifts under the BPS locus towards the lower left corner and keeps this tangent point at
\begin{equation} 
a=-1+\sqrt{1+r_+^2} =  \frac{1}{3+4n}\,,
\label{eq:tangent}
\end{equation}
which goes to 0 as $n\rightarrow\infty$. The modes at higher Kaluza-Klein levels thus gradually destabilize the region below the BPS locus. 
In figure \ref{fig:speccomplete}, we superpose the instability regions for the $t^+$ scalars from all Kaluza-Klein levels in order to exhibit the leftover teal region in which all fluctuations are BF stable. Also, all other simple scalars which we have analyzed remain stable in this region.
Zooming in to the lower left corner, we note that the jumps of the tangent points~(\ref{eq:tangent}) are discrete because the level index $n$ is an integer. As a result, this leaves small ``stability dents'' between two successive tangent points where the scalar field remains stable below the BPS locus. This becomes visible in the zoomed figure~\ref{fig:speczoom}.

Verifying full perturbative stability of the background in the region exhibited in figure \ref{fig:speccomplete} would require to extend this analysis beyond the simple sectors and including all Kaluza-Klein fluctuations.

\section{Conclusions and Outlook} \label{sec:ccl}

In this paper we have developed the perturbation theory of ${\rm E}_{6(6)}$ exceptional field theory
around generalized Scherk-Schwarz backgrounds to first order in fluctuations. 
These fluctuations encode in particular all massive Kaluza-Klein modes of ten- or eleven-dimensional supergravity. Moreover, we presented, for a subset of backgrounds, the details of the Higgs mechanism rendering the higher Kaluza-Klein modes massive, i.e., the rearrangement of the generally gauge redundant fields into gauge invariant or physical modes and pure gauge or St\"uckelberg modes. We employed the framework of homotopy algebras such as $L_{\infty}$ algebras, in which the passing over to gauge invariant field variables can be interpreted as homotopy transfer. 
While in previous studies for AdS backgrounds the mass spectra were determined in an ad hoc 
fashion by discarding certain eigenvalues of the naive mass matrices, 
the homotopy transfer interpretation allowed us, for the first time, to determine the complete mass matrices in which the projection operators that are part of the homotopy data enter. 
Finally, as a first application for more general backgrounds, we analyzed a Kerr-Newman black hole that in the near horizon regime becomes a (fibred) product of AdS$_2$ and a squashed three-sphere. This background is special because it switches on a U$(1)$ gauge field. We determined part of the mass spectrum of Kaluza-Klein modes corresponding to all ten dimensions of type IIB supergravity and studied the perturbative stability of fluctuations around the background.

 \medskip

These results should be extended along the following lines of research.

\begin{itemize}
    \item 
    The analysis of the Higgs mechanism via homotopy transfer 
    should be extended  to all generalized Scherk-Schwarz backgrounds in 
    ${\rm E}_{6(6)}$ exceptional field theory. 
    This requires including background covariant derivatives and $p$-form curvatures, 
    which can plausibly be achieved by perturbing around the present analysis where $p$-form curvatures vanish in the background. 
    This would allow us to determine mass matrices for a 
    very general class of backgrounds, even though the notion of mass needs to be defined on a case by case basis.
     \item 
     The analysis of the Kaluza-Klein spectrum for the black hole background was 
     so far restricted to a small subsector, where the mass matrices can be inferred 
     form a naive analysis without a detailed understanding of the Higgs mechanism.
     It remains to complete this black hole analysis  by including all modes up 
     to a given Kaluza-Klein level. 
    \item 
    So far our analysis is restricted to backgrounds that have an origin in five-dimensional 
    gauged supergravity such as AdS$_5\times S^5$. It will be important to generalize 
    this to other dimensions and ExFTs, notably the E$_{7(7)}$ theory carrying  
    the AdS$_4\times S^7$ vacuum of eleven-dimensional supergravity, but also to those with three external dimensions such as the E$_{8(8)}$ ExFT, which could be used to describe the warped ${\rm AdS}_{3}$ solutions of \cite{Deger:2024xnd,Maurelli:2025iba}, or to theories with less than 
    maximal supersymmetry.
    \item 
    The main reason that the homotopy transfer interpretation of the Higgs mechanism 
    is powerful, apart from 
    enabling the Higgs analysis  in the first place for general backgrounds, is that it becomes 
    a tool for higher order 
    perturbation theory. Indeed, once the homotopy transfer data of the free theory has been 
    established, the homotopy transfer theorem provides an algorithm to determine the gauge invariant field variables and hence the physical modes to any order in 
    perturbation theory, which allows one to determine the $n$-point couplings of the 
    physical fields for any $n$. 
    This will provide a systematic framework extending and completing the initial ExFT results from \cite{Duboeuf:2023cth}.
    It constitutes the first step for the core AdS/CFT application: to compute boundary 
    observables via Witten diagrams, which are conjectured to yield the dual 
    CFT correlation functions. Remarkably, this step also has an interpretation 
    in terms of homotopy transfer to the boundary of, say, AdS$_5$ \cite{Chiaffrino:2023wxk}.

\end{itemize}

 \section*{Acknowledgments} 

We thank Roberto Bonezzi and Christoph Chiaffrino for helpful discussions. 

\noindent
This work is funded by the Deutsche Forschungsgemeinschaft (DFG, German Research Foundation), ``Rethinking Quantum Field Theory", Projektnummer 417533893/GRK2575.   

\newpage
\appendix

\section{Useful formulas}
\label{app:Formulas}
We list here useful formulas and prove some of the identities we used in the main part of the paper.

\paragraph{}
The ${\rm E}_{6(6)}$ currents~\eqref{eq:Gamma} satisfy the following identities~\cite{Malek:2020yue}: 
\begin{equation} \label{eq:Gammad}
    \begin{aligned}
        \Gamma_{\underline{MN}}{}^{[\underline{K}} d^{\underline{L}]\underline{MN}} &= -\frac{1}{5}\,X_{\underline{MN}}{}^{\underline{K}} d^{\underline{LMN}}, \\
        \Gamma_{\underline{MN}}{}^{(\underline{K}} d^{\underline{L})\underline{MN}} &= -\frac{1}{2}\,\Gamma_{\underline{NM}}{}^{\underline{N}} d^{\underline{MKL}}.
    \end{aligned}
\end{equation}

The embedding tensor $X_{\underline{MN}}{}^{\underline{P}}$ has the following expression in terms of $Z^{\underline{MN}}$:
\begin{equation} \label{eq:usefulformXZ}
    \begin{aligned}
        X_{(\underline{MN})}{}^{\underline{P}} &= -\frac{1}{2}\,d_{\underline{MNQ}}\,Z^{\underline{PQ}}, \\
        X_{[\underline{MN}]}{}^{\underline{P}} &= -5\,d_{\underline{MKL}}d_{\underline{NRS}}d^{\underline{PKR}}\,Z^{\underline{LS}},
    \end{aligned}
\end{equation}
and satisfies  the useful identity
\begin{equation} \label{eq:usefulformula}
    \bigg[6\,Z^{\underline{PQ}}  \delta_{\underline{K}}{}^{\underline{L}} - 10\,d^{\underline{PLR}} \, X_{\underline{RK}}{}^{\underline{Q}} - 30\, d^{\underline{PQR}} \, X_{\underline{RK}}{}^{\underline{L}} \bigg] \, t_{\alpha\,\underline{Q}}{}^{\underline{K}} = 9\,Z^{\underline{LQ}}\,t_{\alpha\,\underline{Q}}{}^{\underline{P}},
\end{equation}
with $t_{\alpha\,\underline{M}}{}^{\underline{N}}$ the generators of the ${\rm E}_{6(6)}$ Lie algebra. This last formula has been checked using Mathematica and a generic parametrization of the embedding tensors $X_{\underline{KL}}{}^{\underline{P}}$ and $Z^{\underline{KL}}$.

\paragraph{}
The proof of the identity~\eqref{eq:IZ1} goes as follows:
\begin{equation}
    \begin{aligned}
        Z^{\underline{MN}}\dd_{\underline{N}} & \underset{\eqref{eq:Ztensor}}{=} -2\,d^{\underline{NPQ}}X_{\underline{PQ}}{}^{\underline{M}} \dd_{\underline{N}} \\
        & \underset{(\ref{eq:embeddingtensor})}{=} 2\,d^{\underline{NPQ}}\Gamma_{\underline{PQ}}{}^{\underline{M}} \dd_{\underline{N}} - 12\,d^{\underline{NPQ}}\mathbb{P}_{\underline{L}}{}^{\underline{K}}{}_{\underline{Q}}{}^{\underline{M}}\Gamma_{\underline{KP}}{}^{\underline{L}}\dd_{\underline{N}}+3\,d^{\underline{NPQ}}\mathbb{P}_{\underline{P}}{}^{\underline{L}}{}_{\underline{Q}}{}^{\underline{M}}\Gamma_{\underline{KL}}{}^{\underline{K}} \dd_{\underline{N}} \\
        & \underset{\substack{\eqref{eq:projadj}\\\eqref{eq:norm}}}{=} -5\,d^{\underline{MNP}}\Gamma_{\underline{KP}}{}^{\underline{K}} \dd_{\underline{N}} + 20\,d^{\underline{NPQ}}d^{\underline{MKR}}\,\Gamma_{\underline{KP}}{}^{\underline{L}}d_{\underline{LQR}}\,\dd_{\underline{N}} \\
        & \underset{\substack{\eqref{eq:Gamma}\\ \eqref{eq:tinv}}}{=} -5\,d^{\underline{MNP}}\Gamma_{\underline{KP}}{}^{\underline{K}} \dd_{\underline{N}} - 10\,d^{\underline{NPQ}}d^{\underline{MKR}}\,\Gamma_{\underline{KR}}{}^{\underline{L}}d_{\underline{LPQ}}\,\dd_{\underline{N}} \\
        & \underset{\eqref{eq:norm}}{=} -5\,d^{\underline{MNP}}\Gamma_{\underline{KP}}{}^{\underline{K}} \dd_{\underline{N}} - 10\,d^{\underline{MKR}}\,\Gamma_{\underline{KR}}{}^{\underline{N}}\,\dd_{\underline{N}} \\
        & \underset{\eqref{eq:Gammad}}{=} 5\,d^{\underline{MNP}}\Gamma_{\underline{KP}}{}^{\underline{K}} \dd_{\underline{N}} - 10\,d^{\underline{NPQ}}\,\Gamma_{\underline{PQ}}{}^{\underline{M}}\,\dd_{\underline{N}} \\
        & \underset{\substack{\eqref{eq:sectioncond}\\\eqref{eq:Gamma}\\\eqref{eq:GammaTr}}}{=} 0.
    \end{aligned}
\end{equation}
A similar computation can be done to prove that $\dd_{\underline{M}}Z^{\underline{MN}}=0$. Given this, it is easy to extend the proof to $\dd_{\underline{M}}\mathcal{Z}^{\underline{MN}}=0=\mathcal{Z}^{\underline{MN}}\dd_{\underline{N}}$ using the section constraint.

\paragraph{}
We now prove the equation (\ref{Nilpotentcyrel}). Let us first use the equation~\eqref{eq:embeddingtensor} to write the embedding tensor as $X_{\underline{MK}}{}^{\underline{L}} = X_{\underline{M}}{}^\alpha t_{\alpha\,\underline{K}}{}^{\underline{L}}$, such that the tensor $\Pi_{\underline{MN},\underline{K}}$ of equation~\eqref{PiOp} can be expressed as
\begin{equation}
    \Pi_{\underline{MN},\underline{K}} = 2\,M_{\underline{L}(\underline{M}}\,t_{\alpha\,\underline{N})}{}^{\underline{L}}\,\Pi_{\underline{K}}{}^\alpha, \quad {\rm with} \quad \Pi_{\underline{K}}{}^\alpha = X_{\underline{K}}{}^\alpha - 6\,\kappa^{\alpha\beta} t_{\beta\,\underline{K}}{}^{\underline{L}}\,\dd_{\underline{L}},
\end{equation}
with the inverse Cartan-Killing metric of ${\rm E}_{6(6)}$ $\kappa^{\alpha\beta}$. Now, with equation~\eqref{eq:Zoperator}:
\begin{equation}
    \begin{aligned}
        \Pi_{\underline{K}}{}^{\alpha}\mathcal{Z}^{\underline{KP}} &= \Big(X_{\underline{K}}{}^\alpha - 6\,\kappa^{\alpha\beta}\,t_{\beta\,\underline{K}}{}^{\underline{Q}}\,\dd_{\underline{Q}}\Big)\Big(Z^{\underline{KP}}-10\,d^{\underline{KPR}}\,\dd_{\underline{R}}\Big) \\
        &\underset{\eqref{eq:QCZ}}{=} 6\,Z^{\underline{PK}} \,\kappa^{\alpha\beta}\,t_{\beta\,\underline{K}}{}^{\underline{Q}} \,\dd_{\underline{Q}} -10\,X_{\underline{K}}{}^\alpha  d^{\underline{KPR}}\,\dd_{\underline{R}}+60\,\kappa^{\alpha\beta}\,t_{\beta\,\underline{K}}{}^{\underline{Q}}\,d^{\underline{KPR}} \,\dd_{\underline{Q}}\dd_{\underline{R}}\\
        &\underset{\substack{\eqref{eq:commd}\\\eqref{eq:tinv}\\\eqref{eq:IZ2}}}{=} 6\,Z^{\underline{PK}} \,\kappa^{\alpha\beta}\,t_{\beta\,\underline{K}}{}^{\underline{Q}} \,\dd_{\underline{Q}} -10\,X_{\underline{K}}{}^\alpha  d^{\underline{KPR}}\,\dd_{\underline{R}}+30\,\kappa^{\alpha\beta}\,t_{\beta\,\underline{K}}{}^{\underline{Q}}\,d^{\underline{KPR}} \,X_{\underline{QR}}{}^{\underline{S}} \,\dd_{\underline{S}} \\
        &\underset{\substack{\eqref{eq:usefulformXZ}\\\eqref{eq:IZ1}}}{=} \bigg[6\,Z^{\underline{PK}}  \delta_{\underline{Q}}{}^{\underline{L}} - 10\,d^{\underline{PLR}} \, X_{\underline{RQ}}{}^{\underline{K}} - 30\, d^{\underline{PKR}} \, X_{\underline{RQ}}{}^{\underline{L}} \bigg] \, \kappa^{\alpha\beta}\,t_{\alpha\,\underline{K}}{}^{\underline{Q}} \\
        & \underset{\eqref{eq:usefulformula}}{=}9\,Z^{\underline{LK}}\,\kappa^{\alpha\beta}\,t_{\alpha\,\underline{K}}{}^{\underline{P}}\,\dd_{\underline{L}} \\
        & \underset{\eqref{eq:IZ1}}{=} 0.
    \end{aligned}
\end{equation}

\newpage

\bibliographystyle{utphys}
\bibliography{refs}

\end{document}